\documentclass[twocolumn,showkeys,preprintnumbers,prd,amsmath,amssymb,aps,floats,floatfix]{revtex4}

\usepackage{mathrsfs}
\usepackage{epsfig}
\usepackage{graphicx}% Include figure files
\usepackage{dcolumn}% Align table columns on decimal point
\usepackage{bm}% bold math
\usepackage[usenames]{color}
\usepackage{tabularx}
\usepackage[caption=false]{subfig}
\usepackage{booktabs}

\def\be{\begin{equation}}
\def\ee{\end{equation}}
\def\bea{\begin{eqnarray}}
\def\eea{\end{eqnarray}}
\def\barr{\begin{array}}
\def\earr{\end{array}}
\def\ben{\begin{enumerate}}
\def\een{\end{enumerate}}

\def\bea{\begin{eqnarray}}
\def\eea{\end{eqnarray}}
\def\bsub{\begin{subequations}}
\def\esub{\end{subequations}}

\def\ttbar{t\bar{t}}

\newcommand{\mttb}{\ensuremath{m_{t\bar{t}}\, }}
\newcommand{\dyy}{\ensuremath{\left\vert\Delta y \right\vert\, }}
\newcommand{\afbt}{\ensuremath{A_{_{FB} }^{t \bar t}\, }}

\newcommand{\spincorr}{\ensuremath{C^{t \bar t}\, }}
\newcommand{\crtt}{\ensuremath{\sigma^{t \bar{t}}\, }}

\begin{document}
%\draft
%\baselineskip 0.7cm
\title{Top quark physics in the vector color-octet model} 
\author{Sukanta Dutta$^a$}%\email{sukanta.dutta@gmail.com}
\author{Ashok Goyal$^b$}\email{agoyal45@yahoo.com}
\author{Mukesh Kumar$^{a,b}$}\email{mkumar@physics.du.ac.in}
\affiliation{$^a$SGTB Khalsa College, University of Delhi. Delhi-110007. India.}
\affiliation{$^b$Department of physics and Astrophysics, University of Delhi. Delhi-110007. India.}
\begin{abstract}
%%%%%%%%%%%%
We  study and constrain the  parameter space of the vector color-octet model from the observed data at the Tevatron by studying the  top quark pair production and associated observables \afbt and  spin-correlation. In particular we study the invariant mass and  rapidity dependence  of \afbt at the Tevatron. In addition to the flavor conserving (FC) couplings we extend our study to include the flavor violating (FV) coupling involving the first and third generation quarks for both these processes. In order to ensure that we remain within the constraints imposed by the LHC data, we analyze the charge asymmetry, $p_T$ spectrum and invariant mass in the $t\bar t$ production data at the LHC. The constraints from the dijet resonance searches performed by the LHC are also considered. We also explore the contribution of this model to the single top quark  production mediated by charged and neutral color-octet vector bosons. FV couplings introduced then induce the same-sign top-pair production process which is  analyzed for both the hadron colliders. We have incorporated the effect of finite decay width of color octets on these processes.
We find that  it is possible to explain the observed \afbt anomaly in the color-octet vector model without transgressing the production cross sections of all these processes both through FC and FV couplings at the Tevatron. We predict best  point sets  in the model parameter space for specific choices of color-octet  masses corresponding to $\chi^2_{\rm min}$ evaluated  using  the  \mttb ~ and \dyy spectrum of \afbt  from the observed data set of Run II of the Tevatron at the integrated luminosity  $8.7$~fb$^{-1}$.  We find that the single top quark production is more sensitive to the FC and FV couplings in comparison to the top-pair production. We provide 95 \% exclusion contours on the plane of FV chiral couplings  from the recent data at the Tevatron, CMS and ATLAS corresponding to the nonobservability of large same-sign dilepton events. The four observed point sets are consistent with the cross section, charge asymmetry and spin-correlation measurements for $t\bar t$ production and dijet searches at the LHC.  
\end{abstract}

%\pacs{}
\maketitle
\section{Introduction}
\label{intro}
Top quark production at high luminosity achieved at the recently shut down the Tevatron has thrown tantalizing hints of physics beyond the Standard Model (SM). The combined analysis of the CDF and D\O \, collaboration has given results for top quark mass $m_t = 173.3 \pm 1.1$ GeV \cite{Group:2010ab}. 
The current measured cross section from all channels with 4.6 fb$^{-1}$ data is \crtt
 $= 7.5\pm 0.31$ (stat) $\pm 0.34$ (syst) $\pm 0.15 \,$ (Z theory) pb for $m_t$ = 172.5 GeV \cite{cdf:9913}.  For the same top mass D\O \, collaboration reported  $\sigma_{t\bar t}$ = 7.36$^{+0.90}_{-0.79}$ (stat+syst) pb  using dilepton events \cite{Abazov:2011cq} . The leading order process for $t\bar t$ production at the Tevatron is $q\bar q \to t \bar t$. The top-pair production cross section with the QCD corrections at the next-to-next-to-leading order (NNLO) level is computed  to be $\sigma(t
\bar{t})^{\rm NNLO}_{\rm SM} = 7.08^{+0.00+0.36}_{-0.24-0.27}$~pb for $m_t =
173$~GeV~\cite{Kidonakis:2012db}. These corrections are not only significant but are also   in  agreement with the experiment.
\par The CDF and D\O \, collaborations have  reported top quark forward-backward asymmetry \afbt for large $t\bar t$ invariant mass \mttb  which shows a deviation of about two sigma from the SM prediction \cite{{Aaltonen:2011kc},{Abazov:2011rq}}.  Recently  CDF  observed the parton level \afbt to be $0.296\pm .067$  for \mttb $>$ 450 GeV based on   the full Run II data set with luminosity of 8.7 fb$^{-1}$ \cite{CDF:10807}. 
In the SM, this asymmetry arises only at the next to leading order through the interference between the Born term and higher order of QCD terms and is found to be $0.1$ \cite{AFBth} which is  too small to fit the data. In the literature the \afbt anomaly has been attributed to new physics (NP) beyond SM \cite{Dahiya:2012ka,Shu:2011au,AguilarSaavedra:2011ug,Cao:2010zb}. In this article we analyze the \mttb and \dyy distributions of \afbt  induced by the  color-octet vector bosons and compare with the distribution simulated from the observed data given in Ref. \cite{CDF:10807}. In light of the recent observations at the LHC, we constrain the new physics model parameter space  from the  invariant mass distribution of the $t\bar t$ production cross section and the measured associated charge asymmetry.
\par Single top quark production is an important process at hadron colliders in providing an opportunity to probe the electroweak (EW) interactions of the top quark. Although in the SM, single top quark is produced  at the EW scale, it is noteworthy that the production cross section is comparable and only a little less than half of the  $t\bar t$ pair production. The considerable background however makes the extraction of the signal quite challenging. Recent analysis of the CDF collaboration  uses 7.5 fb$^{-1}$ of data and measures the single top quark total cross section of $\sigma_{s+t} = 3.04^{+0.57}_{-0.53}$ pb \cite{CDF:10793}. Using $5.4\;\rm fb^{-1}$ of collected data, D\O \, at the Fermilab Tevatron Collider measured the combined  single top quark production cross section $\sigma_{s+t} = 3.43^{+0.73}_{-0.74}$ pb \cite{Abazov:2011pt}.
The predicted next-to-next-to-next-to-leading order (NNNLO) approximate calculation for both the modes are $\sigma_s = 0.523^{+0.001+0.030}_{-0.005-0.028}$ pb and $\sigma_t = 1.04^{+0.00}_{-0.02}\pm0.06 $ pb \cite{Kidonakis:2012db} for $m_t$ = 173 GeV. Recently CMS collaboration reported the $t$-channel EW single top quark cross section to be $83.6 \pm 29.8$ (stat. + syst.) $\pm$ 3.3 (lumi) pb at $\sqrt s$ = 7 TeV at the LHC \cite{CMS} corresponding to integrated luminosity of 36 pb$^{-1}$, which agrees with the next-to-leading order (NLO) and resummation of collinear and soft-gluon corrections NNLO \cite{Kidonakis:2012db}. However the involved experimental and theoretical uncertainties allow us to explore the new physics consequences in the determination of the production cross section mediated by the new exotic vector bosons. Therefore it is worthwhile to study the effect of the couplings induced by the color-octet vector model in the $s$ and $t$ channel single top quark production.
\par The production of same-sign top quark pair is a fascinating process and would furnish unmistakable signature of physics beyond the SM. In the SM this process is highly suppressed and involves higher order flavor changing neutral current (FCNC) interactions. The search for the same-sign top quark pair involves searching for events with same-sign isolated leptons accompanied by hadron jets and missing transverse energy in the final sate. The CDF collaboration has set a limit on same-sign top-pair production at the Tevatron using a luminosity of 6.1 fb$^{-1}$, $\sigma(tt+\bar t \bar t)\times $[BR($W\to l \nu$)]$^2$ $<$ 54 fb with a 95 $\%$ confidence level \cite{CDF:10466}. This limit puts severe constraints on physics beyond the SM which allows for FCNC interactions. The 7 TeV data from CMS also disfavors the  same-sign top-pair production at the LHC  mediated through $Z^\prime $ in $t$ and $u$ channels which otherwise was the potent model to explain the \afbt anomaly observed in $\ttbar $ production at the Tevatron \cite{Chatrchyan:2011dk,Aad:2012bb}. We probe the effect of the flavor violating couplings in the same-sign top-pair production both at the Tevatron and the LHC. These couplings are introduced to contest the \afbt anomaly.
\par In Sec. \ref{sec:model}, we introduce the $\bf{ 3} \otimes \bf{\bar 3} $ vector color-octet model. In Sec. \ref{sec:toppair} we compute the  $t\bar t$ cross section and associated  top quark forward-backward asymmetry \afbt and spin-correlation coefficient in this model. Single top quark production is studied in Sec. \ref{sec:singletop} and in Sec. \ref{sec:samesigntop} we probe the same-sign top quark production through FV couplings.  In Sec. \ref{lhc} we consider the constraints obtained by analyzing the LHC top quark data with reference to the resonance searches in $t\bar t$ production, dijet resonance searches and from the charge asymmetry data. Section \ref{sec:summary} is devoted to the discussion of the results and conclusion.
\section{${\bf 3}\otimes \bar{\bf 3}$ Vector color-octets}
\label{sec:model}
The exotic  bosonic states that can couple to a quark $(q)$ and antiquark  $(\bar q)$ pair in physics scenarios beyond the Standard Model are the scalar/vector color singlets, triplets, sextets  and octets. 
Color singlet vector states are the $Z^\prime$s \cite{Zp}, $W^\prime$s \cite{Wp}, unparticles \cite{Dahiya:2012ka}  and Kaluza-Klein gravitons $G_{KK}$ \cite{Agashe:2006hk}. 
Grinstein et. al. \cite{Grinstein:2011dz} considered various scenarios of new physics models which contain scalars as well as vector representation of SM quark flavor symmetry group to study $t\bar t$ forward-backward anomaly and D$\O$ dimuon anomaly at the Tevatron. They also studied the constraints  on the flavor symmetric models and their collider signatures at the LHC  with exotic bosons having masses around the electroweak scale. 
\begin{widetext}
\begin{center}
\begin{table}[h!]
\begin{tabular}{|c|c|cccc|} \hline
$\sqrt s$ @ LHC & $M_{V_8^{\pm,0}}$ GeV & N($u\bar d \to V_8^{+}$) & N($d\bar u \to V_8^{-}$) & N($u\bar u \to V_8^{0}$) & N($d\bar d \to V_8^{0}$) \\ \hline
7 TeV                          & 200 & 2.2$\times 10^{8}$ & 1.2$\times 10^{8}$ & 2.1$\times 10^{8}$ & 1.3$\times 10^{8}$ \\
 $\mathscr L =$5 fb$^{-1}$    & 500 & 8.1$\times 10^{6}$ & 3.5$\times 10^{6}$ & 7.0$\times 10^{6}$ & 4.2$\times 10^{6}$ \\
                               & 900 & 6.9$\times 10^{5}$ & 2.3$\times 10^{5}$ & 5.3$\times 10^{5}$ & 3.0$\times 10^{5}$ \\ \hline
8 TeV                          & 200 & 1.0$\times 10^{9}$ & 5.8$\times 10^{8}$ & 9.6$\times 10^{8}$ & 6.2$\times 10^{8}$ \\
 $\mathscr L =$20 fb$^{-1}$   & 500 & 4.6$\times 10^{7}$ & 1.8$\times 10^{7}$ & 3.6$\times 10^{7}$ & 2.1$\times 10^{7}$ \\
                               & 900 & 3.7$\times 10^{6}$ & 1.3$\times 10^{6}$ & 3.0$\times 10^{6}$ & 1.7$\times 10^{6}$ \\ \hline
14 TeV                         & 200 & 9.9$\times 10^{9}$ & 6.0$\times 10^{9}$ & 9.6$\times 10^{9}$ & 6.3$\times 10^{9}$ \\
 $\mathscr L =$100 fb$^{-1}$  & 500 & 4.6$\times 10^{8}$ & 2.4$\times 10^{8}$ & 4.3$\times 10^{8}$ & 2.6$\times 10^{8}$ \\
                               & 900 & 5.4$\times 10^{7}$ & 2.4$\times 10^{7}$ & 4.8$\times 10^{7}$ & 2.8$\times 10^{7}$ \\ \hline
\end{tabular}
\caption{\small \em{ Production event rate  of the charged and neutral color-octet vector bosons for three different representative masses at three different collider energy and luminosity.  To obtain these events we have set the product of the coupling constant squared and the branching fraction  to unity.}}
\label{productioncrosssection}
\end{table}
\end{center}
\end{widetext}
%%%%%% fig. ttbar in 33bar model
\begin{figure}
\vskip 1 cm
\subfloat[]{\label{fig:ttbarfc}\includegraphics[scale=1.0]{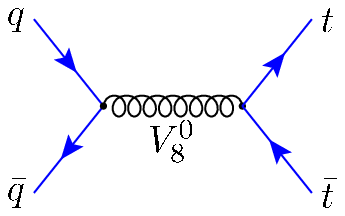}} \qquad \qquad \quad
\subfloat[]{\label{fig:ttbarfv}\includegraphics[scale=1.0]{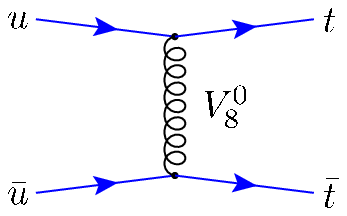}} \qquad \qquad \qquad
\caption{Diagrams for $q\bar q \to t\bar t$ production through $V_8^0$ in (a) $s$-channel FC and (b) $t$-channel FV cases. 
 \label{ttbarNP}}
\end{figure}
\par Some of the colored exotic states are already present in some of theoretical models beyond the SM, for example R-parity violating super-symmetric theories \cite{Barbier:2004ez}, excited quarks in composite models \cite{Cabibbo:1983bk,Baur:1987ga}, diquarks in E$_6$ grand unified theories \cite{Hewett:1988xc},   in theories of extra dimension \cite{Dobrescu:2007xf}, color-triplet and sextets  \cite{Shu:2009xf} and in low scale string resonances \cite{Cullen:2000ef}. Color-octet  scalars  have  been studied in Ref. \cite{Hill:2002ap} while   
color-octet vector states coupled to $q\bar q$ are analyzed in Refs. \cite{axig,Wang:2011hc} for axigluons and in Ref. \cite{wang:2011taa,colorons,Hill:1991at} for colorons. Some of these exotic states have been involved in the literature to explain the top quark forward-backward asymmetry and the CDF dijet resonances. These particles if they exist can be produced at the LHC with their masses and couplings constrained by the measurement of the dijet cross section at the Tevatron and the LHC. The ATLAS and CMS collaborations \cite{Chatrchyan:2011ns}  have reported stringent bounds on these colored states.
 
\par In this article, we  investigate the contribution of color neutral and charged vector states $V_8^{0,\mu}$ and $V_8^{\pm,\mu}$ on  $t\bar t$ production, top-quark forward-backward asymmetry, single top quark production and same-sign top-pair production. The interaction of color-octet vector states $V_8^{0,\pm,\mu}$ with quark is given by 
    \begin{align}
{\mathscr L}_{q\bar q^\prime V} &=  g_s \Big[{V_{8}^{0a}}^{\mu} \,\bar u_i \, T^a \gamma_\mu \, (g_L^U \, P_L + g_R^U\, P_R)\, u_j \notag\\  
&\quad + {V_8^{0a}}^{\mu} \,\bar d_i \,T^a \,\gamma_\mu \,(g_L^D \, P_L + g_R^D \, P_R)\, d_j \notag \\
&\quad + \left.\left({V_{8}^{+a}}^{\mu} \, \bar u_i \, T^a \, \gamma_\mu \,(C_L \, P_L + C_R  \, P_R)\, d_j + \text{h.c.}\right) \right] 
\end{align}
\begin{figure*}[!ht]
  \centering
\subfloat[$\lambda_{LL} = \lambda_{RR} =\lambda_{RL}=\lambda_{LR}=\lambda_{VV} $]{\label{fig:TP_FC_V_sigma}\includegraphics[width=0.5\textwidth]{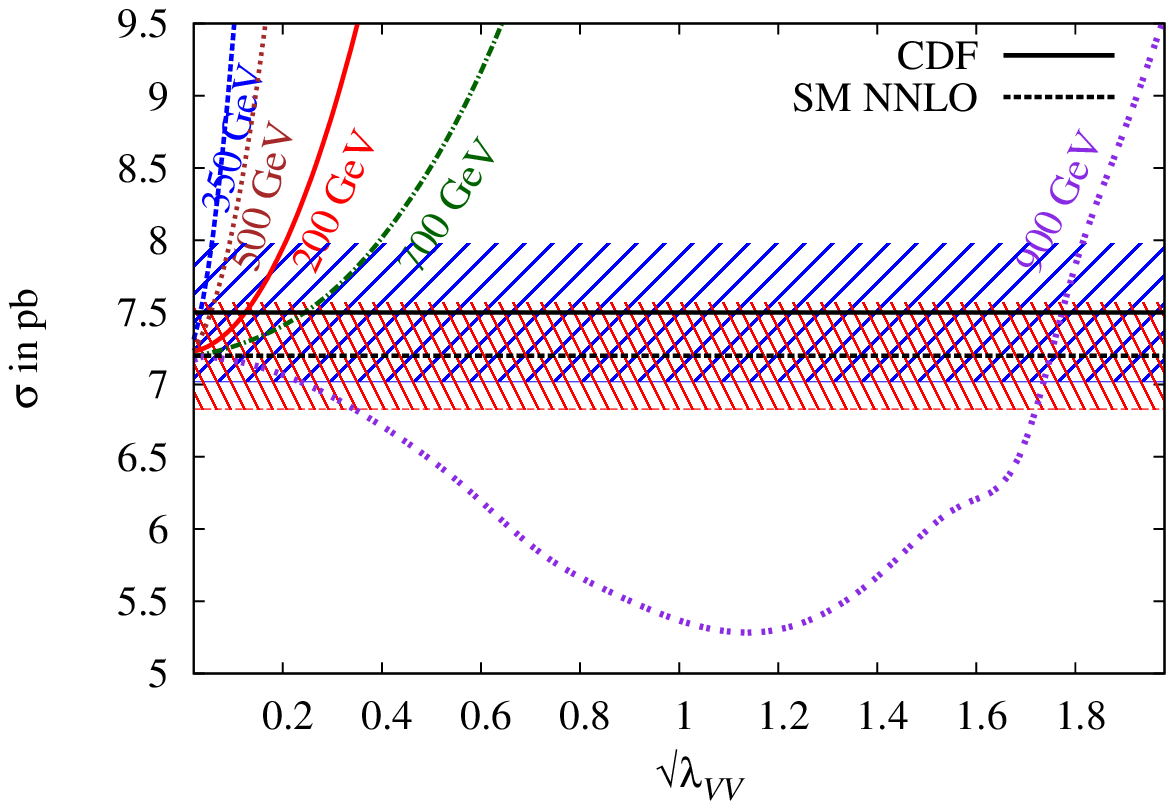}}
  \subfloat[$\lambda_{LL} = \lambda_{RR} =-\lambda_{RL}=-\lambda_{LR}=\lambda_{AA} $]{\label{fig:TP_FC_A_sigma}\includegraphics[width=0.5\textwidth]{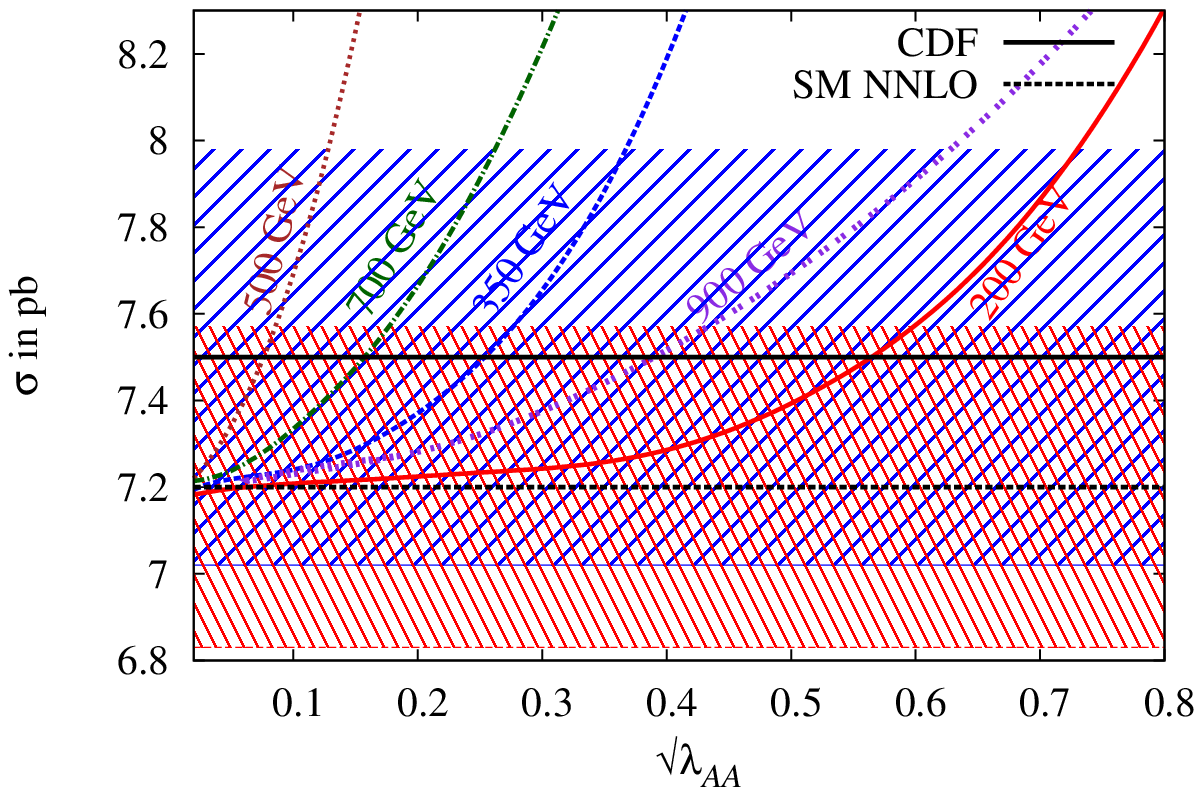}}\\
  \subfloat[$\lambda_{RR} \neq 0= \lambda_{LL} =\lambda_{RL}=\lambda_{LR}$]{\label{fig:TP_FC_R_sigma}\includegraphics[width=0.5\textwidth]{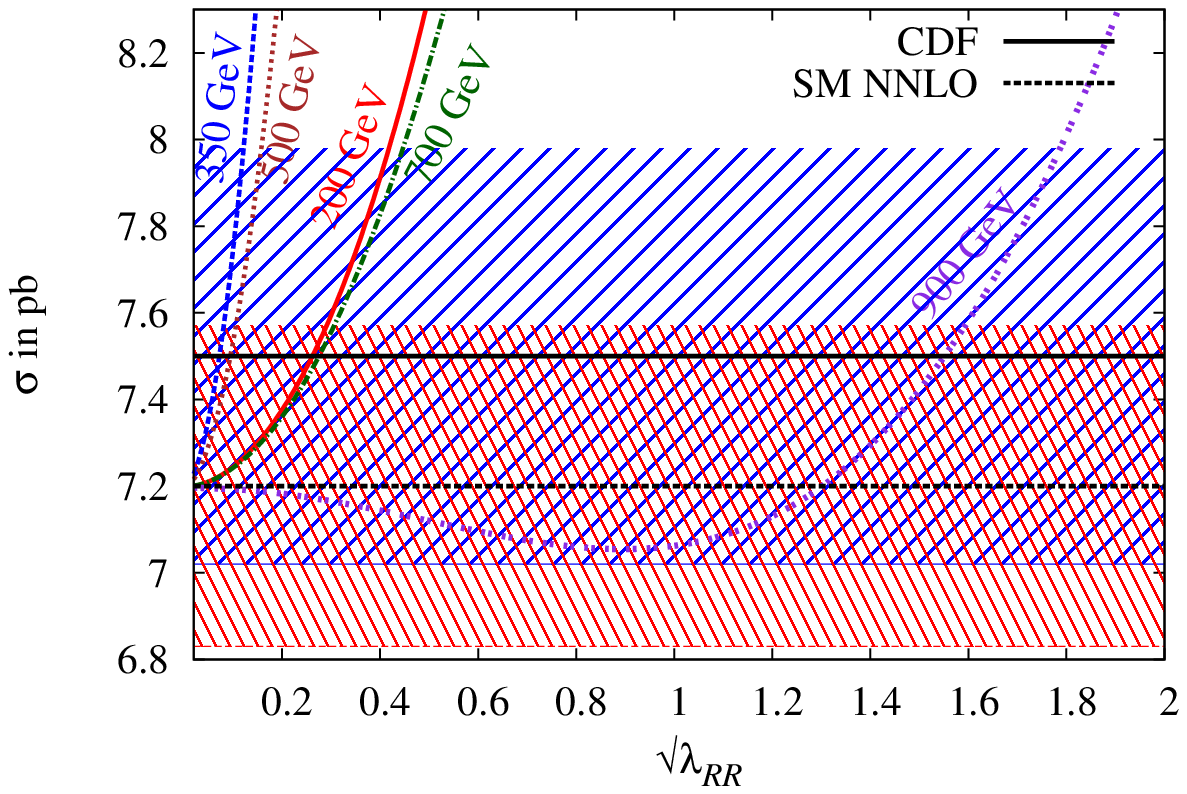}} 
\caption{\small \em {Variation of the cross section $\sigma \left( p\bar p \to t\bar t\right)$ with couplings $\sqrt\lambda_{ij}$ for
flavor conserving vector color-octets corresponding to  different values of  $M_{V_8^0}$. The upper black dotted line associated with a blue band  depicts the  cross section  $7.50 \pm 0.48$ pb from the CDF (all channels) ~\cite{Abazov:2011cq}, while the lower black dot-dashed line associated with a red band show theoretical estimate $7.2\pm 0.37$ pb at NNLO  ~\cite{Kidonakis:2012db}. Figures (a), (b) and (c) show the variation of $\sigma$ for vector, axial and right-handed cases, respectively (the cases (a), (b) and (c) as in the text).}}
\label{ttbarFC}
\end{figure*}
%%%%%%%%%%%%%%%%%%%

Production of these color-octet resonants are studied in \cite{Han:2010rf} where significant bounds for these resonances based on the preliminary CMS data have been obtained. We   estimate  the  production event rates at  a given luminosity  for 200, 500 and 900 GeV  vector color-octets in Table \ref{productioncrosssection} corresponding to the centre of mass energy of 7, 8 and 14 TeV, respectively. For all these subprocesses the coupling constant of the octets with the light quarks along with the  branching ratio  are set to unity.  
\par The color-octet vectors can be detected either  through the resonant dijet  and  top-pair production. The measurement of the dijet events at the Tevatron and the LHC  restricts the coupling  of these exotic colored states to light quark sector in the narrow resonance approximation (in units of $g_s = \sqrt{4\pi \alpha_s}$). However, the coupling of these states to the top quark sector essentially remains unexplored.  It is to be noted that  the $t\bar t$ cross section depends on the product of the couplings of $V_8$ with light and top quarks whereas the dijet cross section depends on the coupling of these exotics with light quarks only. We will revisit the resonant dijet production cross section  in Sec. \ref{dijetlhcsection}.

\par Since there exist stringent constraints from flavor physics on FCNC, we  consider a non universal FCNC couplings  between the up quarks of the first and  third generation only. However, after a rotation from weak eigenstates to mass eigenstates these FCNC interactions would contribute to $B_q^0$-$\bar B_q^0$ and $D^0$-$\bar D^0$ mixing. Consequently, by making a suitable  choice of the left and right-handed mixing matrices along with the relevant couplings, we can evade  the most restrictive constraint on these couplings from the low energy $B$ and $D$ phenomenology \cite{Bai:2011ed}. It has been discussed in the literature see for example \cite{Haisch:2011up} that in a scenario involving non-universal flavor coupling, if the generic flavor violation is confined to up quark sector and is so aligned as to induce minimal flavor violation, the constraints on color-octet vector boson with axial coupling of QCD strength from $D$-meson sector are rather weak $M_{V_8^0}> 0.22$ TeV. The bounds on right-handed couplings are even weaker. On the other hand generic flavor violation in the down quark sector requires $M_{V_8^0}$ with axial couplings to be greater than several TeV from the neutral meson mixing data. Thus by keeping the u-t coupling large and simultaneously making c-t and u-c couplings small, we will not get any strong bounds from flavor physics. This reference also shows  that the contribution to $B_q^0$-$\bar B_q^0$ mixing from the charge current sector can be controlled whenever the mixing matrices are aligned with the SM Cabibbo-Kobayashi-Maskawa resulting in  the couplings $C_{L,R}$, $g_{L,R}$  naturally to be of the  order one. Assuming minimal flavor violation breaking for a particular choice of flavor symmetry subgroup, authors of the Ref. \cite{Grinstein:2011dz} were able to pin point the vector boson models consistent with FCNC constraints from $B^0-\bar B^{0}$ mixing and at the same time able to give right enhancement of $A_{FB}^{t\bar t}$ for a benchmark point say, $M_{V^0_8}=$ 300 GeV.

\par The couplings $g_{L,R}$ and $C_{L,R}$  are free parameters in our model and taken to be diagonal. $g_s = \sqrt{4\pi \alpha_s}$ is the QCD coupling; $i,j$ are flavor indices and $a$ represents the color index. We perform our calculations for top quark mass $m_t = 172.5$ GeV and bottom quark mass $m_b = 4.7 $ GeV at the $p\bar p$ center of mass energy $\sqrt s = 1.96$ TeV, with fixed QCD coupling $\alpha_s = 0.13$ and using CTEQ6L1 parton distribution functions keeping factorization and renormalization scales $\mu_F = \mu_R = m_t$. We have incorporated the $\bf{3}\otimes \bf{\bar 3}$ model  in MadGraph/MadEvent V4 \cite{Alwall:2007st} and generated a total of 10000 events for all the  processes. We do not take into account the effects from parton showering, hadronization and detector conditions in our studies. 
\par The width of the new resonances are computed and taken into consideration for all the phenomenological observables presented in this study.
\section{top-pair production}
\label{sec:toppair}
In this section we make a detailed study of $t\bar t$ production at the Tevatron to put bounds on the parameters of ${\bf 3}\otimes \bar{\bf 3}$ model. The new physics contribution to $t\bar t$ production in the present model arises through the exchange of flavor conserving  and flavor violating   neutral current (NC) $V_8^0$  as shown in Figs. ~\ref{fig:ttbarfc}  and ~\ref{fig:ttbarfv}, respectively.  Here we assume that the flavor changing couplings are present only between the first and third generation quarks. We examine the sensitivity of both these couplings for the   associated observables  namely
the forward-backward  asymmetry and the spin-correlation coefficient along with the total production cross section.
\par At the Tevatron $t\bar t$ pair is predominantly produced through the quark pair annihilation $q\bar q \to t\bar t$, where the quarks (anti-quarks) are mainly moving along the proton (anti-proton) direction. The FB asymmetry can be defined as
\begin{eqnarray}
A_{FB} =&& \frac{N(\Delta y > 0)-N(\Delta y <0)}{N(\Delta y > 0)+N(\Delta y <0)} 
\end{eqnarray}
where N is the number of events and $\Delta y = y_t - y_{\bar t}$ is the difference in rapidities of top and the anti-top quarks along the proton momentum direction in the lab frame. Recent measurements from the CDF and D0 Collaborations at the Tevatron report positive asymmetries ~\cite{Aaltonen:2011kc}~\cite{Abazov:2011rq}, $A_{FB}^{CDF} = 0.158 \pm 0.075$, $A_{FB}^{D0} = 0.196 \pm 0.065$ at the subprocess/parton level after correcting for backgrounds and detector effects corresponding to integrated luminosity of 5.3 and 5.4 fb$^{-1}$, respectively whereas the SM prediction at NLO QCD level is 0.051 \cite{AfbTeva1}. A recent report from the CDF, studied FB asymmetry and its mass and rapidity dependence with an integrated luminosity of 8.7 fb$^{-1}$ at the Tevatron resulting in an inclusive parton level $A_{FB} = 0.162 \pm 0.047$~\cite{CDF:10807}.
\par  Spin-correlation of the the top-antitop pair, in the "helicity basis", (i.e; choosing the direction of the top quark momentum as our spin quantization axis) is  described by four independent helicity states $\bar t_L t_R, \bar t_R t_L, \bar t_L t_L, \bar t_R t_R$.  The spin-correlation parameter is defined as
\begin{eqnarray}
\spincorr =&& \frac{\left[ \sigma(\bar t_R t_L) + \sigma(\bar t_L t_R) \right] - \left[ \sigma(\bar t_R t_R) + \sigma(\bar t_L t_L) \right]}{\left[ \sigma(\bar t_R t_L) + \sigma(\bar t_L t_R) \right] + \left[ \sigma(\bar t_R t_R) + \sigma(\bar t_L t_L) \right]} \cr \cr
=&& \frac{N_O - N_S}{N_O + N_S}
\end{eqnarray}
where $N_S=\Uparrow\Uparrow+\Downarrow\Downarrow$ and $N_O=\Uparrow\Downarrow+\Downarrow\Uparrow$ are the number of top and antitop quarks with their spins parallel and anti-parallel, respectively.  Recent studies have shown strong spin-correlation in the top quark pair production which means that the top quark and antiquark have preferential spin polarizations. At the Tevatron the dominant parton level top-pair production is $q\bar q \to t \bar t$ and  its value for SM   in the helicity basis is 0.299 \cite{Bernreuther:2010ny}. The CDF reported the measurement of the spin-correlation coefficient $\spincorr_{\rm helicity} = 0.60 \pm 0.50 (stat) \pm 0.16 (syst)$ \cite{Aaltonen:2010nz} in the helicity basis and  $0.042^{+0.563}_{-0.562}$ \cite{CDFNote10719} in the beam basis.  Any deviation in the measurement of $\spincorr$ would give an indirect idea of the new $t\bar t$ production mechanism and also models of new physics appeal to the spin-correlation for signal identification and discrimination \cite{Ko:2012gj,np-polarization,spin-corr,Kane:1991bg,Arai:2007ts}.
\begin{figure*}[!ht]
  \centering
  \subfloat[$\lambda_{LL} = \lambda_{RR} =-\lambda_{RL}=-\lambda_{LR}=\lambda_{AA}$]{\label{fig:TP_FC_afb_A}\includegraphics[width=0.5\textwidth]{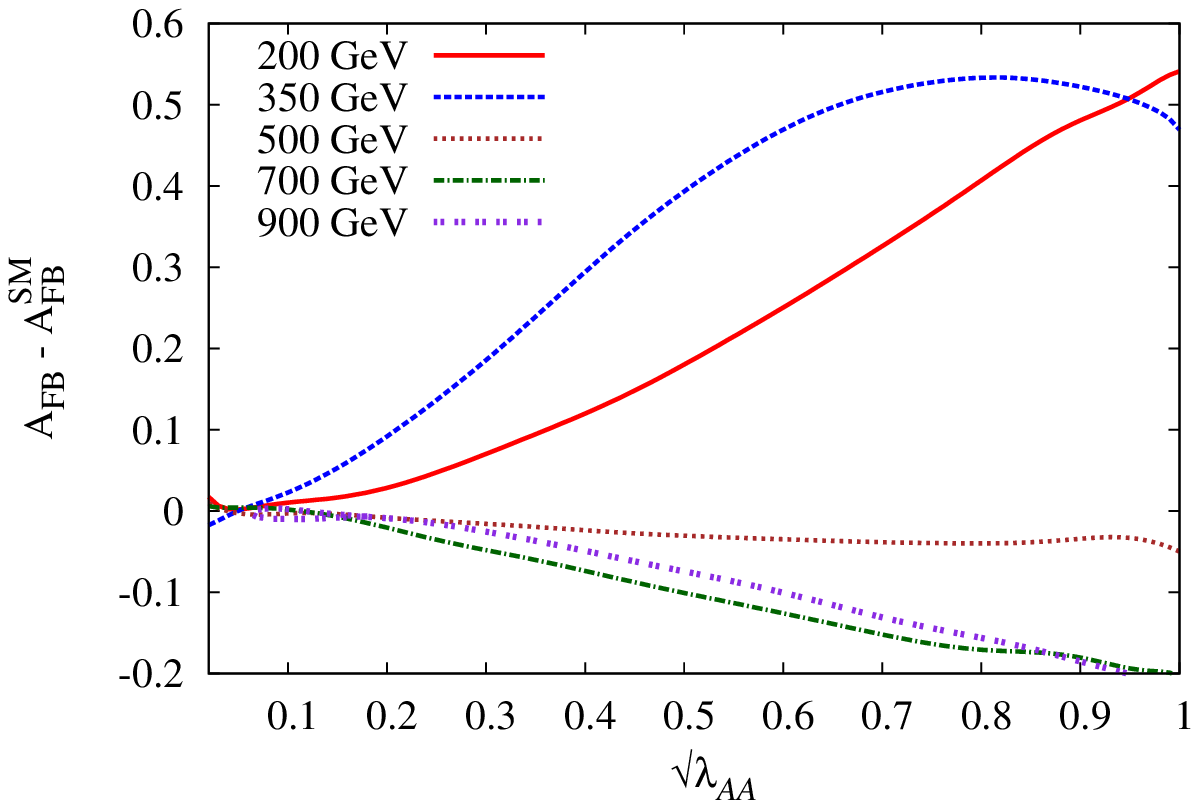}}
 \subfloat[$-\lambda_{LL} = -\lambda_{RR} =\lambda_{RL}=\lambda_{LR}=\lambda_{NA}$]{\label{fig:TP_FC_afb_NA}\includegraphics[width=0.5\textwidth]{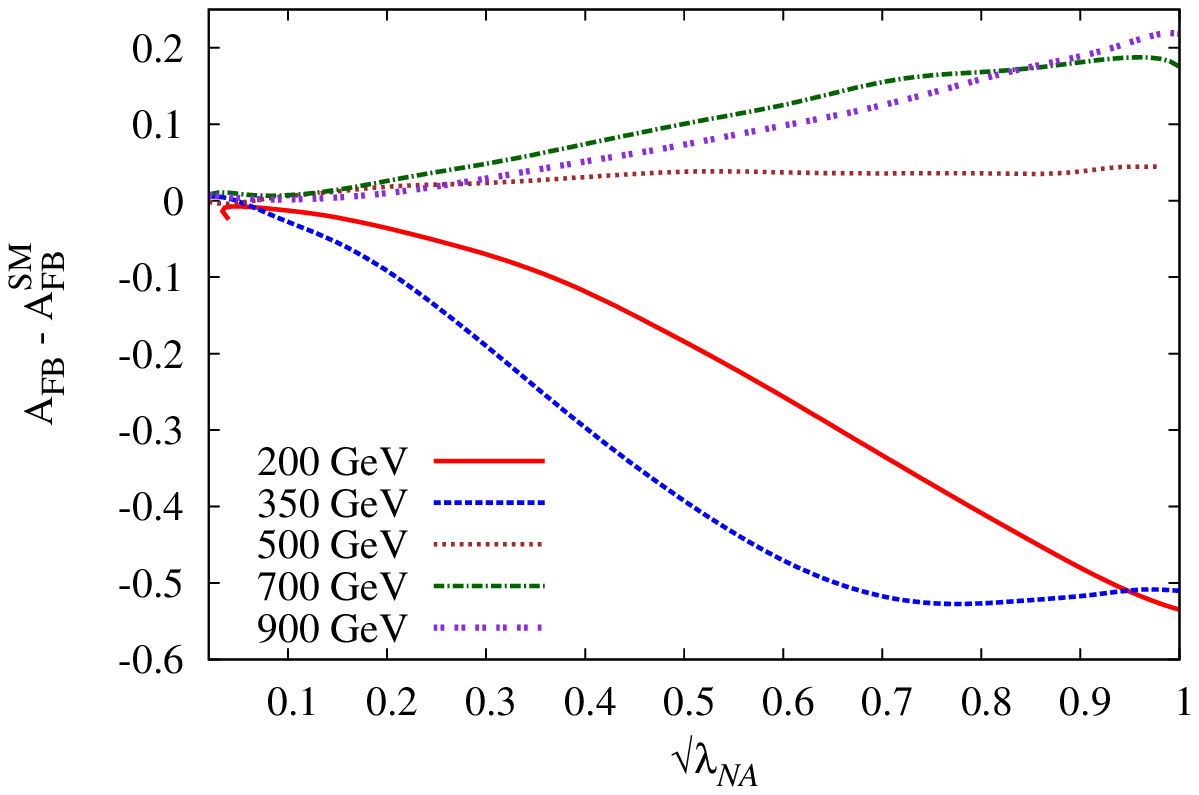}} \\
  \subfloat[$\lambda_{RR} \neq 0 = \lambda_{LL} = \lambda_{LR} =\lambda_{RL}= 0$]{\label{fig:TP_FC_afb_R}\includegraphics[width=0.5\textwidth]{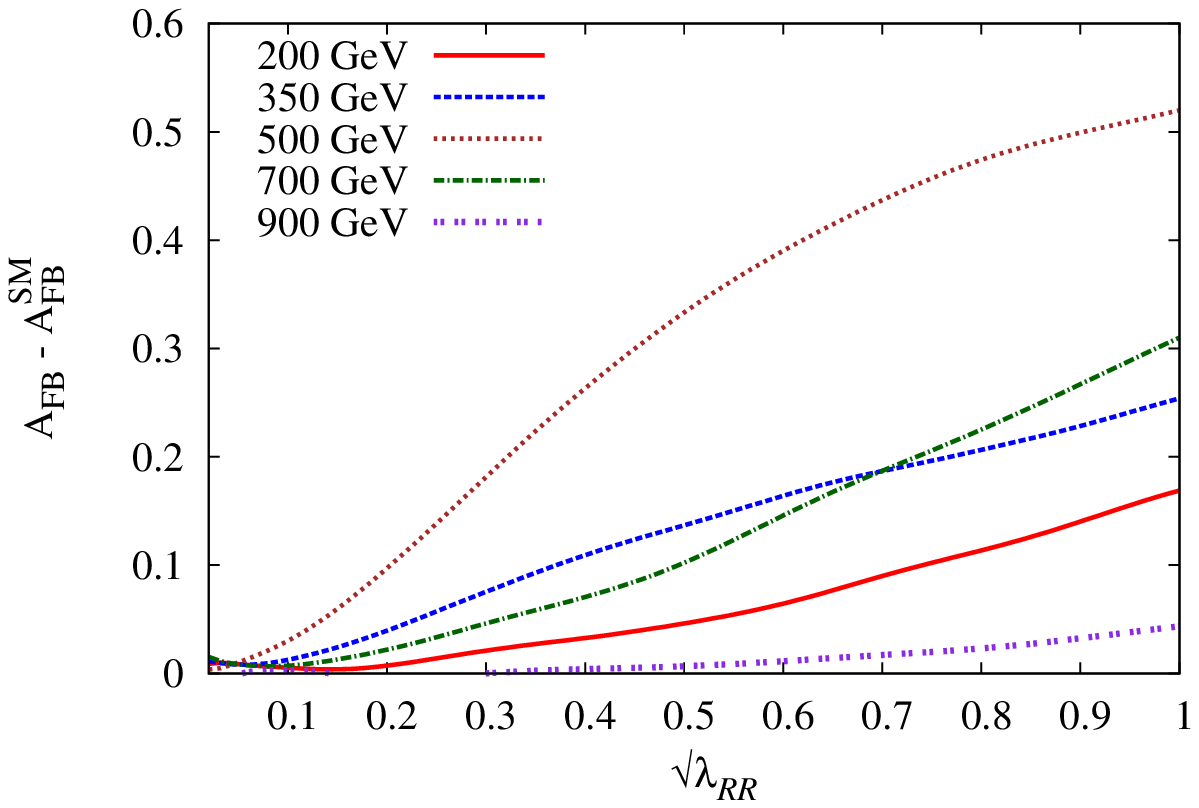}}
\caption{\small \em {Variation of the $A_{FB} - A_{FB}^{SM}$ with couplings $\sqrt\lambda_{i j}$ for flavor conserving  vector color-octets corresponding to  different values of  $M_{V_8^0}$. Figures (a), (b) and (c) corresponds to the positive product of axial couplings, negative product of axial couplings and purely right-handed cases (the first two corresponds to cases (b) and the third corresponds to case (c) in the text, respectively). The \afbt for case (a) vanishes identically for all mass regions.}}
\label{ttbarFCAFB}
\end{figure*}
\begin{figure*}[!ht]
  \centering
  \subfloat[$\lambda_{LL} = \lambda_{RR} =\lambda_{RL}=\lambda_{LR}=\lambda_{VV} $]{\label{fig:TP_FC_spcr_V}\includegraphics[width=0.5\textwidth]{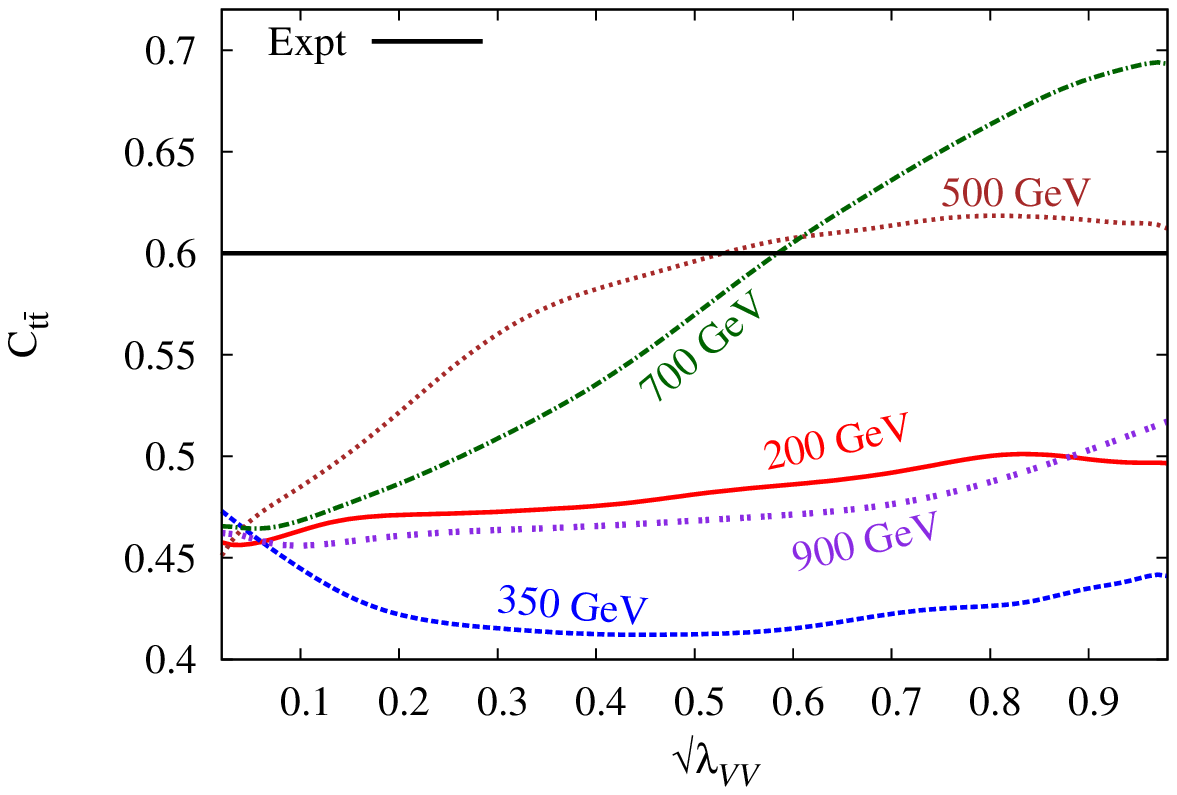}}
  \subfloat[$\lambda_{LL} = \lambda_{RR} =-\lambda_{RL}=-\lambda_{LR}=\lambda_{AA}$]{\label{fig:TP_FC_spcr_A}\includegraphics[width=0.5\textwidth]{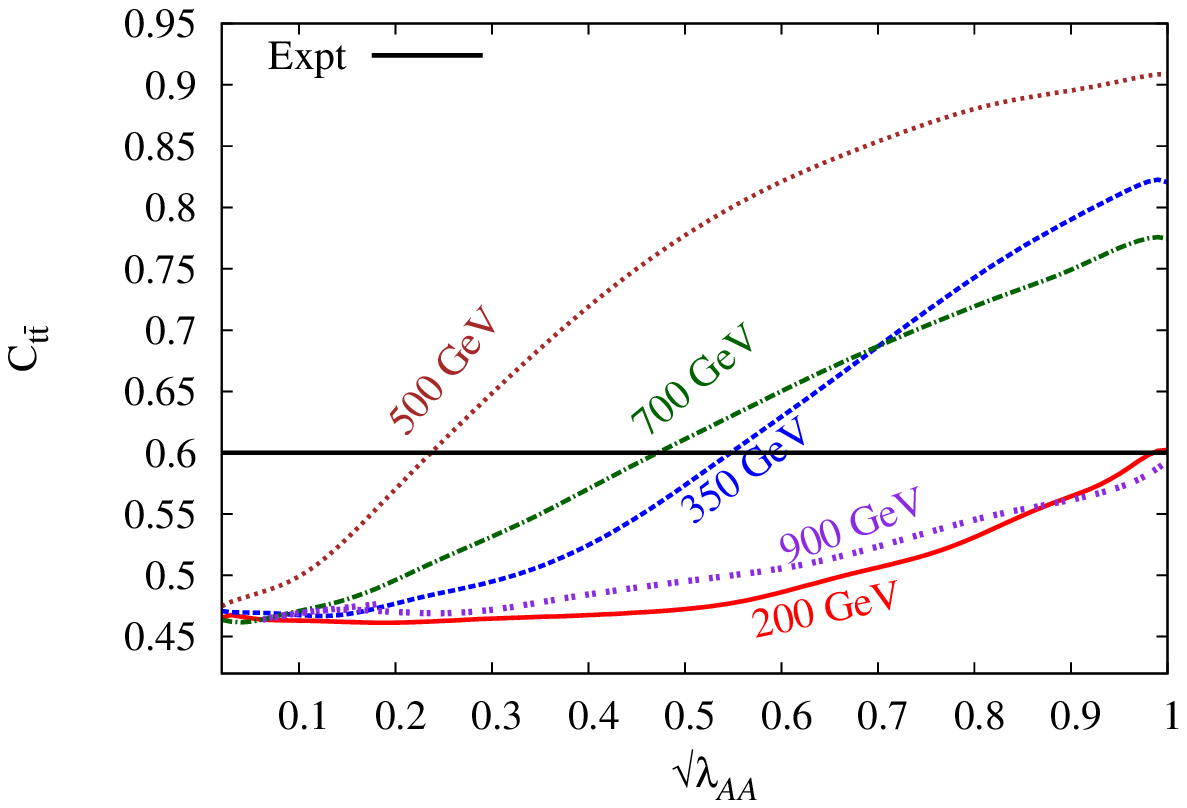}} \\
  \subfloat[$\lambda_{RR} \neq 0=\lambda_{LL}=\lambda_{LR} =\lambda_{RL}$]{\label{fig:TP_FC_spcr_R}\includegraphics[width=0.5\textwidth]{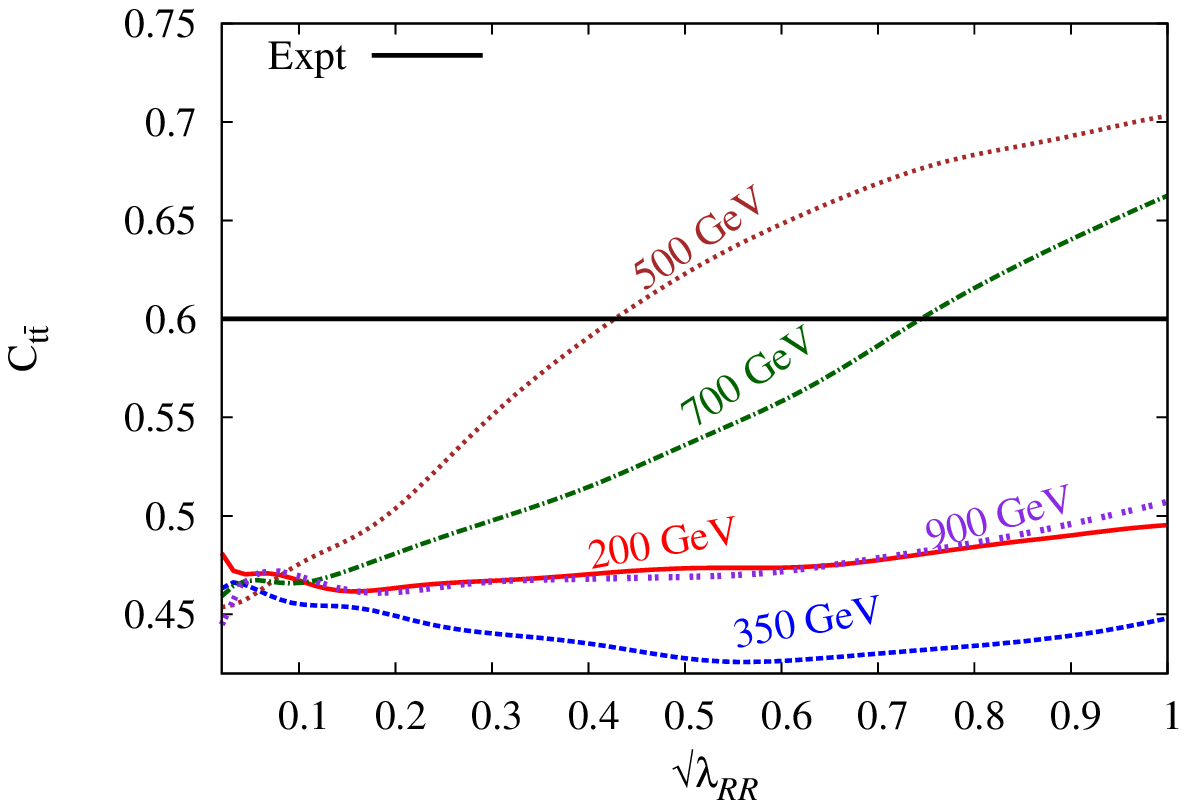}}
\caption{\small \em {Variation of the spin-correlation coefficient $C^{t\bar t}$ with couplings $\sqrt\lambda_{i j}$ for flavor conserving  vector color-octets corresponding to  different values of  $M_{V_8^0}$. Figures (a), (b) and (c) correspond to vector, axial and right-handed cases, respectively (cases (a), (b) and (d) in the text, respectively). The spin-correlation variation for cases (c) and (b) as in text are identical.}}
\label{ttbarFCSC}
\end{figure*}
\subsection{Flavor Conserving}
The additional Feynman diagrams induced by the color-octet vector bosons depicted in Fig. \ref{ttbarNP}, interfere  with  the SM tree annihilation process. We take the color-octet vector boson couplings with the first two generations i.e. light quarks $g^q_i$ (q=u,d,s,c)  to be typically an order of magnitude smaller than the third generation heavy quarks $g^{t,b}_i$ with $i,j = L/ R$. This is consistent with the dijet measurements at the Tevatron. The corresponding NP matrix element is proportional to  the product of light and heavy quarks couplings. For convenience we choose  the product of two couplings $\sqrt{\lambda_{ij}} = \sqrt{g^q_i g^t_j}$  and the mass $M_{V_8^0}$ as a parameter of our model in FC case. Since the total matrix element squared is left-right symmetric, we have only three independent choices of the combinations of couplings which includes the (a) vector, (b) axial-vector and (c) right-chiral interactions.   All  three cases can be explicitly expressed as 
\begin{eqnarray}
&&(a) \text{Vector:} \quad g_L^q = g_R^q, g_L^t = g_R^t, \cr 
&&\qquad   \Rightarrow \lambda_{LL} = \lambda_{RR} =\lambda_{RL}=\lambda_{LR}=\lambda_{VV} \cr
&&(b) \text{Axial:} \quad  g_L^q =  -g_R^q, g_L^t = -g_R^t, \cr
&&\qquad   \Rightarrow \lambda_{LL} = \lambda_{RR} =-\lambda_{RL}=-\lambda_{LR}=\lambda_{AA} \cr
&&\text{or}  \quad  g_L^q = -g_R^q, g_R^t = -g_L^t \, or \,  g_R^q = -g_L^q, g_L^t = -g_R^t, \cr
&&\qquad   \Rightarrow -\lambda_{LL} = -\lambda_{RR} =\lambda_{RL}=\lambda_{LR}=\lambda_{NA} \cr
&&(c) \text{Right-chiral:} \quad \, g_{L}^q = g_{L}^t = 0 , \cr
&&\qquad   \Rightarrow \lambda_{LL} = \lambda_{LR} =\lambda_{RL}= 0 \neq \lambda_{RR}  \notag
\end{eqnarray}
\par These massive color-octets are likely to decay to the lighter quarks as well as to the heavier one if kinetically allowed both through the flavor conserving as well as flavor violating couplings. The non-zero decay width for the heavy color-octets can have appreciable effects not only in the $\ttbar$ cross section but also in the observables like \afbt and \spincorr. The decay width of color-octet vector boson is given by
\begin{eqnarray}
\Gamma_{V_8} =&& \tfrac{1}{6}\alpha_s [(g_L^2+g_R^2)\Big\{ \frac{M_{V_8}^2}{2} - \frac{m_q^2 + m_{q^\prime}^2}{4} - \Big(\frac{m_q^2 - m_{q^\prime}^2}{2 M_{V_8}}\Big)^2 \Big\}  \cr
&&\quad +\, 3\, m_q m_{q^\prime}\, g_L\, g_R]\, \frac{\lambda^{\tfrac{1}{2}}(M^2_{V_8},m_q^2,m_{q^\prime}^2)}{M^3_{V_8}}
\label{deacywidth}
\end{eqnarray}  
where, $\lambda(x,y,z) = x^2 + y^2 + z^2 -2x\cdot y - 2y\cdot z - 2 z\cdot x $. In flavor conserving case, for $M_{V_8} \le 2 m_t$, $q(=q^\prime) = u,d,s,c,b$ while for $M_{V_8} \ge 2 m_t$  top quarks also contribute. The only flavor violating mode we have is $V_8^{0} \to u\bar t+\bar u t$. The decay of the charged color-octet vector boson $V_8^{\pm} \to q q^\prime$ proceeds through the exotic charge current (CC) interactions which are assumed to diagonal but with non-universal coupling to the third generation quark sector. 
\par Throughout our analysis we have taken into account the effect of the finite decay width in evaluating the cross-sections and other associated observables.

\par The analytical expression of differential cross section in terms of $\lambda_{ij}$ for $q\bar q \to t \bar t$ with respect to the cosine of the top quark polar angle $\theta$ in the $t\bar t$ center-of-mass (c.m.) frame is given as
\begin{widetext}
\begin{eqnarray}
\frac{d\hat \sigma}{d\,\cos\theta} =&& \frac{\pi \beta \alpha_s^2}{9\hat s} \Bigg[
f(\theta,\beta^2)+ \left\{ \left(1-   \frac{M_{V_8^0}^2}{\hat s} \right)^2 + \frac{M_{V_8^0}^2}{\hat s} \frac{\Gamma^2_{V_8^0}}{\hat s}\right\}^{-1} \Bigg\{   (\lambda_{LR}^2+\lambda_{RL}^2+\lambda_{LL}^2+\lambda_{RR}^2) (2-\sin^2\theta )\,\frac{\beta^2}{4}\cr
&& + \left((\lambda_{LL}+\lambda_{LR})^2 +(\lambda_{RL}+\lambda_{RR})^2)\right)\,\frac{1-\beta^2}{4}+\frac{1}{2}\left(1-   \frac{M_{V_8^0}^2}{\hat s} \right) (\lambda_{LL}+\lambda_{RR}+\lambda_{LR}+\lambda_{RL})\,f(\theta,\beta^2) \cr
&&+\left(\left(1-   \frac{M_{V_8^0}^2}{\hat s} \right) (\lambda_{LL}+\lambda_{RR}-\lambda_{LR}-\lambda_{RL}) +\frac{1}{4} (\lambda_{LL}^2+\lambda_{RR}^2-\lambda_{LR}^2-\lambda_{RL}^2)\right)\beta \cos\theta
\Bigg\}\Bigg]\label{ttbarxsec}
\end{eqnarray}
\end{widetext}
where $\hat s = (p_q + p_{\bar q})^2$ is the squared c.m. energy of the system, $\beta = \sqrt {1-4m_t^2/{\hat s}}$ is the top quark velocity and $f(\theta,\beta^2)= \left(2 - \beta^2 \sin^2\theta \right)$. The terms proportional to $\cos\theta$ in Eq. \eqref{ttbarxsec} are sensitive to the forward backward asymmetry.
%%%%%%%%%%%%%fig ttbar fv cross section
\begin{figure*}[!ht]
  \centering
  \subfloat[]{\label{fig:TP_FV_A_sigma}\includegraphics[width=0.5\textwidth]{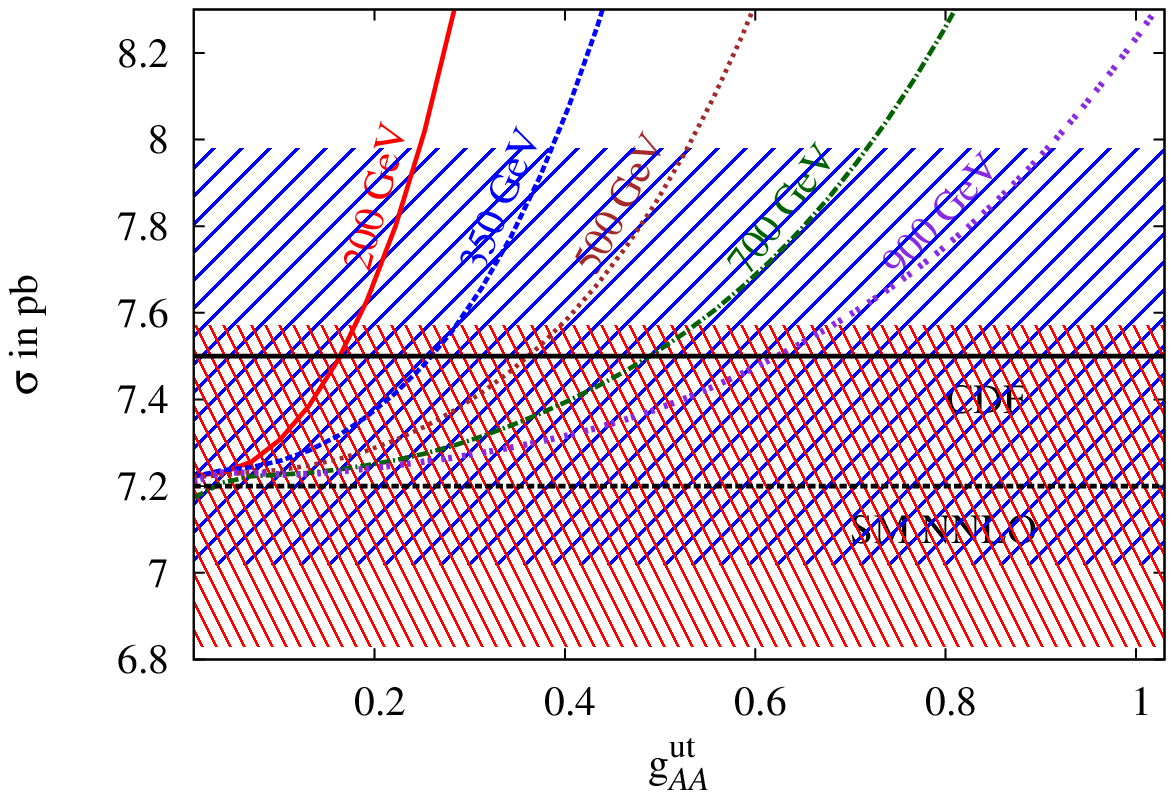}}
  \subfloat[]{\label{fig:TP_FV_R_sigma}\includegraphics[width=0.5\textwidth]{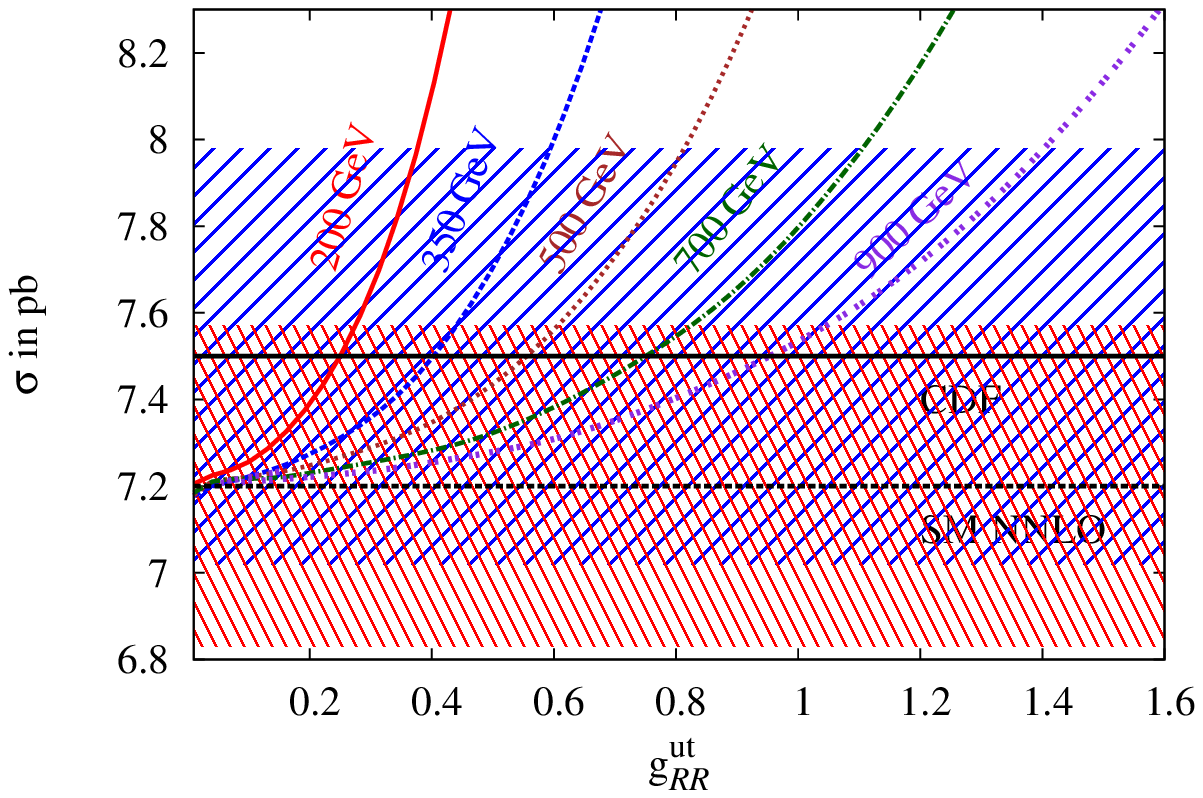}} 
\caption{\small \em {Variation of the cross section $\sigma \left( p\bar p \to t\bar t\right)$ with  couplings $g^{ut}_{i j}$ for flavor violating vector color-octets corresponding to different values of  $M_{V_8^0}$. The upper black dotted line associated with a blue band depicts the cross section  $7.50 \pm 0.48$ pb from the CDF (all channels) ~\cite{Abazov:2011cq}, while the lower black dot-dashed line associated with a red band show theoretical estimate $7.2\pm 0.37$ pb at NNLO  ~\cite{Kidonakis:2012db}. Figures (a) and (b) show the variation of $\sigma$ for axial vector and right-handed cases (the cases (a) and (b) of the text, respectively).}}
\label{ttbarFV}
\end{figure*}
In Figs. \ref{ttbarFC}, \ref{ttbarFCAFB} and \ref{ttbarFCSC} we plot the variation of the cross section, the forward backward asymmetry $\afbt$ and the spin-correlation $C_{t\bar t}$ , respectively as a function of coupling $\sqrt{\lambda_{ij}}$.
\par We look at three different regions based on the octet masses and the threshold of top-pair production, region I where $M_{V_8^0} \ll   2 m_t$,  region II  $M_{V_8^0}-\Gamma_{V_8^0} \le  2 m_t     \le  M_{V_8^0}+\Gamma_{V_8^0}$  and  region III where   $M_{V_8^0}\gg 2 m_t$ , respectively. Fig. \ref{fig:TP_FC_V_sigma} corresponds to the case (a), which implies pure vector interactions. For $M_{V_8^0}$ = 200 GeV which lies in region I,  the cross section grows with coupling due to the  
positive interference inspite of the $s$ channel suppression. For $M_{V_8^0}$ = 350-500 GeV which lies in the resonant  region II, we have a much sharper rise in the cross section with the increasing  coupling. For higher octet masses we observe the effect of negative interference with SM  as long as the coupling $\left\vert \lambda_{ij}\right\vert \le 1$ and then it gradually grows with the couplings due to the dominance of the new physics squared term. However, pure vector interactions fail to generate the \afbt.
\par Figure \ref{fig:TP_FC_A_sigma} corresponds to phenomenologically interesting case (b), which is purely an axial interaction and thus generates a large  forward-backward asymmetry. Since the interference terms contributing to the cross section vanish in this case and only the squared term grows with coupling, we observe that this generates an increasing \afbt without enhancing the cross section for the couplings  $\left\vert \lambda_{ij}\right\vert \le 1$. However for the higher masses in region III, the \afbt becomes negative due to the negative interference with SM as shown in Fig. \ref{fig:TP_FC_afb_NA}. 
\par The fair amount of  \afbt can also be generated through axial current for large masses by taking the negative product of the axial couplings  of light quark and the top quark, keeping the cross section same as before. We depict the  \afbt contribution for the negative product of axial couplings in Fig. \ref{fig:TP_FC_afb_A}.
\par The variation of cross section for case (c) is given in Fig. \ref{fig:TP_FC_R_sigma}  which is similar to case (a). In this case we get positive $A_{FB}$ for all the cases but it grows faster for $M_{V_8^0}$ = 500 GeV with respect to $\sqrt{\lambda_{RR}}$ comparing to other two masses of ${V_8^0}$ as shown in Fig. \ref{fig:TP_FC_afb_R}.
\par The behavior of the variation of the spin-correlation coefficients can be understood by exhibiting  the total    matrix element squared as a combination of the same and opposite helicity amplitudes. The differential cross section corresponding to the same and opposite helicity amplitudes are
\begin{widetext}
\begin{eqnarray}
\frac{d{\hat\sigma}^{\rm FC}_S}{d\cos\theta} =&& g_s^4 (1-\beta^2) \sin^2\theta \Big[ 8 + \frac{2 \hat s (\hat s - M_{V_8^0}^2)}{(\hat s - m_{V_8^0}^2)^2 + \Gamma_{V_8^0}^2 M_{V_8^0}^2}(\lambda_{LL}+\lambda_{RR}+\lambda_{LR}+\lambda_{RL}) \cr 
&&\qquad \qquad \qquad \qquad \,+  \frac{ {\hat s}^2}{(\hat s - M_{V_8^0}^2)^2 + \Gamma_{V_8^0}^2 M_{V_8^0}^2} (\lambda_{LL}^2+\lambda_{RR}^2+\lambda_{LR}^2+\lambda_{RL}^2+2(\lambda_{LL}\lambda_{LR}+\lambda_{RL}\lambda_{RR})) \Big] 
\label{same_hel_FC}
\end{eqnarray}
\begin{eqnarray}
\frac{d{\hat\sigma}^{\rm FC}_O}{d\cos\theta} =&& g_s^4 (1+\cos^2\theta) \Big[ 8 + \frac{\hat s (\hat s - M_{V_8^0}^2)}{(\hat s - M_{V_8^0}^2)^2 + \Gamma_{V_8^0}^2 M_{V_8^0}^2}(\lambda_{LL}+\lambda_{RR}+\lambda_{LR}+\lambda_{RL}) \cr &&\qquad +  \frac{ {\hat s}^2}{(\hat s - M_{V_8^0}^2)^2 + \Gamma_{V_8^0}^2 M_{V_8^0}^2} \Big\{ (1+\beta^2)(\lambda_{LL}^2+\lambda_{RR}^2+\lambda_{LR}^2+\lambda_{RL}^2)+ 2 (1-\beta^2)(\lambda_{LL}\lambda_{LR}+\lambda_{RL}\lambda_{RR}) \Big\} \Big] \cr &&\qquad +\, g_s^4 (2\beta\cos\theta)\Big[ \frac{\hat s (\hat s - M_{V_8^0}^2)}{(\hat s - M_{V_8^0}^2)^2 + \Gamma_{V_8^0}^2 M_{V_8^0}^2}(\lambda_{LL}+\lambda_{RR}-\lambda_{LR}-\lambda_{RL}) \cr &&\qquad +  \frac{ {\hat s}^2}{(\hat s - M_{V_8^0}^2)^2 + \Gamma_{V_8^0}^2 M_{V_8^0}^2} 2 (\lambda_{LL}^2+\lambda_{RR}^2-\lambda_{LR}^2-\lambda_{RL}^2)\Big] 
\label{opp_hel_FC}
\end{eqnarray}
\end{widetext}
Examining Eqs. \eqref{same_hel_FC} and \eqref{opp_hel_FC}, we find  that for  case (a), the contribution from the $ \frac{d{\hat\sigma}^{\rm FC}_S}{d\cos\theta}$ is suppressed by the factor $\left(1-\beta^2\right) \sin^2\theta /\left(1+\cos^2\theta\right) $ with respect to $\frac{d{\hat\sigma}^{\rm FC}_O}{d\cos\theta}$, while for the  case (b), only new physics squared term from $\frac{d{\hat\sigma}^{\rm FC}_O}{d\cos\theta}$ contributes to the spin-correlation coefficient. For the right-chiral current case (c), we observe that  for the   interference term the ratio  is again suppressed by the same  factor as in case (a), while it is further suppressed by the factor $1/\left(1+\beta^2\right)$ for the  squared term. Therefore it is evident that \spincorr is likely to increase with  the increasing coupling products. 
\begin{figure*}[!ht]
  \centering
  \subfloat[]{\label{fig:TP_FV_A_afb}\includegraphics[width=0.5\textwidth]{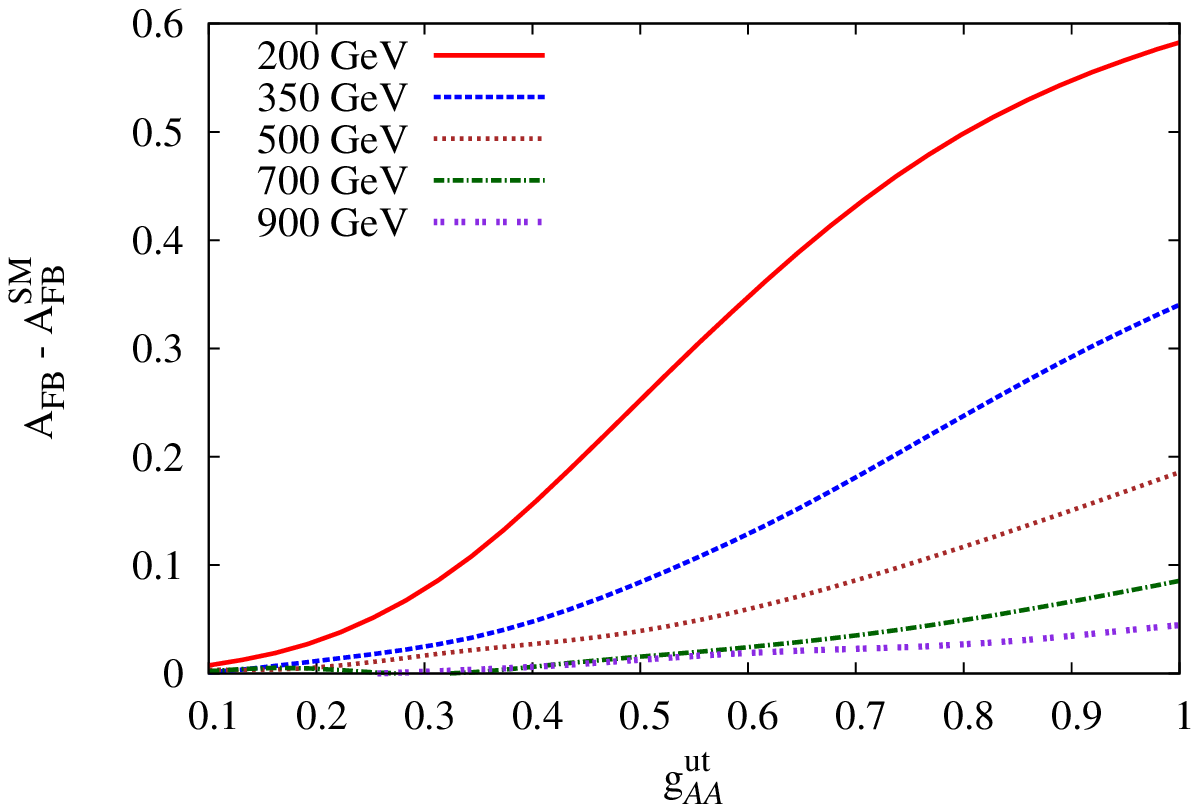}}
  \subfloat[]{\label{fig:TP_FV_R_afb}\includegraphics[width=0.5\textwidth]{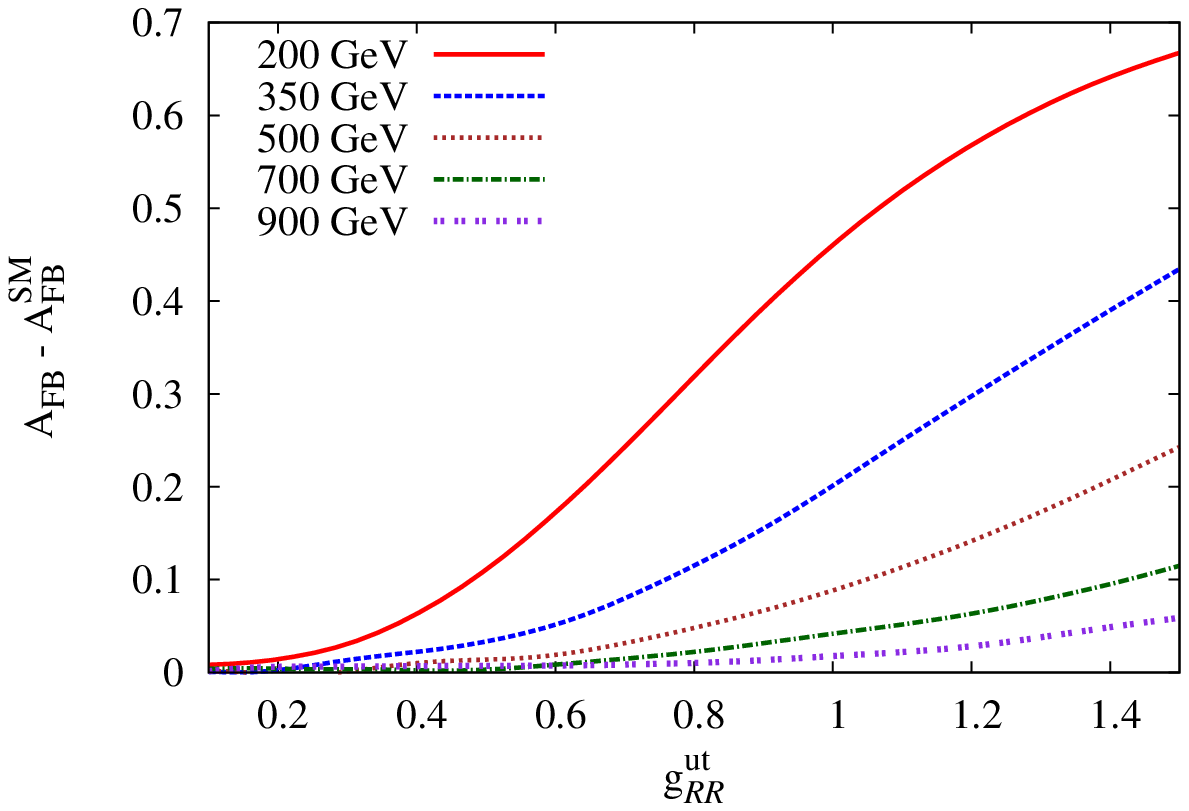}} 
\caption{\small \em {Variation of the $A_{FB} - A_{FB}^{SM}$ with couplings $g^{ut}_{ij}$ for flavor violating  vector color-octets corresponding to  different values of  $M_{V_8^0}$. Figures (a) and (b) corresponds to axial vector and right-handed cases (the cases (a) and ( b) of the text, respectively).}}
\label{ttbarFVAFB}
\end{figure*}
%%%%%%%%%%%%%fig ttbar fv SC 
\begin{figure*}[!ht]
  \centering
  \subfloat[]{\label{fig:TP_FV_A_spcr}\includegraphics[width=0.5\textwidth]{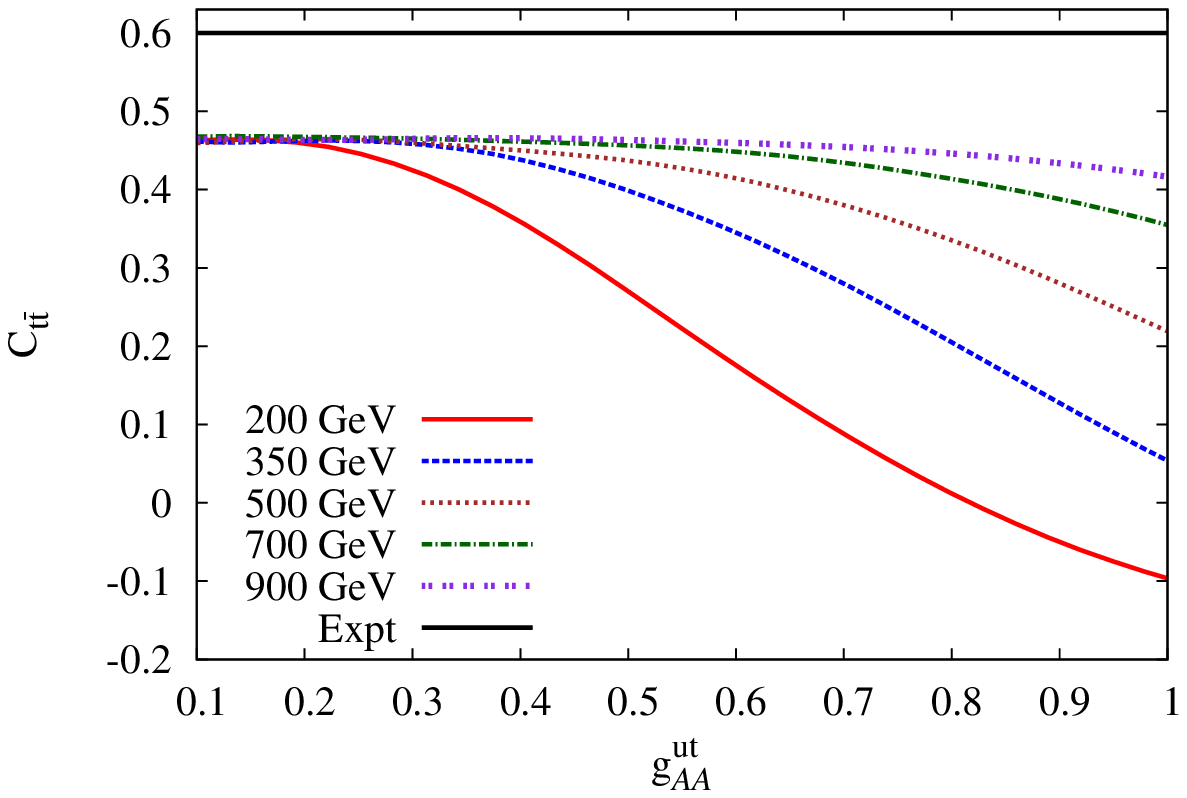}}
  \subfloat[]{\label{fig:TP_FV_R_spcr}\includegraphics[width=0.5\textwidth]{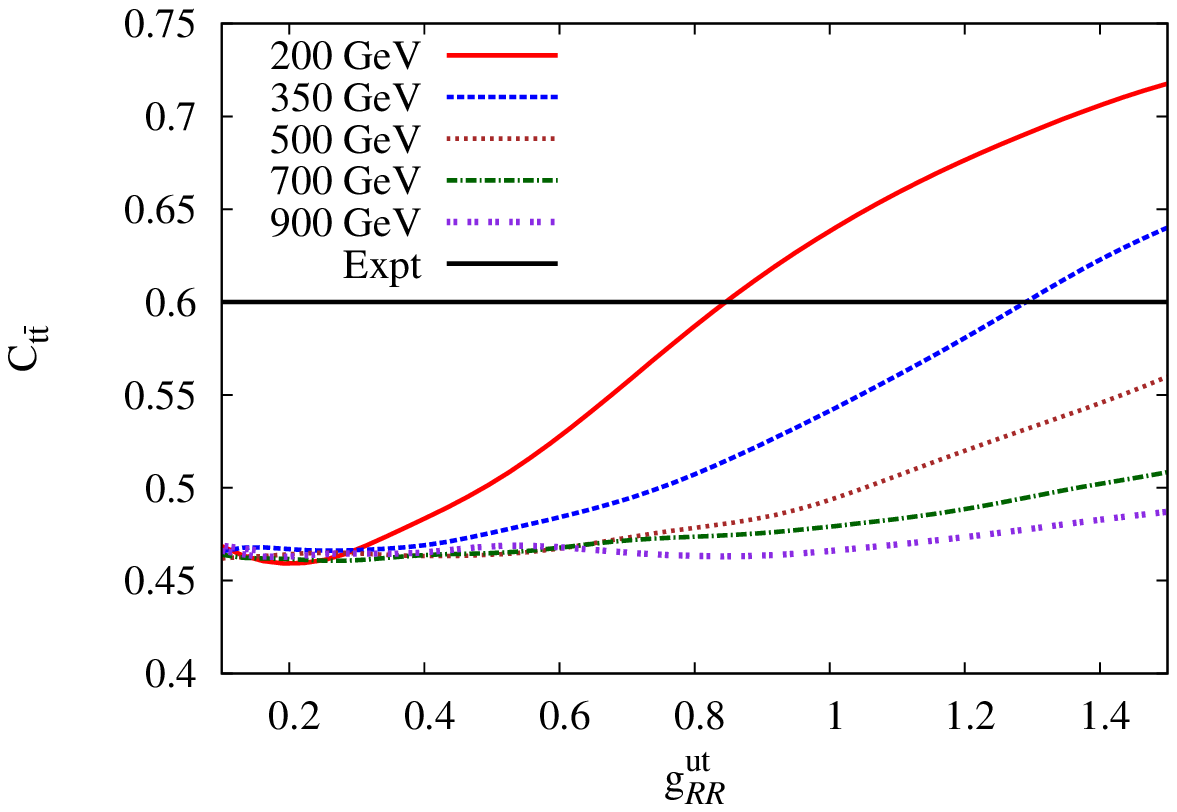}} 
\caption{\small \em {Variation of the spin-correlation coefficient $C^{t\bar t}$ with couplings $g^{ut}_{i j}$ for flavor violating  vector color-octets corresponding to  different values of  $M_{V_8^0}$. Figures (a) and (b) corresponds to axial vector and right-handed cases (the cases (a) and ( b) of the text, respectively).}}
\label{ttbarFVSC}
\end{figure*}
%%%%%%%%%%%%%%%%%%%
\par The variation of spin-correlation coefficient $\spincorr$ with respect to $\sqrt{\lambda_{ij}}$ has almost similar behavior for all the cases considered as shown in Figs. \ref{fig:TP_FC_spcr_V} -\ref{fig:TP_FC_spcr_R}. Here the \spincorr in region I first decreases  with couplings due to negative interference and then increases due to the dominance of the squared term. For the octet mass 350 GeV at the region II the \spincorr registers the minimum value showing that at threshold production it is likely to have an equal number of parallel and antiparallel states. In axial cases (b) and (c), since there is no interference, the \spincorr increases with couplings for all mass regions which is evident from Fig. \ref{fig:TP_FC_spcr_A}.
\subsection{Flavor Violating}
In the flavor violating case, apart from the usual SM diagrams we have $t$-channel diagrams for $u\bar u \to t\bar t$ with $V_8^0$ as shown in Fig. \ref{fig:ttbarfv} which interfere with the corresponding $s$-channel SM diagrams initiated with $u$ and $\bar u$ partons. The flavor violating neutral coupling is considered only among the first and third generation $u$ and $t$ quarks $g^{ut}_{i}$ ($i=L,R$) with $V_8^0$. In contrast to the flavor conserving case, here we consider only two choices of coupling combinations (a) $g^{ut}_L$ = $g^{ut}_R$ or $g^{ut}_L$ = -$g^{ut}_R$ and (b) $g^{ut}_L$ = 0 or $g^{ut}_R$ = 0 to study the three observables because corresponding matrix element square is symmetric in both the cases. The differential cross section in terms of flavor violating neutral coupling $g^{ut}_i$ for $u\bar u \to t \bar t$ with QCD $s$-channel and additional $t$-channel diagram through $V_8^0$ (NP) with respect to the cosine of the top quark polar angle $\theta$ in the $t\bar t$ center-of-mass (c.m.) frame is
\begin{widetext}
\begin{eqnarray}
\frac{d\hat \sigma}{d\,\cos\theta} =&& \frac{\pi \beta \alpha_s^2}{9\hat s}\Big(2 - \beta^2 \sin^2\theta \Big) -\frac{\pi \beta {\alpha_s}^2}{54{\hat s}^2} \frac{ {\hat s}^2 (\hat t -M_{V_8^0}^2) }{ (\hat t -M_{V_8^0}^2)^2 + \Gamma_{V_8^0}^2 M_{V_8^0}^2} (1+\beta \cos\theta)^2 ({g_L^{ut}}^2 + {g_R^{ut}}^2) \cr
&& + \frac{\pi \beta {\alpha_s}^2}{36\hat s} \frac{{\hat s}^2}{(\hat t - M_{V_8^0}^2)^2 + \Gamma_{V_8^0}^2 M_{V_8^0}^2 }  \Big[ ({g_L^{ut}}^4 + {g_R^{ut}}^4) \Big(1+\beta \cos\theta \Big)^2 + 8 {g_L^{ut}}^2 {g_R^{ut}}^2 (1+\beta^2)\Big] \label{tchanttbarcrosssec}
\end{eqnarray}
where $\hat t = (p_u - p_t)^2 = (p_{\bar u} - p_{\bar t})^2$. 
\end{widetext}
Figure \ref{ttbarFV} shows the variation of cross section as a function of coupling $g^{ut}_{ij} , (i,j = L$ or $R)$ for both cases (a) and (b) for different $M_{V_8^0}$ same as in FC case.
\par We observe the growth of the cross section with the couplings due to the overall positive interference in Eq. \eqref{tchanttbarcrosssec} generated from  large negative value of $ \hat t$. It is  also observed that the variations are comparatively flat with respect to the corresponding cases of the flavor conserving scenarios due to the suppressed $\hat t$ channel propagator $\left(\hat t-m_{V_8}^2 \right)^2+  \Gamma_{V_8^0}^2 M_{V_8^0}^2$ in the interference term.
\par $A_{FB}$ as a function of couplings for both  cases is plotted in  Fig. \ref{ttbarFVAFB}. We find that $A_{FB}$ is positive in both cases and increases with coupling, the increase being more rapid for lower mass. 
\par The behavior of \spincorr  can be studied by writing the matrix element squared in terms of the same and opposite helicity contributions as before. From Eqs. \eqref{A3}-\eqref{A5}, we define $\frac{d{\hat\sigma}^{\rm FV}_S}{d\cos\theta}$ and $\frac{d{\hat\sigma}^{\rm FV}_O}{d\cos\theta}$ as 
\begin{widetext}
\begin{eqnarray}
\frac{d{\hat\sigma}^{\rm FV}_S}{d\cos\theta} =&& g_s^4 (1-\beta^2) \sin^2\theta \Big[ 8 + \tfrac{2}{3}\frac{\hat s (\hat t - M_{V_8^0}^2)}{(\hat t - M_{V_8^0}^2)^2+ \Gamma_{V_8^0}^2 M_{V_8^0}^2}({g_L^{ut}}^2 + {g_R^{ut}}^2) + \frac{{\hat s}^2}{(\hat t - M_{V_8^0}^2)^2 + \Gamma_{V_8^0}^2 M_{V_8^0}^2}({g_L^{ut}}^4 + {g_R^{ut}}^4) \Big]  \cr &&+ \,8 \,g_s^4 \,\frac{{\hat s}^2}{(\hat t - M_{V_8^0}^2)^2 + \Gamma_{V_8^0}^2 M_{V_8^0}^2} {g_L^{ut}}^2  {g_R^{ut}}^2 (1+\beta^2) 
\label{same_hel_FV}
\end{eqnarray}
\begin{eqnarray}
\frac{d{\hat\sigma}^{\rm FV}_O}{d\cos\theta} =&& g_s^4 (1+\cos^2\theta) \Big[ 8 + \tfrac{2}{3}\frac{\hat s (\hat t - M_{V_8^0}^2)}{(\hat t - M_{V_8^0}^2)^2 + \Gamma_{V_8^0}^2 M_{V_8^0}^2}({g_L^{ut}}^2 + {g_R^{ut}}^2) + \frac{{\hat s}^2}{(\hat t - M_{V_8^0}^2)^2 + \Gamma_{V_8^0}^2 M_{V_8^0}^2}({g_L^{ut}}^4 + {g_R^{ut}}^4) (1+\beta^2) \Big]  \cr && + \, g_s^4 \,(2\beta\cos\theta) \Big[\tfrac{2}{3}\frac{\hat s (\hat t - M_{V_8^0}^2)}{(\hat t - M_{V_8^0}^2)^2 + \Gamma_{V_8^0}^2 M_{V_8^0}^2}({g_L^{ut}}^2 + {g_R^{ut}}^2) + \frac{2{\hat s}^2}{(\hat t - M_{V_8^0}^2)^2 + \Gamma_{V_8^0}^2 M_{V_8^0}^2}({g_L^{ut}}^4 + {g_R^{ut}}^4) \Big] 
\label{opp_hel_FV}
\end{eqnarray}
\end{widetext} 
Analyzing the expression given in Eqs. \eqref{same_hel_FV} and \eqref{opp_hel_FV} we find that the contribution of  $\frac{d{\hat\sigma}^{\rm FV}_S}{d\cos\theta}$ in vector/axial-vector case dominates over the $\frac{d{\hat\sigma}^{\rm FV}_O}{d\cos\theta}$  due to the presence of the  cross term proportional to $\left(g_L^{ut}\, g_R^{ut}\right)^2$ and hence  spin-correlation decreases with the coupling as shown in the Fig. \ref{fig:TP_FV_A_spcr}. In contrast, this cross term vanishes for the pure right-handed interactions and then the $\frac{d{\hat\sigma}^{\rm FV}_S}{d\cos\theta}$ is suppressed by $(1-\beta^2)\sin^2\theta /  (1+\beta^2)(1+\cos^2\theta )$ with respect to $\frac{d{\hat\sigma}^{\rm FV}_O}{d\cos\theta}$, rendering \spincorr to increase with the coupling as shown in Fig. \ref{fig:TP_FV_R_spcr}.
\par Our results are in broad agreement with the existing results in the literature \cite{axig,Wang:2011hc,wang:2011taa,colorons}. Their study was however based on the earlier results from Tevatron \cite{Aaltonen:2011kc} with the \afbt dependence on \mttb in two  regions  of $\le 450$ GeV and $\ge 450$ GeV, respectively.
\subsection{$\chi^2$ Analysis} 
\label{chi2section}
We have studied the production cross section and also the model  contribution to the observables. We further analyze the one dimensional distribution
 plots and investigate the possibility to explain the observed \afbt anomaly.
Recently  CDF collaboration  have performed  detail seven bin analysis with  invariant mass distribution of \afbt  from the reconstructed top-pairs \cite{CDF:10807}. They observed that the large forward-backward asymmetry comes from the higher invariant mass $M_{t\bar t}$ bins of the top-antitop pair.  The forward-backward asymmetry as a function of $M_{t\bar t}$ is defined as 
\begin{eqnarray}
A_{FB}(M_{t\bar t}) = \frac{N_F(M_{t\bar t}) -N_B(M_{t\bar t})}{N_F(M_{t\bar t}) + N_B(M_{t\bar t})}
\end{eqnarray}
where $N_F$ and $N_B$ are the events in the forward and backward region, respectively. The analysis in Ref. \cite{CDF:10807} also gives the four bin analysis of the \afbt with the top-antitop rapidity difference distribution defined as
\begin{eqnarray}
A_{FB}(|\Delta y|) = \frac{N(\Delta y > 0) -N(\Delta y < 0)}{N(\Delta y > 0) + N(\Delta y < 0)}
\end{eqnarray}
where rapidity difference $\Delta y = y_t - y_{\bar t}$, $N(\Delta y < 0)$ and $N(\Delta y > 0)$ are the number of events with positive and negative rapidity difference, respectively.
\par We scan our model parameter space for a given mass of the color-octet vector boson on the two dimensional plane of two distinct  product of couplings which can provide the matched \mttb  and $\Delta y$  distribution of \afbt   with the observed data. We perform $\chi^2$ analysis  for
both FC and FV cases and predict the set of best parameters which can possibly explain the \afbt anomaly.
To make this $\chi^2$ study we  take into account the \afbt  distribution  over \mttb bins as well as $\Delta y$ bins from the full Run II Tevatron dataset \cite{CDF:10807}. We define the combined $\chi^2$ from the study of the \mttb and  $\Delta y$  distribution.
For these analysis we use standard $\chi^2$ fit, defined as 
\begin{eqnarray}
\chi^2 = \sum_i \frac{({\cal O}_i^{\rm th}-{\cal O}_i^{\rm exp})^2}{(\delta {\cal O}_i)^2}
\end{eqnarray} 
where $i$ is the \mttb or $\Delta y$ bin index, ${\cal O}_i^{\rm th}$  and ${\cal O}_i^{\rm exp}$ are model and experimental estimate of the \afbt in the $i^{\rm th}$ bin, respectively. The model estimate includes the both SM  and new physics contribution. $\delta {\cal O}_i$  is the the experimental error  in the corresponding  $i^{\rm th}$ bin. We have considered seven and four suggested bins for the \mttb and $\Delta y$, respectively. In addition we have also considered the total cross section $\sigma (p\bar p\to t\bar t) =
7.5 \pm 0.31 ({\rm stat}) \pm 0.34 ({\rm syst}) \pm 0.15 ({\rm Z \, theory})$ pb \cite{cdf:9913} as one of the observed data.  Therefore we have used twelve observables to perform the analysis. For ${\cal O}_i^{\rm th}$ we have taken the total cross section for this study as $\sigma^{tot} = K \cdot \sigma^{SM} + \sigma^{NP}$, where $K = \frac{\sigma^{\rm NLO}}{\sigma^{\rm SM}} = 1.046$ and $\afbt = \afbt^{\rm SM\, \rm NLO} + \afbt^{\rm NP}$. 
\par
\begin{figure*}[ht]
\centering
\subfloat[$\lambda_{LL}=\lambda_{RR}=-\lambda_{RL}=-\lambda_{LR}=\lambda_{AA} =.30$]{\label{fig:TP_FC_mtt_A}\includegraphics[width=0.5\textwidth]{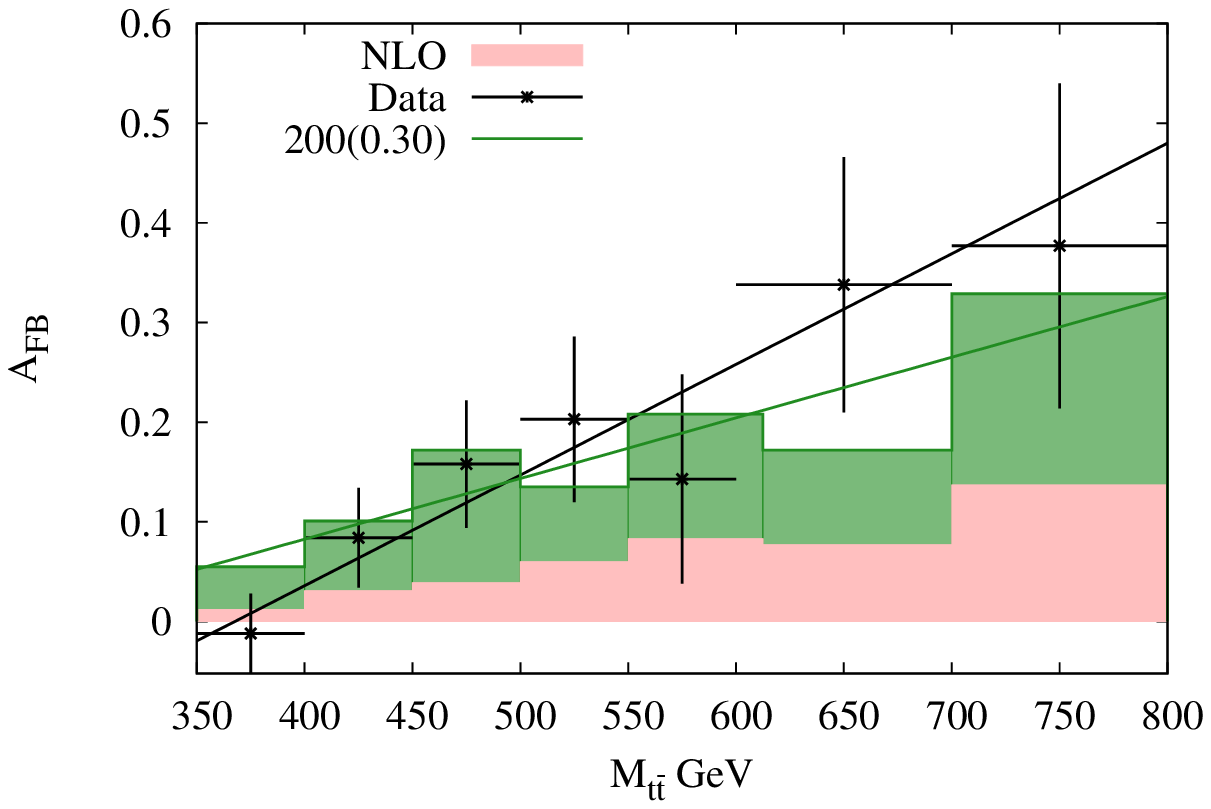}}
\subfloat[$\lambda_{LL}=\lambda_{RR}=-\lambda_{RL}=-\lambda_{LR}=\lambda_{AA}=.30 $]{\label{fig:TP_FC_dy_A}\includegraphics[width=0.5\textwidth]{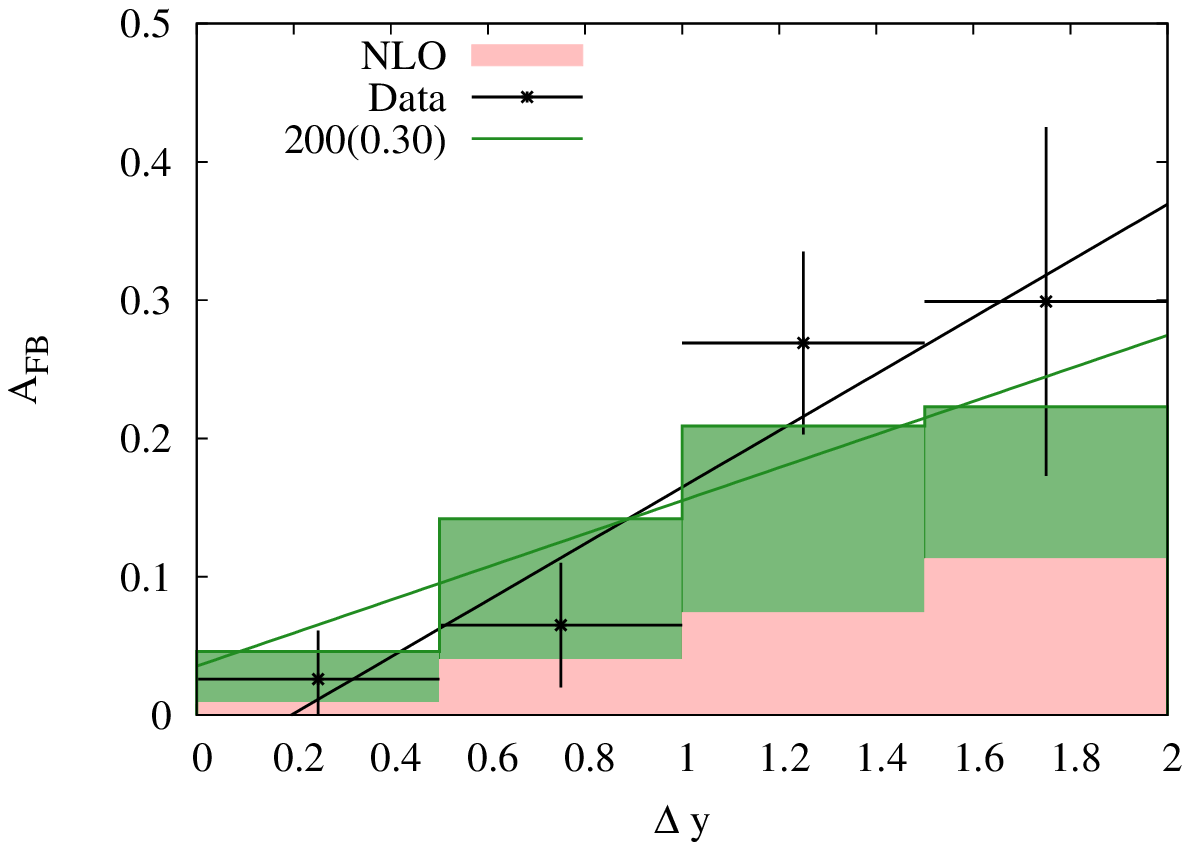}}  \\
\subfloat[$\lambda_{RR} =.11 \ne \lambda_{LR}=\lambda_{RL}=\lambda_{LL} =0 $]{\label {fig:TP_FC_mtt_R}\includegraphics[width=0.5\textwidth]{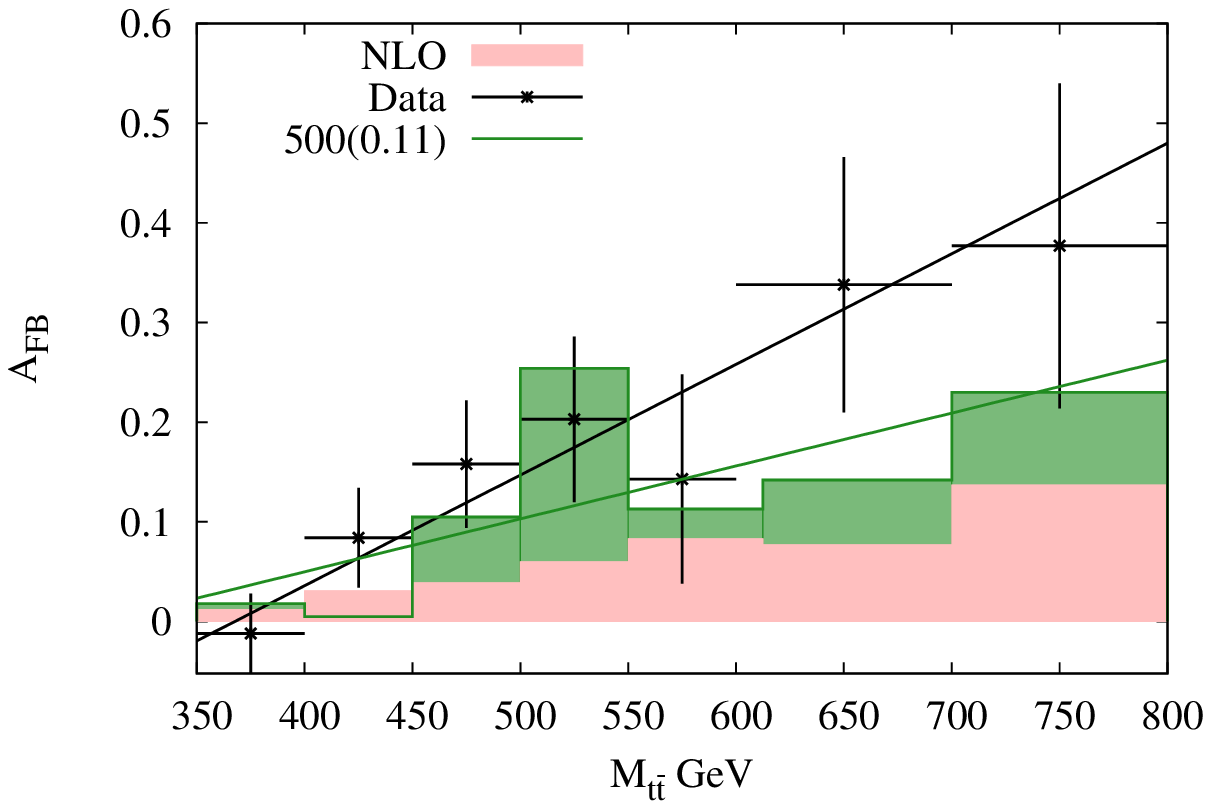}}
\subfloat[$\lambda_{RR} =.11  \ne \lambda_{LR}=\lambda_{RL}=\lambda_{LL} =0 $]{\label {fig:TP_FC_dy_R}\includegraphics[width=0.5\textwidth]{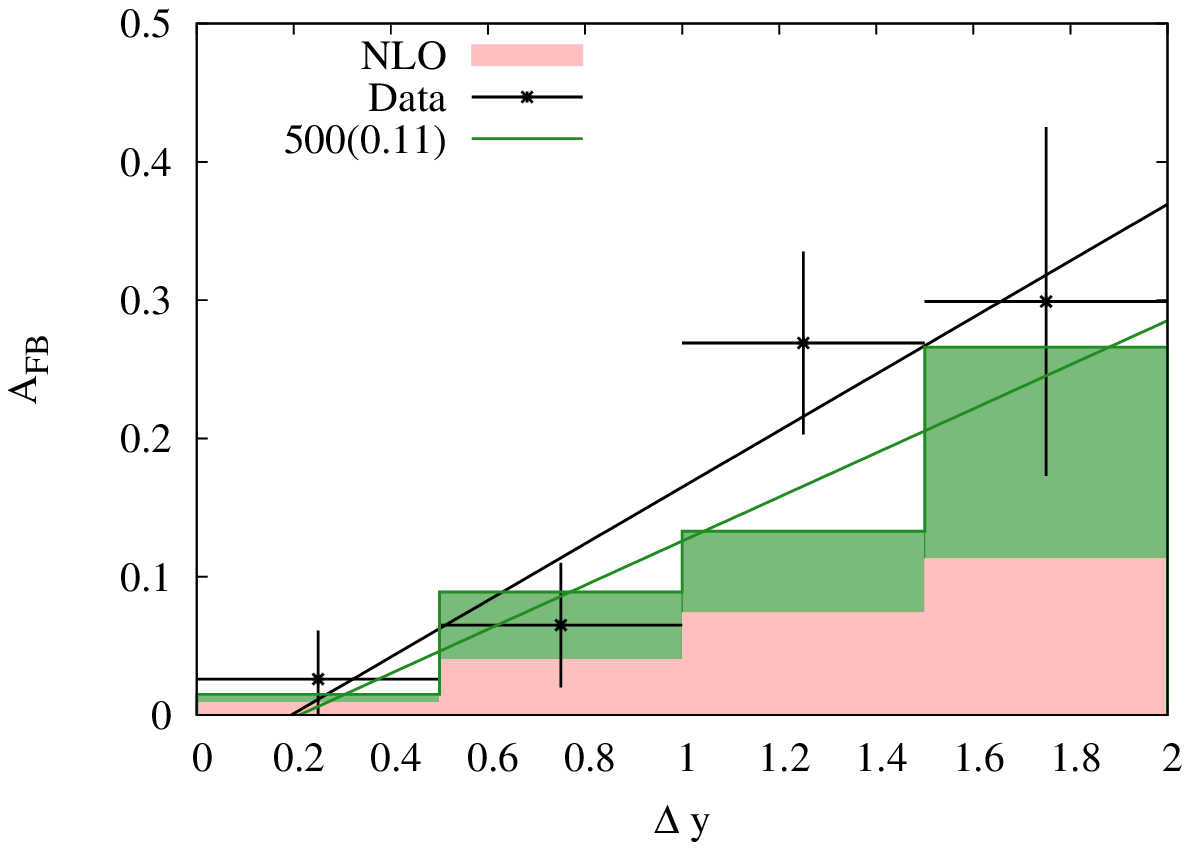}} \\
\subfloat[$-\lambda_{LL}=-\lambda_{RR}=\lambda_{RL}=\lambda_{LR}=\lambda_{NA}=.35 $]{\label {fig:TP_FC_mtt_NA}\includegraphics[width=0.5\textwidth]{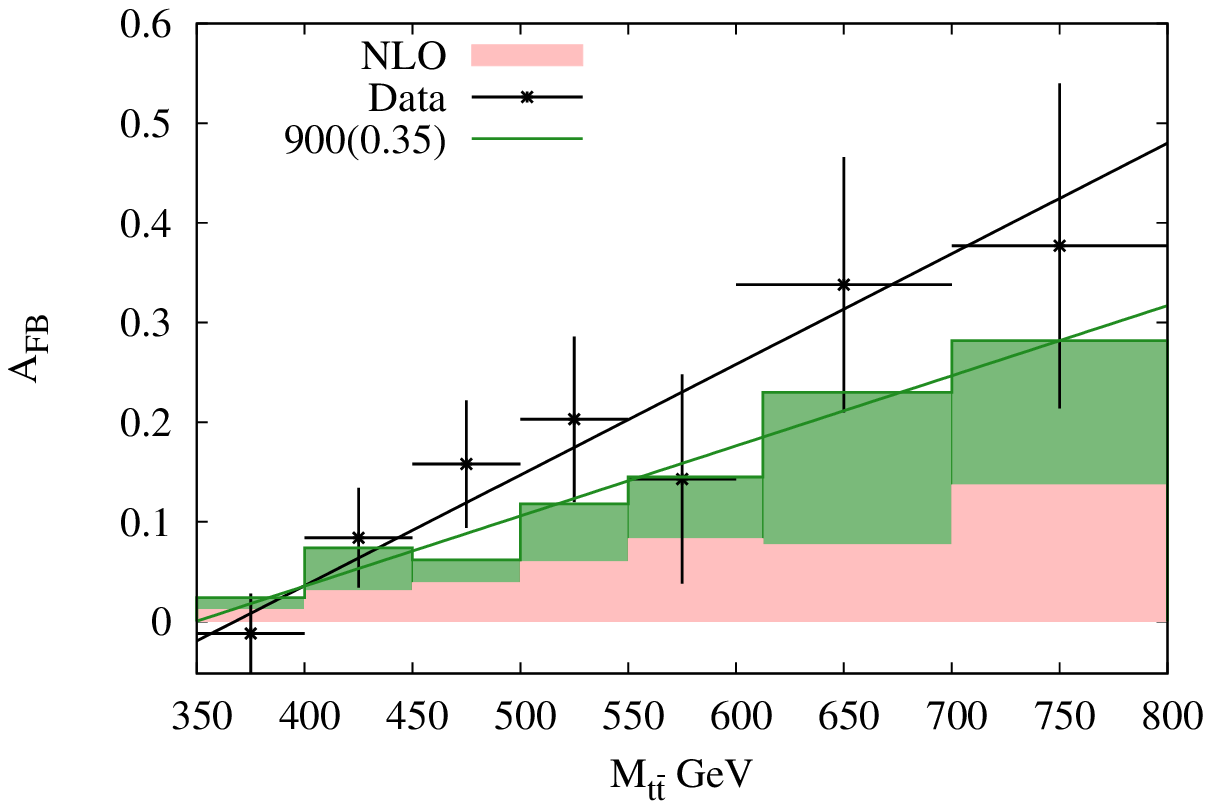}}
\subfloat[$-\lambda_{LL}=-\lambda_{RR}=\lambda_{RL}=\lambda_{LR}=\lambda_{NA} =.35$]{\label {fig:TP_FC_dy_NA}\includegraphics[width=0.5\textwidth]{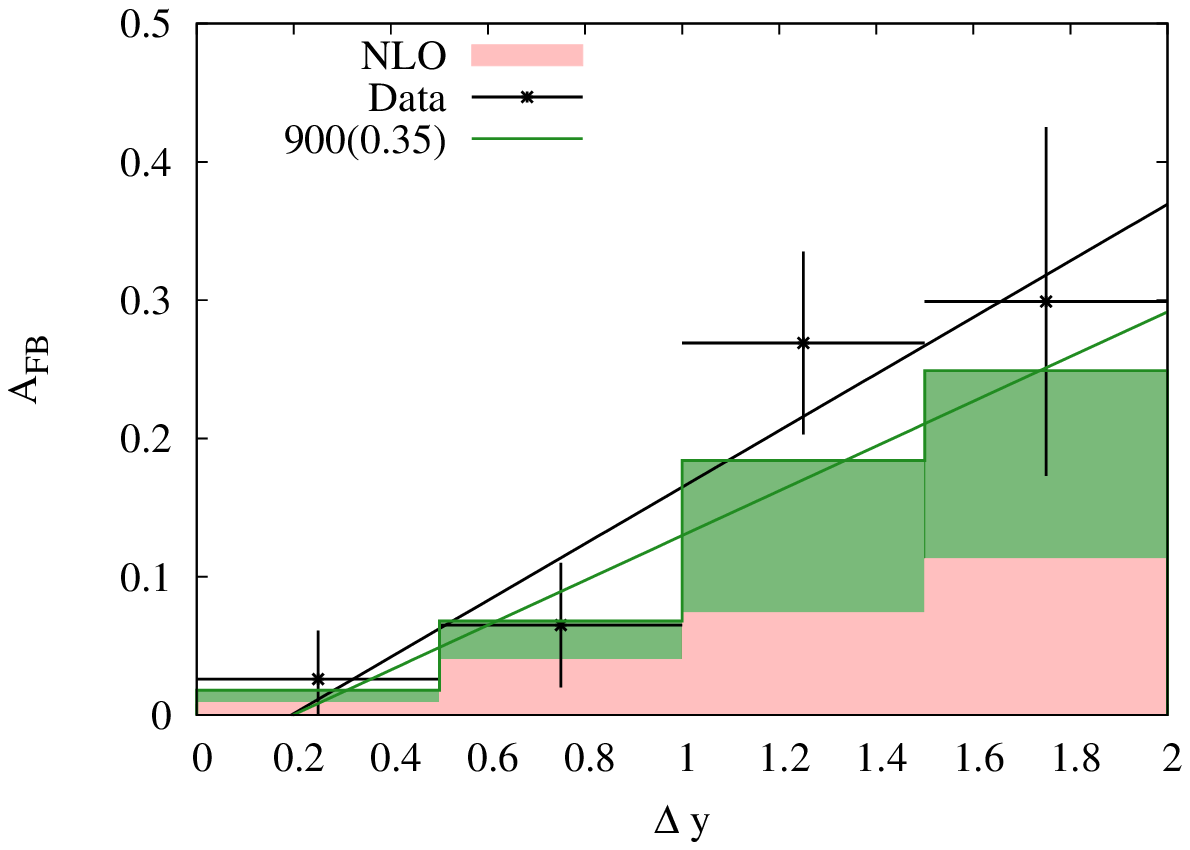}}
  \caption{\small \em{ $\mttb$ and $|\Delta y|$ distribution of \afbt at $\chi^2_{\rm min.}$ for the three favorable point-sets in parameter space at fixed $M_{V_8^0} = 200,500$ and 900 GeV for flavor conserving case, shown in the shaded green histogram. The experimental data point is shown with its error in black, while the SM (NLO+QCD) with background subtracted are shown in the shaded pink histogram. The black line in all  graphs is the best-fit line with the experimental data while the green line  depicts   the best-fit  line with  the model data.}}
\label{afbmttdata_FC}
\end{figure*}

\begin{figure*}[ht]
\centering
\subfloat[$g_R^{ut}=-g_L^{ut}=.26 $]{\label{fig:TP_FV_mtt_A1}\includegraphics[width=0.5\textwidth]{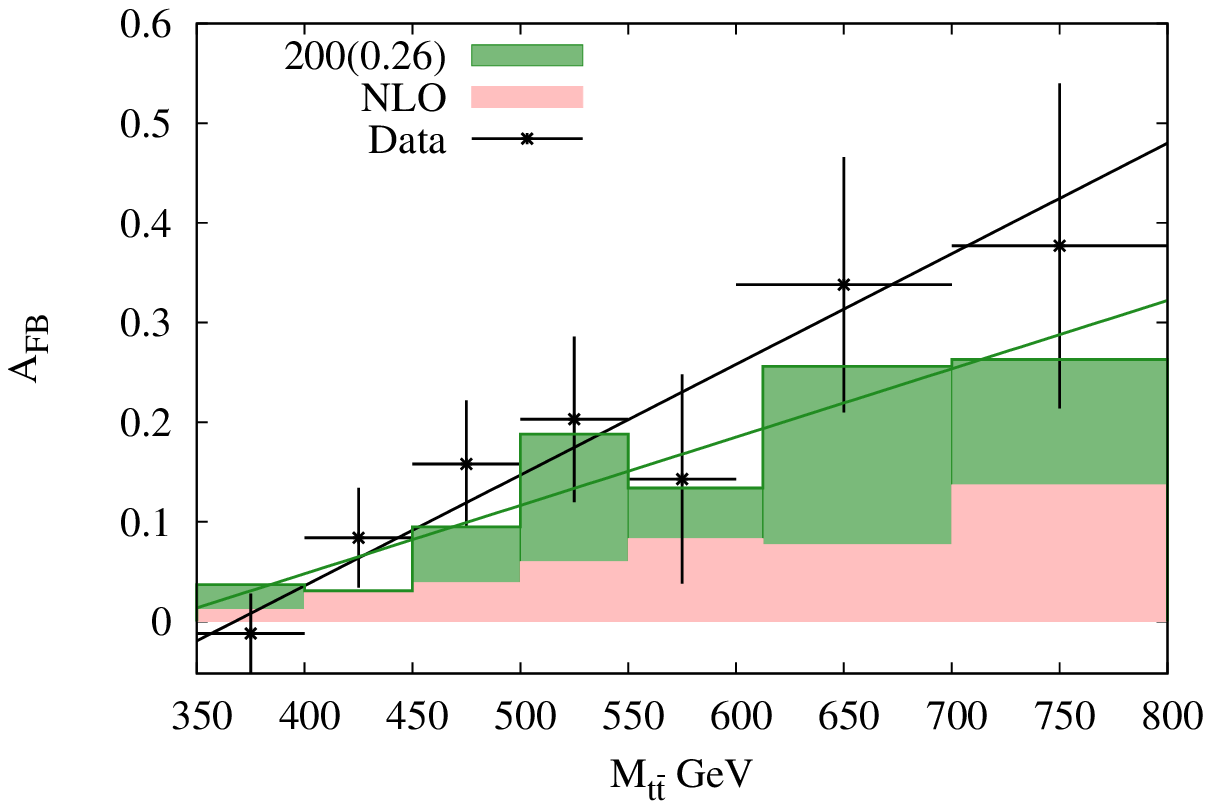}}
\subfloat[$g_R^{ut}=-g_L^{ut}=.26 $]{\label{fig:TP_FV_dy_A1}\includegraphics[width=0.5\textwidth]{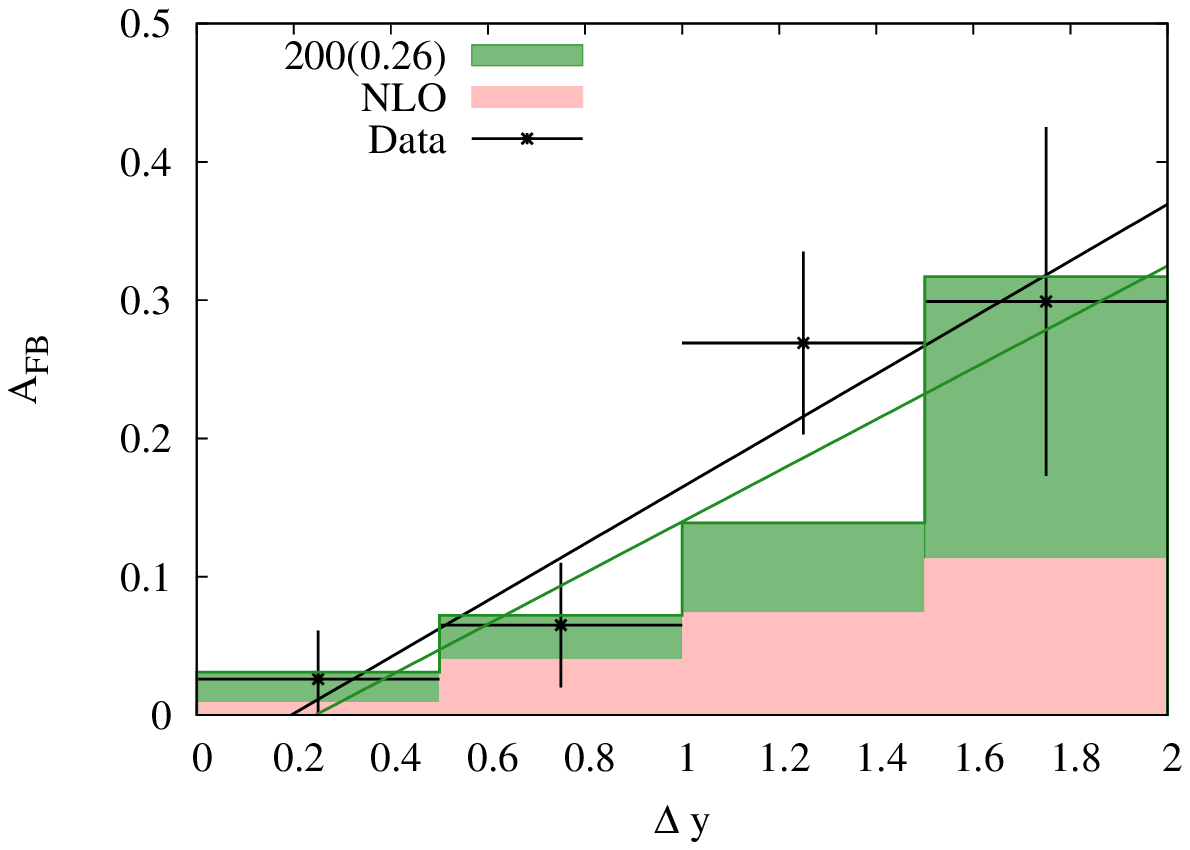}} \\
\subfloat[$g_R^{ut}=-g_L^{ut}=.53 $]{\label{fig:TP_FV_mtt_A2}\includegraphics[width=0.5\textwidth]{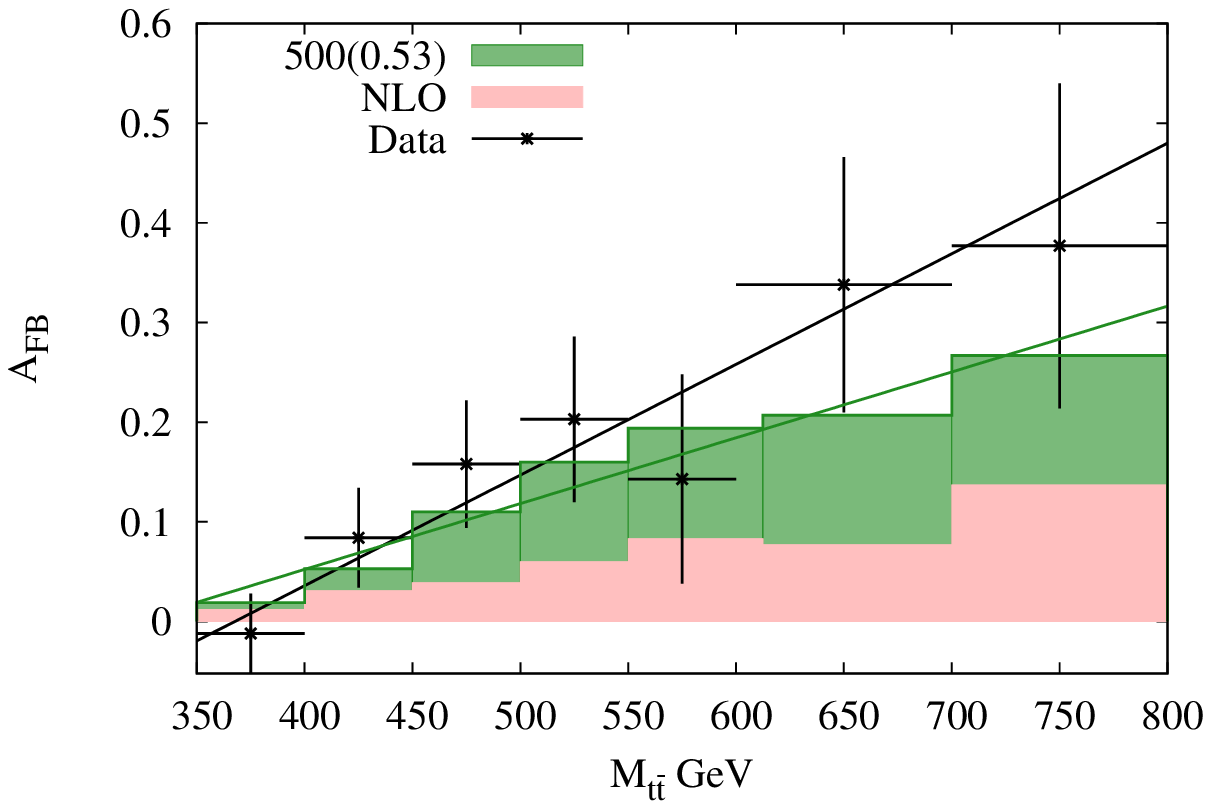}}
\subfloat[$g_R^{ut}=-g_L^{ut}=.53 $]{\label{fig:TP_FV_dy_A2}\includegraphics[width=0.5\textwidth]{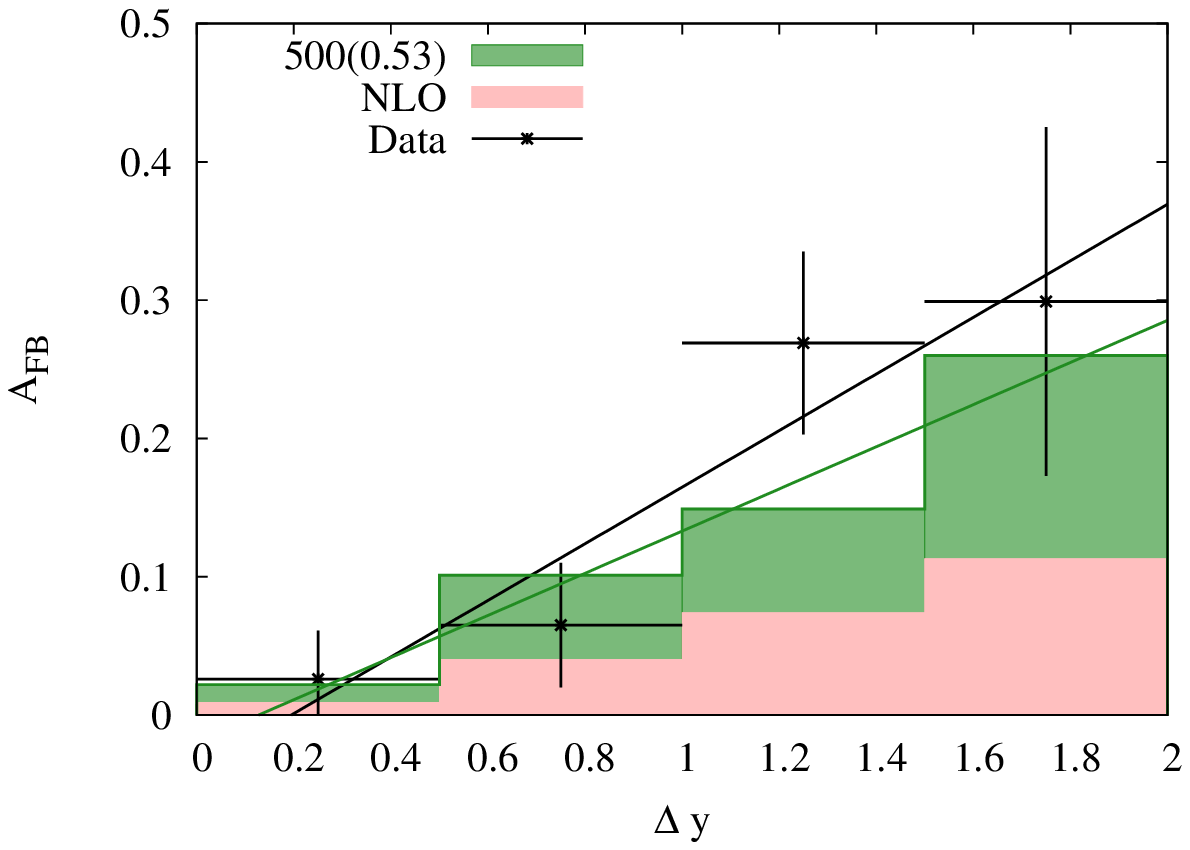}} \\
\subfloat[$g_R^{ut}=1.26\ne g_L^{ut}=0$]{\label {fig:TP_FV_mtt_R}\includegraphics[width=0.5\textwidth]{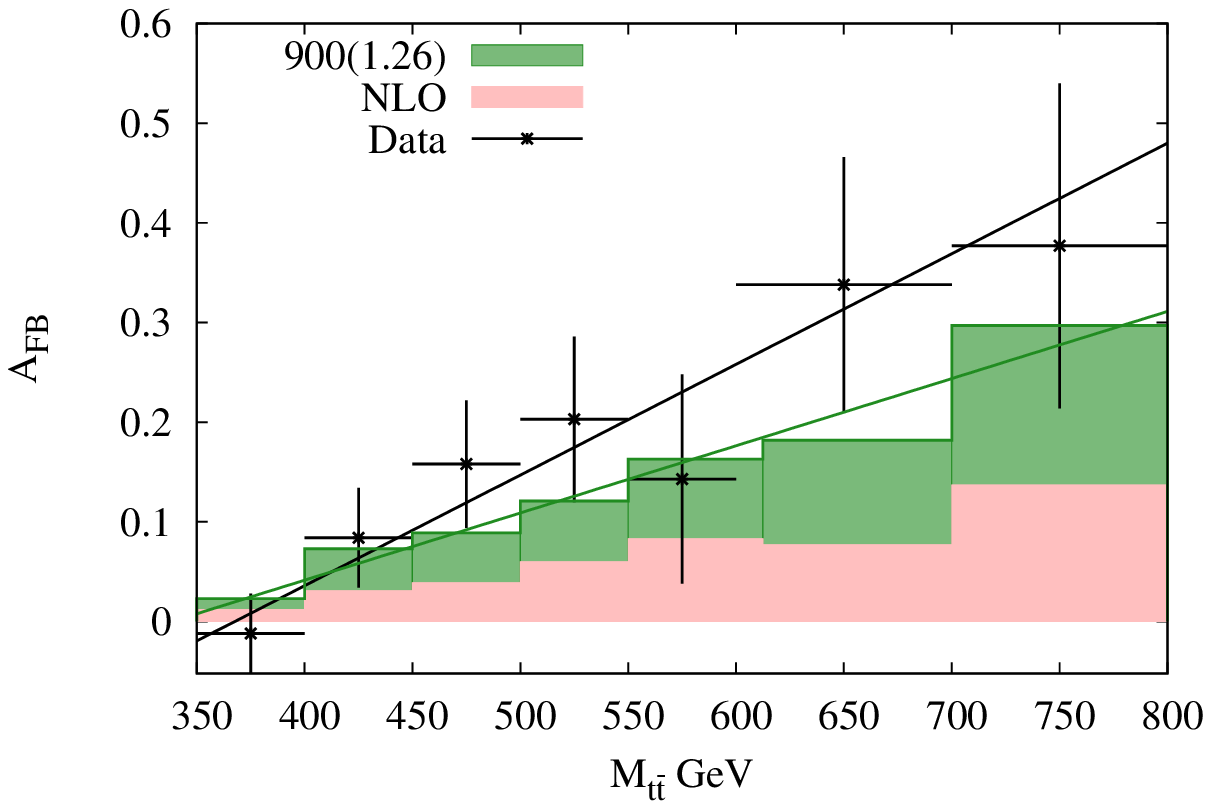}}
\subfloat[$g_R^{ut}=1.26\ne g_L^{ut}=0 $]{\label {fig:TP_FV_dy_R}\includegraphics[width=0.5\textwidth]{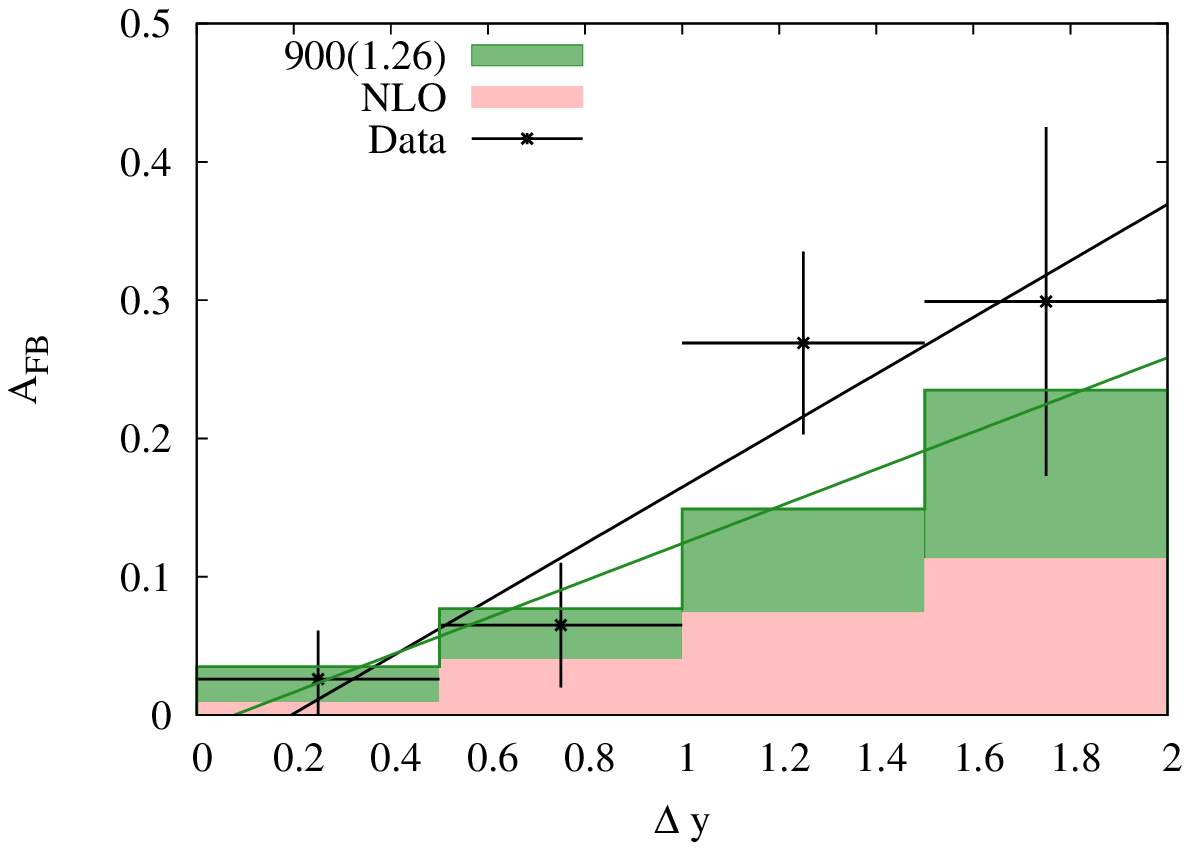}} 
  \caption{\small \em{ $\mttb$ and $|\Delta y|$ distribution of \afbt at $\chi^2_{\rm min.}$ for the three favorable point-sets in parameter space at fixed $M_{V_8^0} = 200,500$ and 900 GeV for flavor violating case, shown in the shaded green histogram. The experimental data point is shown with its error in black, while the SM (NLO+QCD) with background subtracted are shown in the shaded pink histogram. The black line in all  graphs is the best-fit line with the experimental data while the green line  depicts   the best-fit  line with  the model data.}}
\label{afbdydata}
\end{figure*}

  The two dimensional parameter space (($\sqrt{\lambda_{LL}}$,\, $\sqrt{\lambda_{RR}}$) for FC cases and ($g^{ut}_L,\, g^{ut}_R$) for the FV case)  with a given fixed mass   is scanned leading to the  minimum value of the $\chi^2 \equiv\chi^2_{\rm min.}$. We plot  histograms   showing   the $\mttb$ spectrum of \afbt at combined $\chi^2_{\rm min.}$ in  Figs. 
\ref{fig:TP_FC_mtt_A}, \ref{fig:TP_FC_mtt_R}, \ref{fig:TP_FC_mtt_NA} for flavor conserving cases and Figs. \ref{fig:TP_FV_mtt_A1}, \ref{fig:TP_FV_mtt_A2} and \ref{fig:TP_FV_mtt_R} for flavor violating cases, respectively. We have shown and compared the slope of our best-fit line with that from the experimental data in these figures and Tables \ref{afbmttdata_FC1} and \ref{afbmttdata_FV} . Similarly Figs. \ref{fig:TP_FC_dy_A}, \ref{fig:TP_FC_dy_R}, \ref{fig:TP_FC_dy_NA} for flavor conserving cases and Figs. \ref{fig:TP_FV_dy_A1}, \ref{fig:TP_FV_dy_A2} and \ref{fig:TP_FV_dy_R} for flavor violating cases exhibit the $\Delta y$ spectrum of \afbt at combined $\chi^2_{\rm min.}$ along with  the slope of the best fit line. These values for the $\Delta y$ distribution of \afbt are also summarized in Tables \ref{afbdydata_FC} and \ref{afbdydata_FV}. 
\begin{widetext}
\begin{center}
\begin{table}[h]
\begin{tabular}{ccc|c|c|c}\hline\hline
$M_{t\bar t}$ & \afbt ($\pm$ stat.) & {\small NLO} & 200 GeV & 900 GeV & 500 GeV \\
&&$t\bar t$-Bkg & $\sqrt{\lambda_{AA}}$ = 0.30  & $\sqrt{\lambda_{NAA}}$ = 0.35 & $\sqrt{\lambda_{RR}}$ = 0.11  \\\hline
$<$ 400 GeV   & -0.012 $\pm$ 0.040 & 0.012 & 0.055 & 0.024 & 0.018   \\
400 $-$ 450 GeV &0.084 $\pm$ 0.050 & 0.031 & 0.101 & 0.074 & 0.005   \\
450 $-$ 500 GeV &0.158 $\pm$ 0.064 & 0.039 & 0.172 & 0.062 & 0.105   \\
500 $-$ 550 GeV &0.203 $\pm$ 0.083 & 0.060 & 0.135 & 0.118 & 0.254   \\
550  $-$600 GeV &0.143 $\pm$ 0.105 & 0.083 & 0.208 & 0.145 & 0.113   \\
600 $-$ 700 GeV &0.338 $\pm$ 0.128 & 0.077 & 0.172 & 0.230 & 0.142   \\
$\geq$  700 GeV &0.377 $\pm$ 0.163 & 0.137 & 0.329 & 0.282 & 0.230   \\
\hline\hline
$\chi^2_{\rm min.}$ & -& -& 6.05 & 5.27 & 7.43  \\ 
\hline\hline 
Slope of &&&&&\\
  Best-Fit Line & $\left(11.1\pm2.9\right)\times 10^{-4}$ & $ 3.0 \times 10^{-4}$ & $6.08\times 10^{-4}$ & $7.04\times 10^{-4}$ & $5.31\times 10^{-4}$ \\ 
\hline\hline
\end{tabular}
\caption{\small \em{The first three columns give the bin limits of the $M_{t\bar t}$, the observed \afbt  with error and the NLO (QCD+EW) generated \afbt, respectively \cite{CDF:10807}. The next three consecutive columns provide the differential \afbt  corresponding to the model parameters (given in Figs. \ref{fig:TP_FC_mtt_A}, \ref{fig:TP_FC_mtt_R} and \ref{fig:TP_FC_mtt_NA} ) leading to $\chi^2_{\rm min.}$ at fixed coupling and $M_{V_8^0}$ in flavor conserving cases. The penultimate line gives the  $\chi^2_{\rm min.}$ for  respective cases. The last line in the table gives the slope of the best fit line with the simulated data.}}
\label{afbmttdata_FC1}
\end{table}
\begin{table}[h]
\begin{tabular}{ccc|c|c|c}\hline\hline
$|\Delta y|$ & \afbt ($\pm$ stat.) & {\small NLO} & 200 GeV & 900 GeV & 500 GeV       \\
&&$t\bar t$-Bkg & $\sqrt{\lambda_{AA}}$ = 0.30  & $\sqrt{\lambda_{NAA}}$ = 0.35 & $\sqrt{\lambda_{RR}}$ = 0.11   \\\hline
0.0-0.5    & 0.026 $\pm$ 0.035 & 0.009 & 0.046 & 0.018 & 0.015    \\
0.5-1.0    & 0.065 $\pm$ 0.045 & 0.040 & 0.142 & 0.068 & 0.089   \\
1.0-1.5    & 0.269 $\pm$ 0.066 & 0.074 & 0.209 & 0.184 & 0.133    \\
$\geq$ 1.5 & 0.299 $\pm$ 0.126 & 0.113 & 0.223 & 0.249 & 0.266    \\
\hline\hline
$\chi^2_{\rm min.}$ & -& -& 4.41 & 1.87 & 4.73  \\ 
\hline\hline 
Slope of &&&&& \\
  Best-Fit Line & $\left(20.0\pm5.9\right)\times 10^{-2}$ & $ 6.7 \times 10^{-2}$ & $11.96\times 10^{-2}$ & $16.18\times 10^{-2}$ & $15.94\times 10^{-2}$  \\ 
\hline\hline
\end{tabular}
\caption{\small \em{The first three columns give the bin limits of the $|\Delta y|$, the observed \afbt  with error and the NLO (QCD+EW) generated \afbt, respectively \cite{CDF:10807}. The next three consecutive columns provide the differential \afbt  corresponding to the model parameters (given in Figs. \ref{fig:TP_FC_dy_A}, \ref{fig:TP_FC_dy_R} and \ref{fig:TP_FC_dy_NA}) leading to $\chi^2_{\rm min.}$ at fixed coupling and $M_{V_8^0}$ in flavor conserving cases. The penultimate line gives the  $\chi^2_{\rm min.}$ for  respective cases. The last line in the table gives the slope of the best fit line with the simulated data.}}
\label{afbdydata_FC}
\end{table}
\begin{table}[h]
\begin{tabular}{c|ccc}\hline\hline
$M_{t\bar t}$ & 200 GeV    & 500 GeV      & 900 GeV    \\
                & $g^{ut}_{AA}$ = 0.26 & $g^{ut}_{AA}$ = 0.53 & $g^{ut}_{RR}$ = 1.26 \\\hline
$<$ 400 GeV     & 0.037     & 0.019     & 0.023    \\
400 $-$ 450 GeV & 0.031     & 0.053     & 0.073    \\
450 $-$ 500 GeV & 0.095     & 0.110     & 0.089    \\
500 $-$ 550 GeV & 0.188     & 0.160     & 0.121    \\
550 $-$ 600 GeV & 0.134     & 0.194     & 0.163    \\
600 $-$ 700 GeV & 0.256     & 0.207     & 0.182    \\
$\geq$  700 GeV & 0.263     & 0.267     & 0.297    \\
\hline\hline
$\chi^2_{\rm min.}$ & 4.92 & 4.14 & 4.93 \\ 
\hline\hline 
Slope of Best-fit Line & $6.85 \times 10^{-4}$ & $6.6 \times 10^{-4}$ & $6.74 \times 10^{-4}$ \\ 
\hline\hline
\end{tabular}
\caption{\small \em{Same as TABLE I, first column give the bin limits of the $M_{t\bar t}$ and the next three consecutive columns provide the differential \afbt  corresponding to the model parameters (given in Figs. \ref{fig:TP_FV_mtt_A1}, \ref{fig:TP_FV_mtt_A2} and \ref{fig:TP_FV_mtt_R} leading to $\chi^2_{\rm min.}$ in flavor violating cases.}}
\label{afbmttdata_FV}
\end{table}%
\begin{table}[h]
\begin{tabular}{c|ccc}\hline\hline
$|\Delta y|$ & 200 GeV    & 500 GeV      & 900 GeV   \\
           & $g^{ut}_{AA}$ = 0.26 & $g^{ut}_{AA}$ = 0.53 & $g^{ut}_{RR}$ = 1.26 \\\hline
0.0-0.5    & 0.031     & 0.022     & 0.035    \\
0.5-1.0    & 0.072     & 0.101     & 0.077   \\
1.0-1.5    & 0.139     & 0.149     & 0.149    \\
$\geq$ 1.5 & 0.317     & 0.260     & 0.235    \\
\hline\hline
$\chi^2_{\rm min.}$ & 3.97 & 4.03 & 3.71 \\ 
\hline\hline 
Slope of Best-fit Line & $18.5 \times 10^{-2}$ & $15.24 \times 10^{-2}$  & $13.44 \times 10^{-2}$  \\ 
\hline\hline
\end{tabular}
\caption{\small \em{Same as TABLE II, first column give the bin limits of the $|\Delta y|$ and the next three consecutive columns provide the differential \afbt  corresponding to the model parameters (given in Figs. \ref{fig:TP_FV_dy_A1}, \ref{fig:TP_FV_dy_A2} and \ref{fig:TP_FV_dy_R} leading to $\chi^2_{\rm min.}$ in flavor violating cases.}}
\label{afbdydata_FV}
\end{table}
\end{center}
\end{widetext}
\section{Single top}
\label{sec:singletop}
\begin{figure*}[ht]
%\centering
\subfloat[$s$-channel]{\label{stopsmsa}\includegraphics[width=0.13\textwidth]{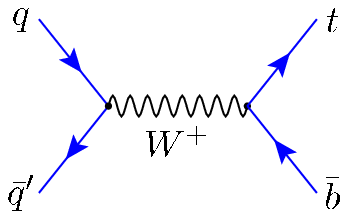}} \qquad \qquad \quad
\subfloat[$Wt$- channel]{\label{stopsmsb}\includegraphics[width=0.13\textwidth]{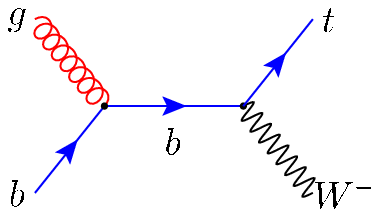}} \qquad \qquad \quad
\subfloat[$2 \to 2 $ $t$-channel]{\label{stopsmta}\includegraphics[width=0.12\textwidth]{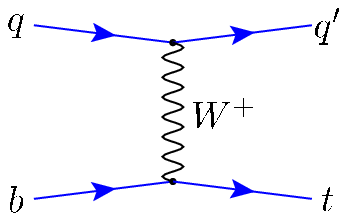}} \qquad \qquad \quad
\subfloat[$2 \to 3$ $t$-channel]{\label{stopsmtb}\includegraphics[width=0.12\textwidth]{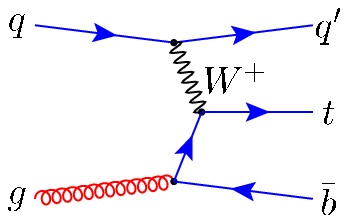}} \qquad \qquad \qquad \qquad 
\caption{\small \em {Leading order single top production channels in SM.  }} 
\label{fig:stopsm}
\vskip 0.5 cm
\end{figure*}
\begin{figure*}[ht]
\centering
\subfloat[$s$-channel CC]{\label{stopV8s}\includegraphics[width=0.12\textwidth]{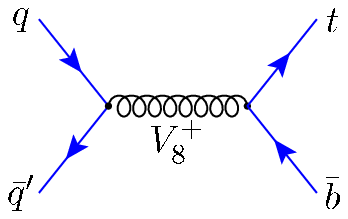}} \qquad \qquad \quad
\subfloat[$t$-channel CC]{\label{stopV8t}\includegraphics[width=0.12\textwidth]{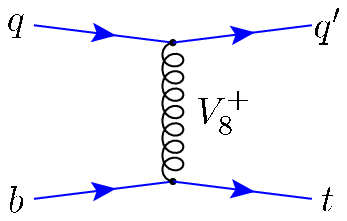}} \qquad \qquad \quad
\subfloat[$s$-channel NC]{\label{stopV0s}\includegraphics[width=0.12\textwidth]{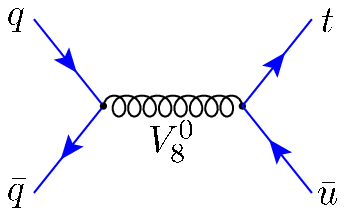}} \qquad \qquad \quad
\subfloat[$t$-channel NC]{\label{stopV0t}\includegraphics[width=0.12\textwidth]{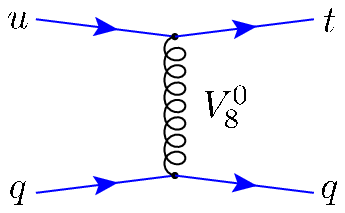}} \qquad \qquad \qquad \qquad 
\caption{\small \em {Leading order single top production channels  mediated by   charged color-octet vector  current are shown in figures (a)-(b) and  flavor violating neutral color-octet vector current are  shown in figures (d)-(e), respectively.}}
\label{fig:stopnp}
\end{figure*}
\par In the SM, single top quark production is studied  through  three different channels  with different final states, respectively  which have their own distinct  kinematics and do not interfere with one another. The $s$-channel process takes place through an off-shell time like $W$ boson which further decay into a top and bottom quark as shown in Fig.~\ref{stopsmsa}. The $t$-channel process is the dominant one and mediates  through the exchange of a virtual $W$ as shown in Fig.~\ref{stopsmta}. The $t$-channel process resembles deep inelastic scattering while $s$-channel process resembles the Drell-Yan process. 
The single top quark production cross section in these two modes have been estimated to be  $\sigma^{\rm NNNLO}_{\rm t-channel}$ = 1.05 $\pm$ 0.11 pb, $\sigma^{\rm NNNLO}_{s-channel}$ = 0.52 $\pm$ 0.03 pb, respectively at the NNNLO approximation for $m_t= 173 $ GeV \cite{Kidonakis:2012db}. Therefore the single top quark production in the $t$-channel is roughly twice of the $s$-channel production cross section  in the SM. The third channel for single top production is the associated $tW$- production as shown in Fig. \ref{stopsmsb} which is estimated to be $\sigma^{\rm NNNLO}_{Wt-channel}$ = 0.11 $\pm$ 0.04 pb for $m_t= 173 $ GeV \cite{Kidonakis:2012db}. We do not consider this process for our analysis. 
The single top quark production cross section in the $t$-channel is thus expected to dominate over the $s$-channel production both at the Tevatron and the LHC while the cross section for $tW$ production  is very small at the Tevatron but significant at the LHC. The three channels discussed above are sensitive to quite different manifestations of physics beyond the SM such that Flavor Changing Neutral Current (FCNC), existence of color singlet / octet vector bosons $W^{\prime\pm},H^\pm, Z^\prime, V_8^{\pm,0}$ etc, fourth generation quarks or detection of more general four fermion interactions.  D\O \, reported the $s$- and $t$- channel cross-sections to be $\sigma_s = 0.68^{+0.38}_{-0.35}$ pb and $\sigma_t = 2.86^{+0.69}_{-0.63}$ pb for $m_t=172.5$ GeV, respectively \cite{Abazov:2011pt} in agreement with the estimated cross-sections in the SM. However, very recently the CDF collaboration \cite{CDF:10793} using 7.5 fb$^{-1}$ of $p \bar p$ collisions data collected by CDF Run II experiment channel as $\sigma_s = 1.81^{+0.63}_{-0.58}$ pb and $\sigma_t = 1.49^{+0.47}_{-0.42}$ pb where in the central values of $s$- and $t$-channel cross-sections are comparable which is in conflict with the SM NNNLO prediction. The total single top cross section measured by this group $\sigma_{total} = 3.04^{+0.57}_{-0.53}$ pb for $m_t=172.5$ GeV is however consistent with the total single top production in SM at the NNNLO approximation.  
\par There are two alternative approaches available in the literature to study $t$- channel single top quark production at the leading order (LO) and NLO. One of the approach is based on $2\to 2$ scattering process Fig.~\ref{stopsmta}, where $b$ quark is taken to be present in the initial state. This is so called five flavor (5F) scheme. In this scheme the presence of $b$ $jet$ and the effect of $b$ mass appears only at NLO. In the second approach the LO (Born process) is the $2\to 3$ scattering process Fig.~\ref{stopsmtb}. In this four flavor (4F) scheme, the $b$ quark does not enter in the QCD evolution of the parton distribution function. For details see \cite{Campbell:2009ss} \cite{Campbell:2009gj}. The approaches are shown to be equivalent and give the same result at all orders in the perturbation expansion. In the present work we treat proton in 5F scheme and study the $b$-initiated $2\to 2$ process since we are only interested in estimation of the total production cross section \cite{Maltoni:2012pa}.
\par In this section we will study single top $s$- and $t$- channel production in ${\bf 3}\otimes \bar{\bf 3}$ model in flavor conserving and flavor violating cases. Throughout our analysis we assume $\left\vert V_{tb}\right\vert= 1$. In the flavor conserving case, the single top production  in the $s$- and $t$- channels proceeds through the charge current interactions through $V_8^{\pm}$ as shown in Figs. \ref{stopV8s} and \ref{stopV8t}, respectively.  The flavor violating mode medaites through flavor changing neutral current $via$ $V_8^0$ as shown in Figs.~ \ref{stopV0s} and \ref{stopV0t}, respectively.
The charged octet vector boson does not contribute to the other processes in the top quark sector addressed in our study. However, the contribution of the FCNC is likely to provide the common global allowed parameter space of the model from all the processes involving the top quark sector.
\subsection{Flavor Conserving}
\par The differential cross-sections with respect to emerging angle of the single massive top quark $\cos\theta_{t}$  the $s$-channel subprocess $u\bar d \to t \bar b$, and the $t$- channel subprocess $ub \to td$ are given  as 
\begin{widetext}
\begin{eqnarray}
\frac{d\sigma_{u\bar d\to t \bar b}}{d\cos\theta_{t}} =&& \frac{\pi\beta^\prime \alpha_s^2}{18 \hat s} \frac{1}{(\hat s - M_{V_8}^2)^2 + M_{V_8}^2 \Gamma_{V_8}^2} \Big[\mathscr C_{+} \hat u (\hat u -m_t^2) + \mathscr C_{-} \hat t (\hat t -m_t^2) \Big] , \\
\frac{d\sigma_{ub\to td}}{d\cos\theta_t} =&& \frac{\pi\beta^\prime \alpha_s^2}{18 \hat s} \frac{1}{(\hat t - M_{V_8}^2)^2 + M_{V_8}^2 \Gamma_{V_8}^2 } \Big[\mathscr C_{+} \hat s (\hat s -m_t^2) + \mathscr C_{-} \hat t (\hat t -m_t^2) \Big],  \\
\text{where} \quad \mathscr C_{\pm} =&& ( |C_L^{tb}|^2 + |C_R^{tb}|^2 )( |C_L^{ud}|^2 + |C_R^{ud}|^2 ) \pm ( |C_L^{tb}|^2 - |C_R^{tb}|^2 )( |C_L^{ud}|^2 - |C_R^{ud}|^2 )\quad \text{and } \beta^\prime = 1-\frac{m_t^2}{\hat s} \label{dist_sing_top}
\end{eqnarray}  
\end{widetext}
\par
\begin{figure*}[ht]
\centering
\subfloat[$s$ channel $\sigma ( p\bar p \to t\bar b + \bar t b)$ : $\lambda_{LL} = \lambda_{RR} =-\lambda_{RL}=-\lambda_{LR}=\lambda_{AA}$]{\label{fig:sST_FC_A}\includegraphics[width=0.5\textwidth]{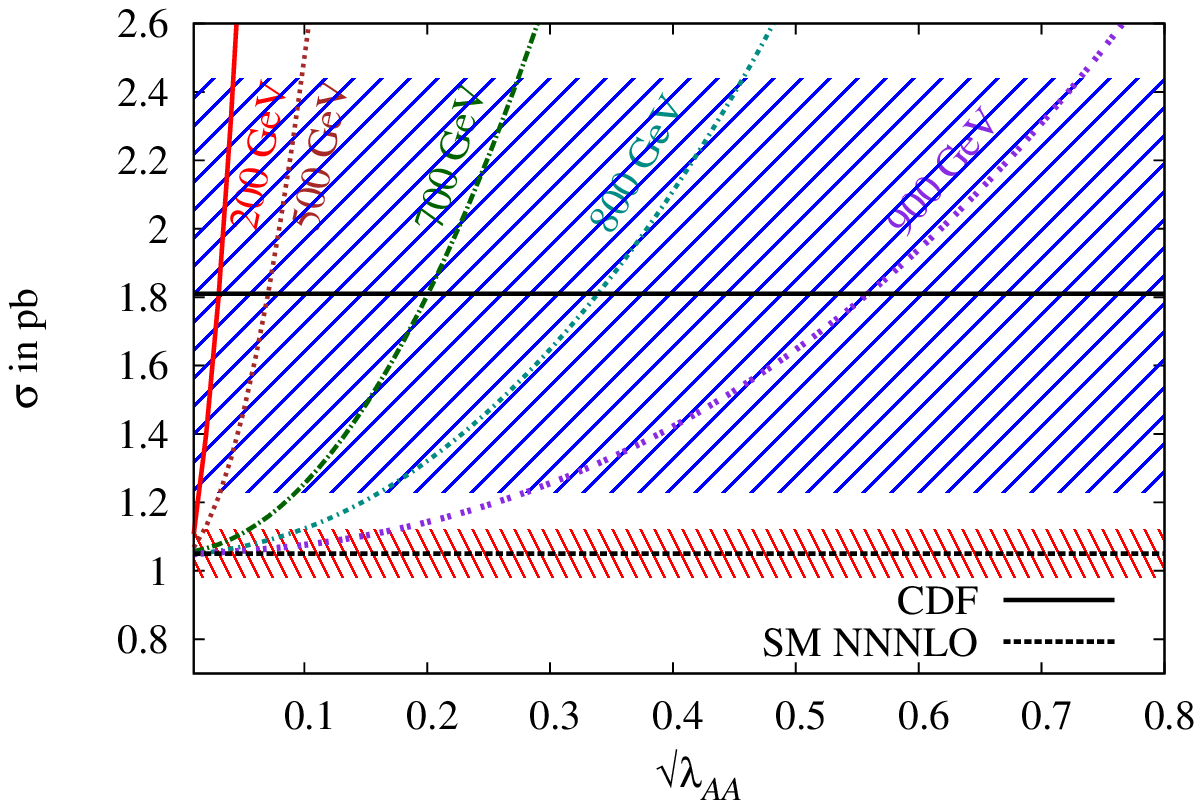}}
\subfloat[$s$ channel $\sigma ( p\bar p \to t\bar b + \bar t b)$: $\lambda_{RR} \ne 0=\lambda_{LR}=\lambda_{RL}=\lambda_{LL}$]{\label{fig:sST_FC_R}\includegraphics[width=0.5\textwidth]{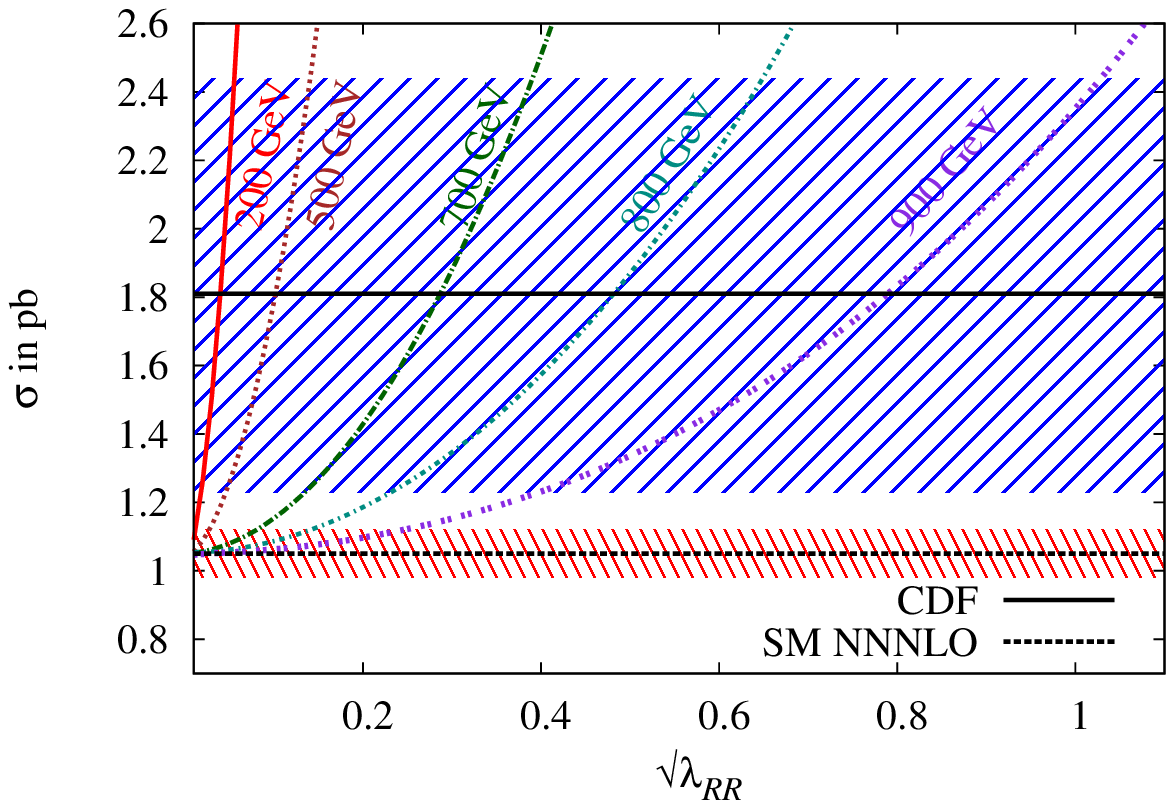}}\\
\subfloat[$t$ channel $\sigma ( p\bar p \to tj + \bar t j )$: $\lambda_{LL} = \lambda_{RR} =-\lambda_{RL}=-\lambda_{LR}=\lambda_{AA}$]{\label{fig:tST_FC_A}\includegraphics[width=0.5\textwidth]{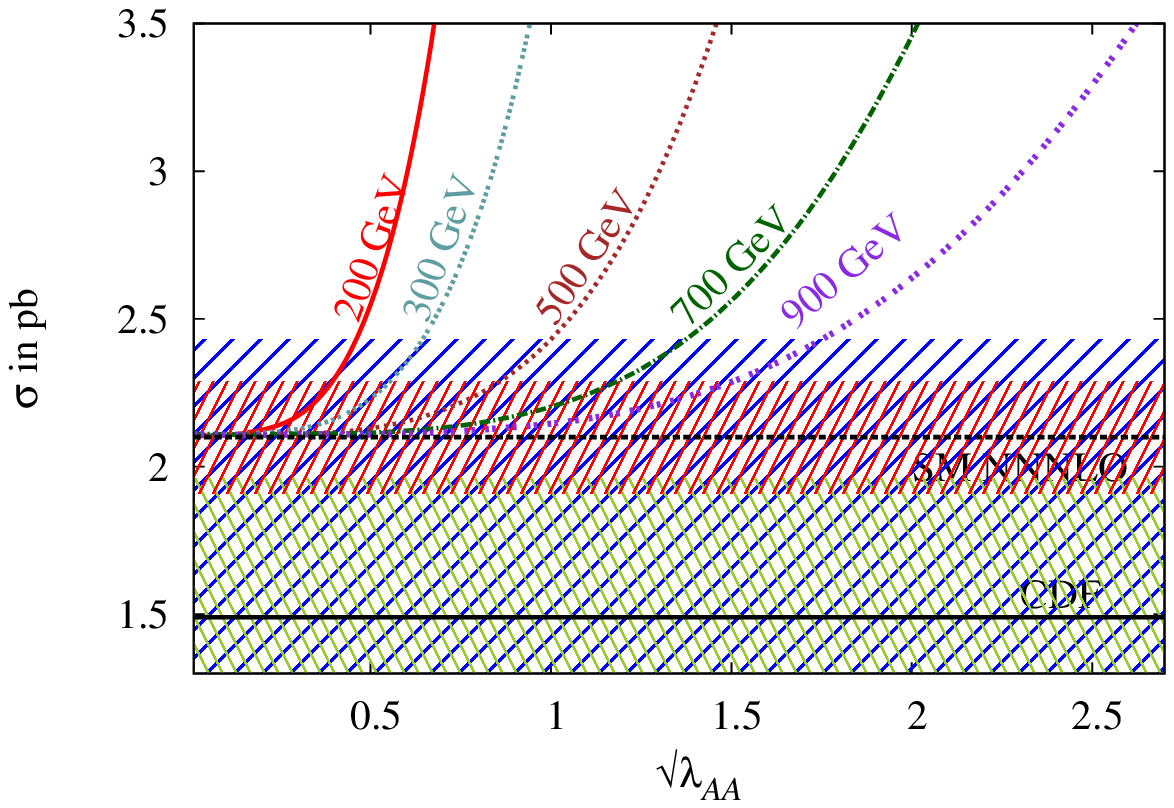}}
\subfloat[$t$ channel $\sigma ( p\bar p \to tj + \bar t j )$: $\lambda_{RR} \ne 0=\lambda_{LR}=\lambda_{RL}=\lambda_{LL}$]{\label{fig:tST_FC_R}\includegraphics[width=0.5\textwidth]{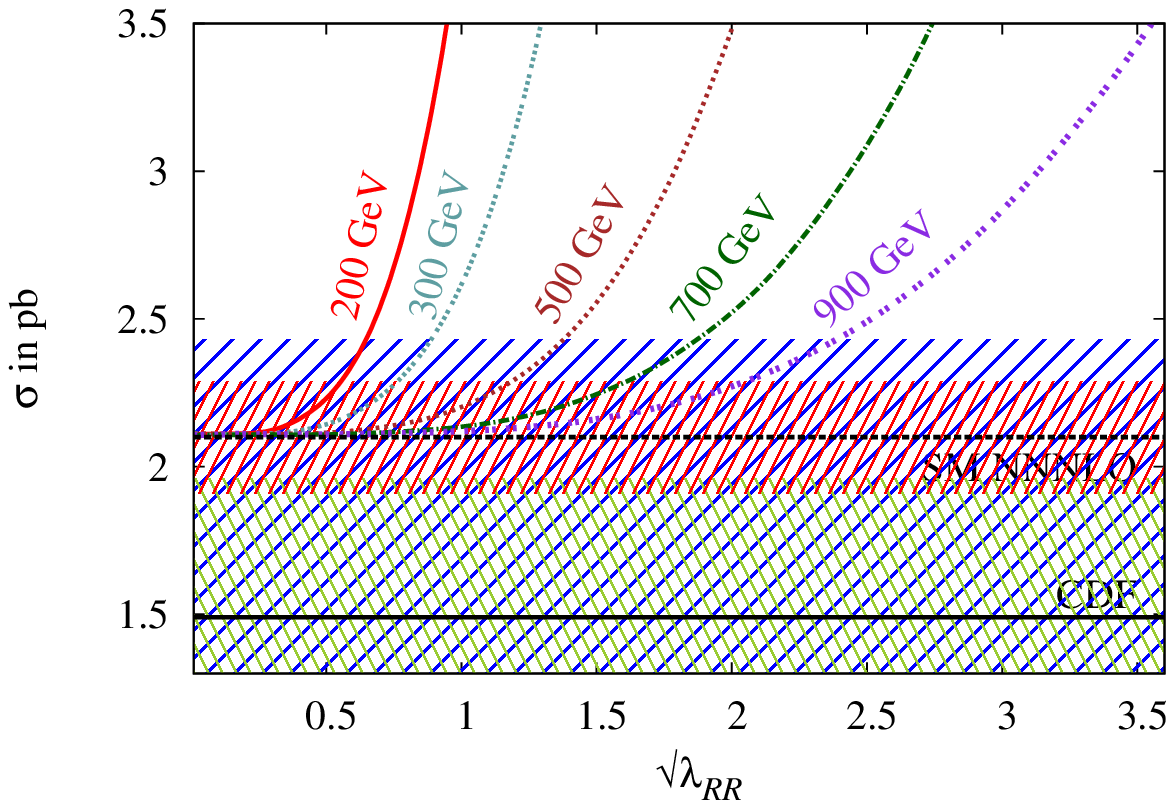}}
\caption{\small \em {Variation of the cross section  with  flavor conserving couplings for various vector color-octet masses $M_{V_8^{\pm}}$.  In the top panel figures (a)  and (b) the upper dotted black line  and associated  blue band  depicts  the CDF central value and the one sigma band, respectively for $s$-channel cross section  $1.81^{+0.63}_{-0.58}$ pb ~\cite{CDF:10793}, while the lower dot-dashed  black line with a red band show theoretical central value  and the  one sigma band, respectively for $1.05\pm 0.07$ pb at NNNLO  ~\cite{Kidonakis:2012db}. Similarly in the lower panel figures (c)  and (d) the experimental central value is shown with lower dotted black line and the associated  1-sigma green band, 2-sigma blue band  corresponds to $t$-channel cross section  $1.49^{+0.47}_{-0.42}$ pb from CDF ~\cite{CDF:10793}. The upper dot-dashed black line associated with with a red band show theoretical central value and one sigma band for $2.10\pm 0.19$ pb at NNNLO  ~\cite{Kidonakis:2012db}.}} 
\label{fig:FCstop}
\end{figure*}

 It is evident from the Eq. \eqref{dist_sing_top} that the contribution for the pure vector current and pure axial current is identical which also hold true between the right and left chiral contributions.
We study the variation of the single top quark production in the $s$ channel with the couplings for various vector octet masses which are shown in Figs. \ref{fig:sST_FC_A} and  \ref{fig:sST_FC_R} corresponding to the charged axial and right-chiral current, respectively. We observe the sharp growth in the cross-sections even with the smaller couplings and later flattens out with the increasing mass of the octet.
\par Figs. \ref{fig:tST_FC_A} and  \ref{fig:tST_FC_R} depicts the variation in the  $t$ channel mode. It is evident that the plugging of decay width in $t$ channel propagator flattens the variation of the curve with respect to couplings.  Since there is no interference between the SM and color-octet model (for both the channels), we find that the cross section grows with the coupling and decreases with the mass of the color-octet vector bosons. 
\subsection{Flavor Violating}
\begin{figure*}[ht]
\centering
\subfloat[$f^{q,\,ut}_{LL} = f^{q,\,ut}_{RR} =-f^{q,\,ut}_{RL}=-f^{q,\,ut}_{LR}=f^{q,\,ut}_{AA}$]{\label{fig:fv_a}\includegraphics[width=0.5\textwidth]{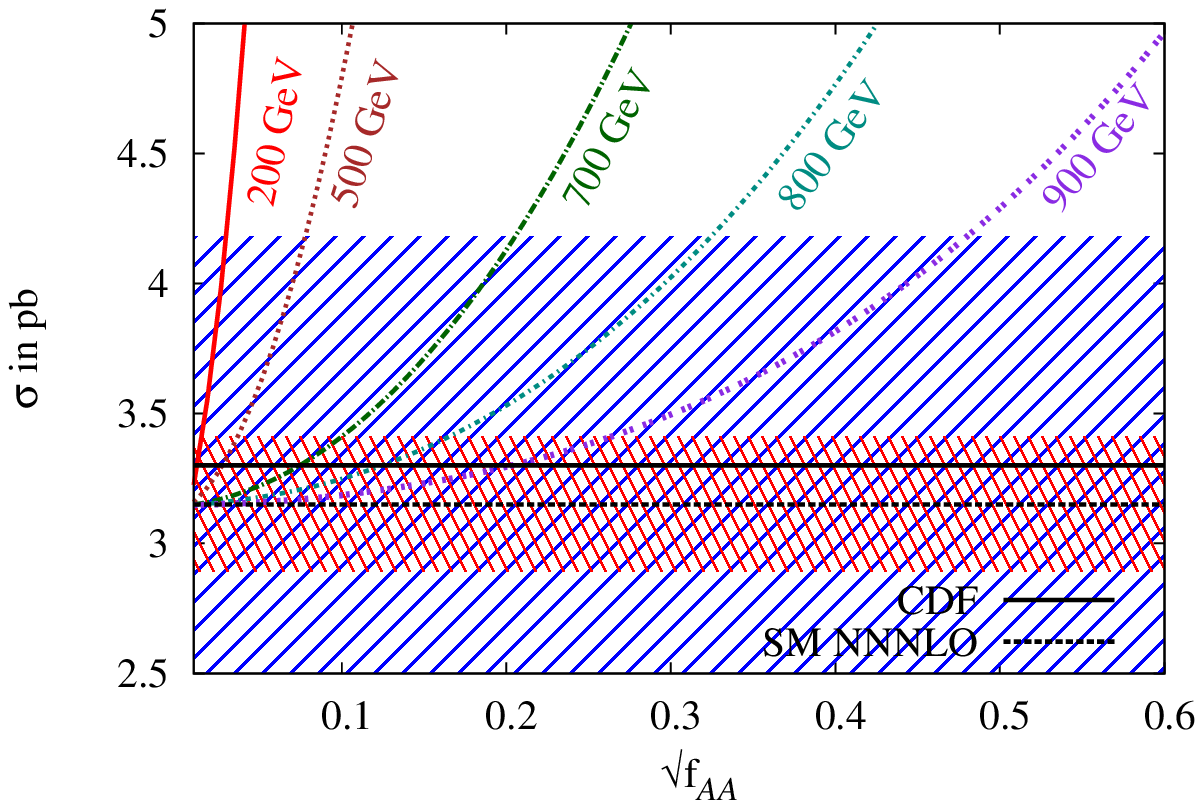}}
\subfloat[$f^{q,\,ut}_{RR} \ne 0=f^{q,\,ut}_{LR}=f^{q,\,ut}_{RL}=f^{q,\,ut}_{LL}$]{\label{fig:fv_r}\includegraphics[width=0.5\textwidth]{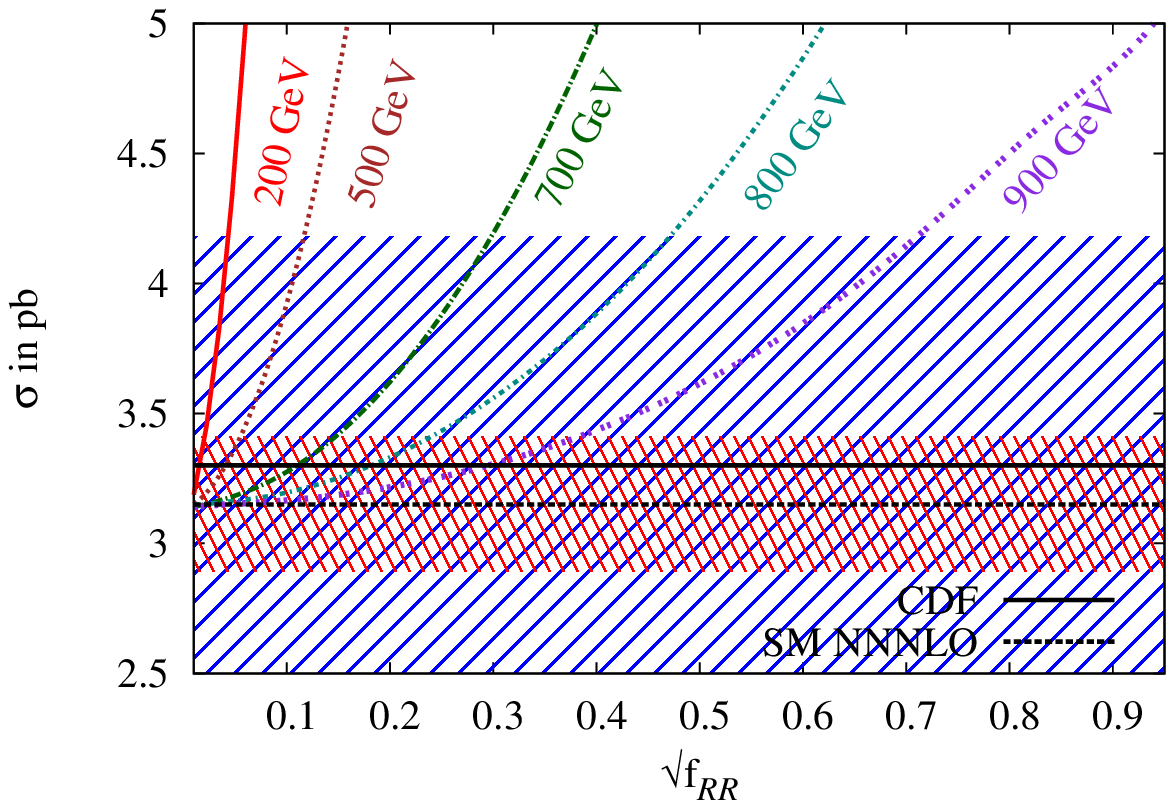}}
\caption{\small \em {Variation of the combined cross section $\sigma ( p\bar p \to t\bar b + \bar t b)$ and  $\sigma ( p\bar p \to tj + \bar t j )$ with  couplings $\surd f_{ij}^{q,\,ut}$ corresponding to  different values of  $M_{V_8^{0}}$. In the figures (a) and (b) the upper dotted black line and associated   blue band  depicts the central value of combined experimental $s+t$-channel cross section and one sigma allowed region  of $3.04^{+0.57}_{-0.53}$ pb from CDF ~\cite{CDF:10793}, while the lower dot-dashed black line with a red band show theoretical estimate and its uncertainty $3.15\pm 0.26$ pb at NNNLO  ~\cite{Kidonakis:2012db}.}} 
\vskip 0.5 cm
\end{figure*}
FCNC induces additional channels for single top quark production. These additional Feynman diagrams are shown in Figs. \ref{stopV0s} and \ref{stopV0t}. This contribution is realized through the flavor changing coupling $g^{ut}_{L,R}$ at one of the vertices only but can be mediated through both $s$ and $t$ channels. This then would make the life uncomfortable for measuring $s$ and $t$ channel separately as the final states for both the diagrams are same.
In this view for this we need to have a combined study of $s$ and $t$ channel with one FV vertex from the new physics sector. Thus the new physics contribution at the amplitude level is proportional to  the product of the flavor conserving coupling $g_{L,R}^q$ and flavor violating coupling $g^{ut}_{L,R}$. We define $\sqrt{f_{ij}^{q,\,ut}}=\sqrt{g_{i}^q \, g_{j}^{ut}}$ for $i\equiv L,R,V,A$. 
The differential cross section with respect to $\cos\theta_{t}$ for the $s$-channel subprocess $q\bar q \to t \bar u (q=d,s,c,b)$, $s+t$-channel subprocess $u\bar u \to t\bar u$ and $t+u$-channel subprocess $uu \to tu$ assuming top quark mass $m_t$ and others to be massless is given as
\begin{widetext}
\begin{eqnarray}
\frac{d\sigma_{q\bar q \to t \bar u}}{d\cos\theta_t} =&& \frac{\pi\beta^\prime \alpha_s^2}{18 \hat s} \frac{1}{(\hat s - M_{V_8^0}^2)^2 + M_{V_8}^2 \Gamma_{V_8^0}^2} \Big[\mathscr V_{+} \hat u (\hat u -m_t^2) + \mathscr V_{-} \hat t (\hat t -m_t^2) \Big] \\
\frac{d\sigma_{u\bar u \to t \bar u}}{d\cos\theta_t} =&& \frac{\pi\beta^\prime \alpha_s^2}{18 \hat s}\Big\{\frac{1}{(\hat s - M_{V_8^0}^2)^2 + M_{V_8^0}^2 \Gamma_{V_8^0}^2}  \Big[\mathscr V_{+} \hat u (\hat u -m_t^2) + \mathscr V_{-} \hat t (\hat t -m_t^2) \Big] \cr
&&\qquad - \frac{2}{3}\frac{(\hat s - M_{V_8^0}^2)}{(\hat s - M_{V_8^0}^2)^2 + M_{V_8^0}^2 \Gamma_{V_8^0}^2} \frac{(\hat t - M_{V_8^0}^2)}{(\hat t - M_{V_8^0}^2)^2 + M_{V_8^0}^2 \Gamma_{V_8^0}^2 } \hat u (\hat u -m_t^2) \Big[ (g_L^{uu}g_L^{ut})^2 +g_R^{uu}g_R^{ut})^2 \Big] \cr
&&\qquad  +  \frac{1}{(\hat t - M_{V_8^0}^2)^2 + M_{V_8^0}^2 \Gamma_{V_8^0}^2 } \Big[\mathscr V_{+} \hat s (\hat s -m_t^2) + \mathscr V_{-} \hat u (\hat u -m_t^2) \Big]\Big\} \\
\frac{d\sigma_{uu \to t u}}{d\cos\theta_t} =&& \frac{\pi\beta^\prime \alpha_s^2}{18 \hat s}\Big\{\frac{1}{(\hat t - M_{V_8^0}^2)^2 + M_{V_8^0}^2 \Gamma_{V_8^0}^2} \Big[\mathscr V_{+} \hat u (\hat u -m_t^2) + \mathscr V_{-} \hat s (\hat s -m_t^2) \Big] \cr
&&\qquad + \frac{2(\hat t - M_{V_8^0}^2)(\hat u - M_{V_8^0}^2)}{((\hat t - M_{V_8^0}^2)^2+ M_{V_8^0}^2 \Gamma_{V_8^0}^2)((\hat u - M_{V_8^0}^2)^2+ M_{V_8^0}^2 \Gamma_{V_8^0}^2)} \hat s (\hat s -m_t^2) \Big[ (g_L^{uu}g_L^{ut})^2 +g_R^{uu}g_R^{ut})^2 \Big] \cr
&&\qquad  +  \frac{1}{(\hat u - M_{V_8^0}^2)^2 + M_{V_8^0}^2 \Gamma_{V_8^0}^2} \Big[\mathscr V_{+} \hat t (\hat t -m_t^2) + \mathscr V_{-} \hat s (\hat s -m_t^2) \Big]\Big\}  \\
\text{where} \quad \mathscr V_{\pm} =&& ( |g_L^{ut}|^2 + |g_R^{ut}|^2 )( |g_L^{qq}|^2 + |g_R^{qq}|^2 ) \pm ( |g_L^{ut}|^2 - |g_R^{ut}|^2 )( |g_L^{qq}|^2 - |g_R^{qq}|^2 )  \notag
\end{eqnarray}
\end{widetext}
Since the amplitude is the quadratic symmetric function of the left and right-handed couplings, the axial and vector currents are identical and so is the case for right and left handed currents. We have also taken into account of the additional decay channels contributing to the total decay width of the color-octet neutral vector boson due to the introduction of FV couplings. We exhibit the variation of the single top quark production with the product of the couplings $\sqrt{f_{ij}^{q,\,ut}}$ for different flavor violating color-octet neutral vector boson masses  in Fig.~\ref{fig:fv_a} and~\ref{fig:fv_r} corresponding to the axial and right-chiral current. We observe the the coupling $f_{i}$ are quite sensitive to the production cross section for the cases. It would then imply that for a given FC coupling the single top production can constrain the FV coupling more severely than the top-pair production.
\section{Same-sign top}
\label{sec:samesigntop}
\begin{figure}[t]
\centering
{\label{fig:tta}\includegraphics[scale=1.0]{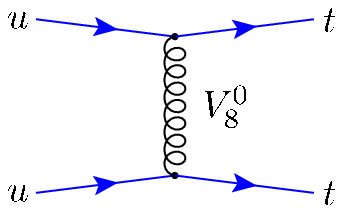}} \qquad \qquad \quad
{\label{fig:ttb}\includegraphics[scale=1.0]{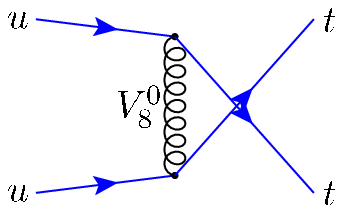}} \qquad \qquad \qquad
\caption{\small\em{Diagrams for same-sign top production through $V_8^0$ in (a) $t$- and (b) $u$-channels.} } 
 \label{tt}
\end{figure}
\begin{figure*}[ht]
\centering
\subfloat[\label{ttFVA}]{\includegraphics[width=0.5\textwidth]{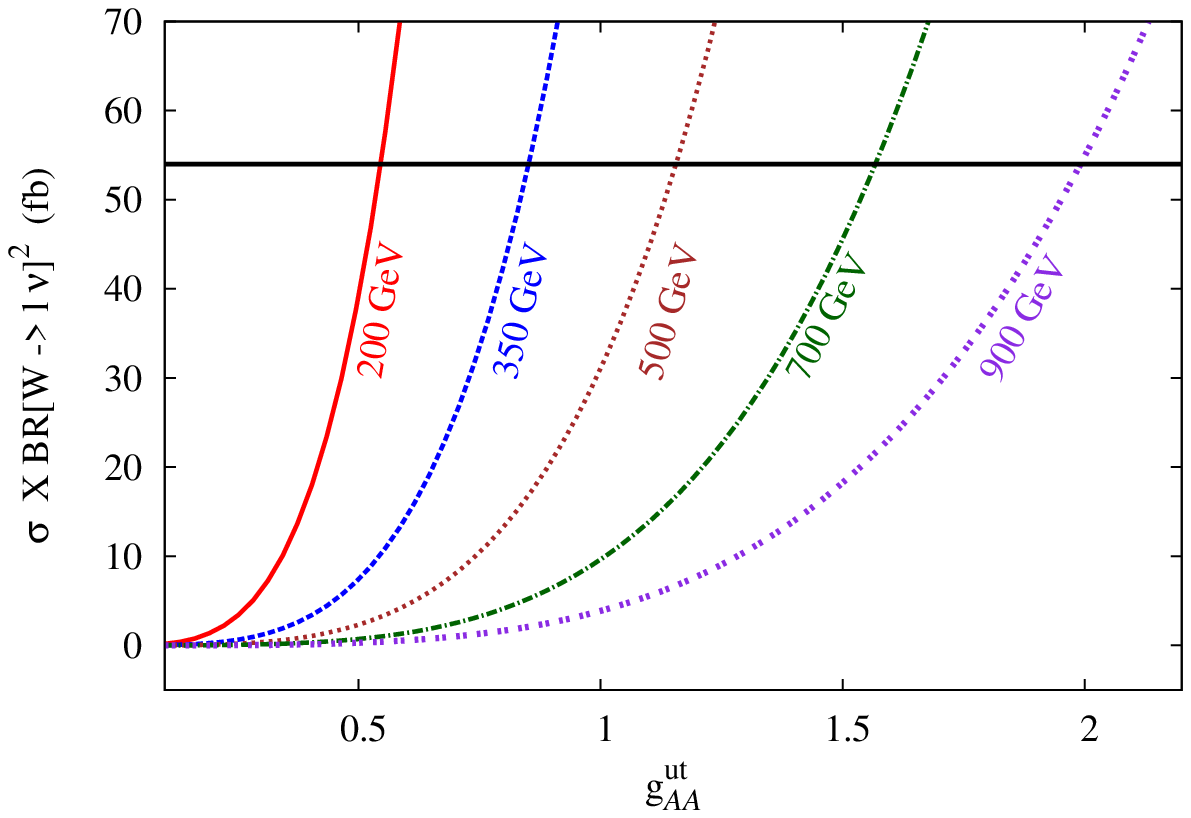}}
\subfloat[\label{ttFVR}]{\includegraphics[width=0.5\textwidth]{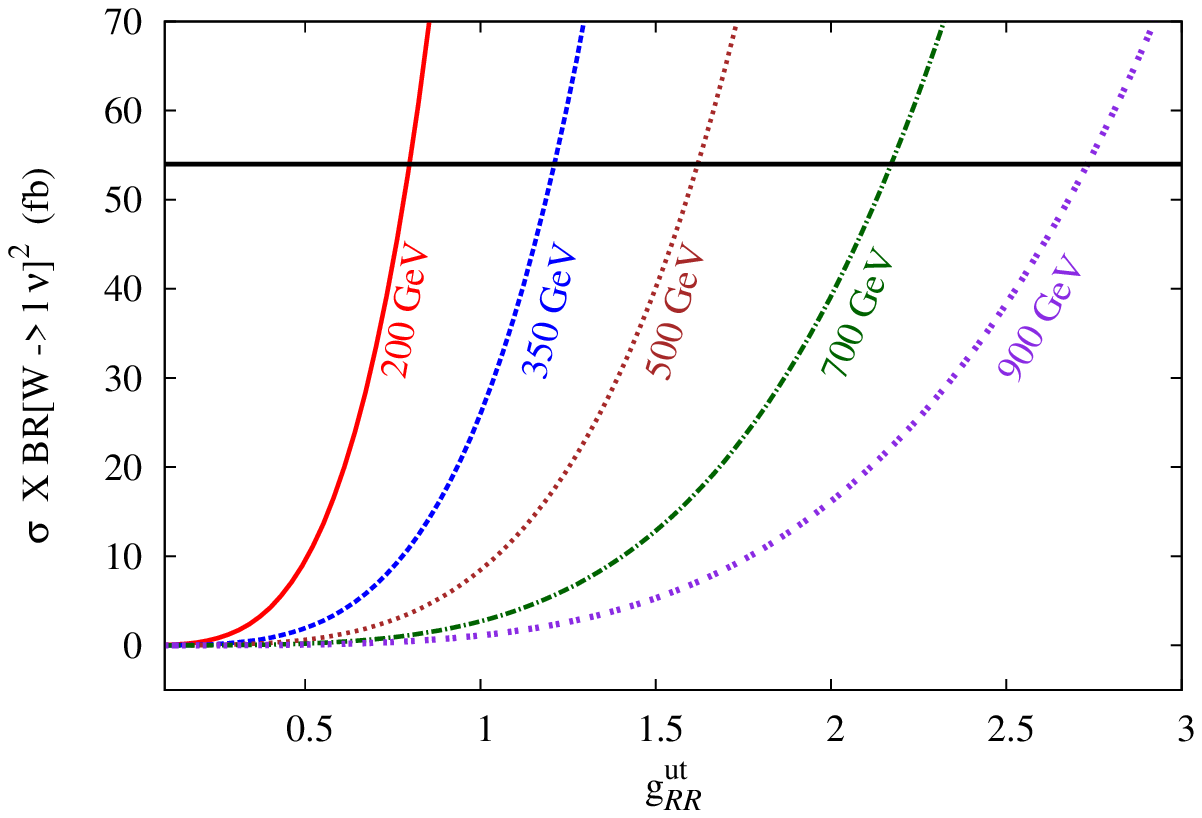}}
\caption{\small \em {Variation of the cross section $\sigma ( p\bar p \to tt + \bar t \bar t)$ times branching ratio BR$(W \to l \nu)^2$ with  couplings $g^{ut}_{i j}$ for flavor violating vector color-octets corresponding to  different values of  $M_{V_8^0}$ GeV for both the cases (a) and (b) given in the text. The  upper dotted line depicts the maximum allowed $\sigma_{tt+\bar t\bar t} \times BR(W \to l \nu)^2$ = 54 fb with a 95 $\%$ confidence level \cite{CDF:10466}.}}
\label{ttFV}
\end{figure*}
\begin{figure*}[ht]
\centering
\subfloat[\label{tevcont}]{\includegraphics[width=0.5\textwidth]{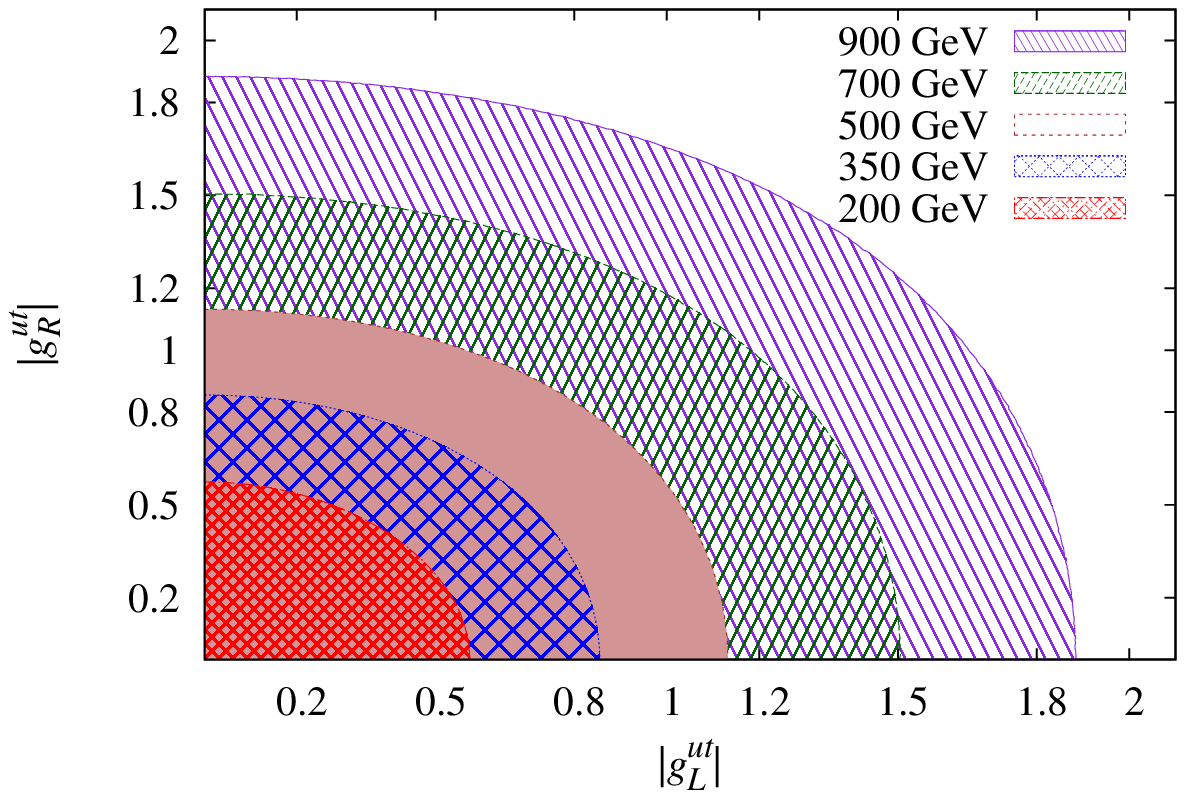}}   \\
\subfloat[\label{cmscont}]{\includegraphics[width=0.5\textwidth]{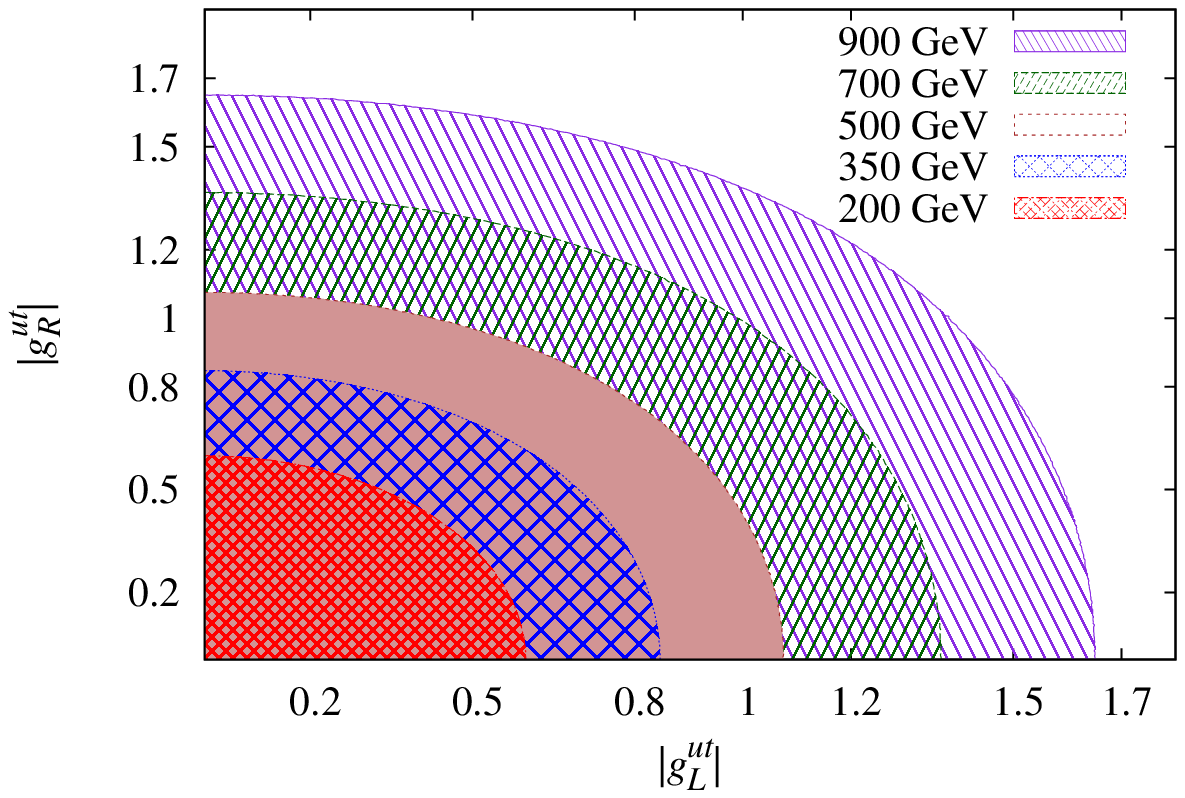}}
\subfloat[\label{atlascont}]{\includegraphics[width=0.5\textwidth]{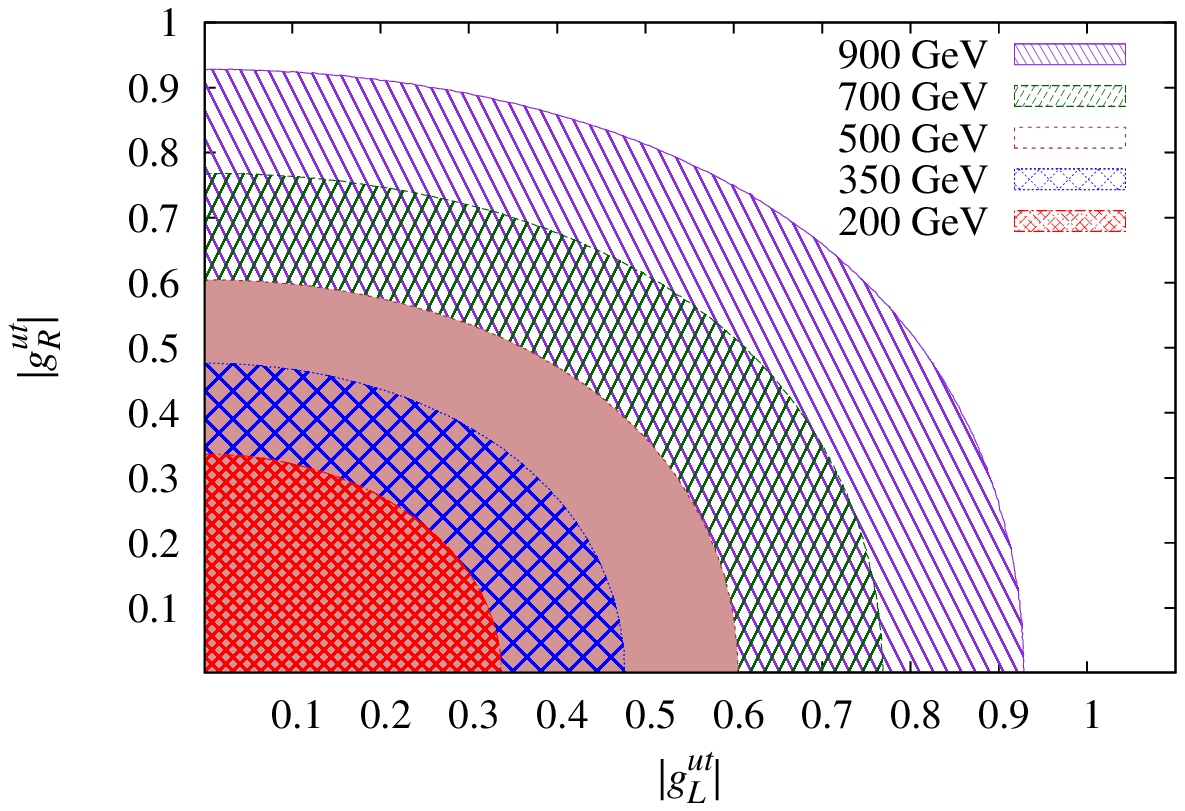}}
\caption{\small \em {95$\%$ CL exclusion contours   on the plane of $g^{ut}_L$  and  $g^{ut}_R$ for varying color-octet masses. Figure  (a) corresponds to $\sigma(p\bar p \to tt+\bar t \bar t) \times {BR} [W \to l\nu]^2\le  54$ fb from the Tevatron at CDF \cite{CDF:10466}. Figure  (b) corresponds to  measurement from CMS at the LHC, constraining   $\sigma(pp \to tt(j)) \le $ 17 pb  \cite{Chatrchyan:2011dk} and Fig. (c) corresponds to $\sigma(pp \to tt) \le  1.7$ pb from ATLAS detector at the LHC \cite{Aad:2012bb}.}   }
\label{ttFVCont}
\end{figure*}
\par Introduction of the flavor violating couplings involving first and third generation for the top-pair production also induces the new channel for same-sign top/ anti-top pair production. The same-sign top production is highly suppressed in SM because it involves higher order flavor changing neutral current interactions. 
\par In the present study the process $uu (\bar u\bar u) \to tt (\bar t \bar t)$ proceeds through the exchange of neutral color-octet vector boson $V_8^0$ with flavor changing neutral current interactions between the first and third generation only in the $t$- channel and the exchange diagram $u$- channel as shown in Fig.~\ref{tt}.
\par The differential cross section for $uu (\bar u\bar u) \to tt (\bar t \bar t)$ with respect to the cosine of the top quark polar angle $\theta$ in the $tt$ center-of-mass (c.m.) frame is
\begin{widetext}
\begin{eqnarray}
\frac{d\hat \sigma}{d\,cos\theta} =&& \frac{\pi \beta \alpha_s^2}{(\hat t - M_{V_8^0}^2)^2+ M_{V_8^0}^2 \Gamma_{V_8^0}^2} \frac{{\hat s}}{18} \Big[2({g_L^{ut}}^4 + {g_R^{ut}}^4) + {g_L^{ut}}^2 {g_R^{ut}}^2 (1 + \beta cos\theta)^2 \Big] \cr 
&&+ \frac{2\pi \beta \alpha_s^2 (\hat t - m_{V_8^0}^2)(\hat u - m_{V_8^0}^2)}{((\hat t - M_{V_8^0}^2)^2+ M_{V_8^0}^2 \Gamma_{V_8^0}^2)((\hat u - M_{V_8^0}^2)^2+ M_{V_8^0}^2 \Gamma_{V_8^0}^2)} \frac{{\hat s}}{9} \Big[({g_L^{ut}}^4 + {g_R^{ut}}^4) - 2 {g_L^{ut}}^2 {g_R^{ut}}^2 \frac{m_t^2}{\hat s} \Big] \cr
&&+ \frac{\pi \beta \alpha_s^2}{(\hat u - M_{V_8^0}^2)^2+ M_{V_8^0}^2 \Gamma_{V_8^0}^2} \frac{{\hat s}}{18} \Big[2({g_L^{ut}}^4 + {g_R^{ut}}^4) + {g_L^{ut}}^2 {g_R^{ut}}^2(1 - \beta cos\theta)^2 \Big] 
\end{eqnarray}
\end{widetext}
where $\hat s = (p_u + p_u)^2$ is the squared c.m. energy of the system with top quark velocity $\beta = \sqrt {1-4m_t^2/{\hat s}}$.
\par We study  the variation of the production cross section $\sigma ( p\bar p \to tt + \bar t \bar t)$ with respect to axial-vector and right-chiral FCNC couplings, respectively. To compare with the experimental results we allow these tops/ antitops to decay through leptonic channels only as shown in Figs. \ref{ttFVA} and \ref{ttFVR}.
  However, the non observability of same-sign dilepton events at the hadronic colliders restricts the parameter space of the model generating such events. In Fig. \ref{tevcont} we depict the constrain on the left and right-chiral couplings from the observed cross section of $\sigma ( p\bar p \to tt + \bar t \bar t)\times {BR} [W \to l\nu]^2\le 0.54$ pb for the combined signature of the same-sign top-pair and same-sign anti-top pair production and then  decaying through the respective leptonic channels \cite{CDF:10466}. CMS \cite{Chatrchyan:2011dk} and ATLAS \cite{Aad:2012bb} data  constrains the parameter space from   $\sigma ( p p \to t t)$ only with observed cross section $\le 17$ pb and $\le 1.7$ pb, respectively. Figs. \ref{cmscont} and \ref{atlascont} provides the 95 $\%$ confidence level exclusion contours in  the two dimensional plane of  flavor violating chiral couplings $g^{ut}_L$  and  $g^{ut}_R$ for a given color-octet mass corresponding to the observed data from CMS and ATLAS, respectively.
\par We observe that these contours severely narrows the allowed parameter space contributing to the top-antitop pair production and generating the positive \afbt.
\section{Consistency with the $t\bar t$ and dijet production at the LHC}
\label{lhc}
In the previous section we investigated the parameter region for color-octet vector bosons and found constraints on the masses and couplings by taking total top quark pair production at the Tevatron including single top quark production and same-sign top quark production cross sections. We then analyzed the exclusion region of the parameters $via$ same-sign top quark pair production at the Tevatron as well as at the LHC. Additionally, fitted data for $\afbt$ at the Tevatron  restricted the parameter space further and gave some favorable parameters in the model considered in this article. Further investigations can then lead to exclusion/acceptance of the parameters by studying transverse momentum of the final states and invariant mass differential distributions for top-pair and dijet production cross sections. The other observables like S, T parameters and the Z decay width effects further constraint the color-octet vector boson model but these studies are beyond the scope of this article, detail studies can be found in \cite{Haisch:2011up}. In this section we investigate the consistency of the favorable parameters found at the Tevatron by studying the cross section, charge asymmetry, spin-correlation, invariant mass differential distributions for top quark pair production and dijets production data at the LHC.
  \subsection{$t\bar t$ production and \mttb distribution }
\label{ttbarlhc}
\par The color-octet vector bosons not only contribute to the $t\bar t$ production cross section both at the Tevatron and the LHC but modify its shape as a function of invariant mass $m_{t\bar t}$ as well. The $t\bar t$ resonance searches will put constraints on the resonance mass. For $t\bar t$ production, the decay width of the color-octet vector boson is relevant for $M_{V_8} > 2 m_t$ so that the top-pair can be produced at resonance. Earlier studies incorporated rather large values of the decay width e.g. $\Gamma_{V_8} \approx 0.1-0.2 M_{V_8}$. We have on the other hand used the width calculated for the parameters employed in our study of the cross sections. 
\par
\begin{figure*}[!ht]
  \centering
  \subfloat[FC]{\label{fig:xsecfc}\includegraphics[width=0.5\textwidth]{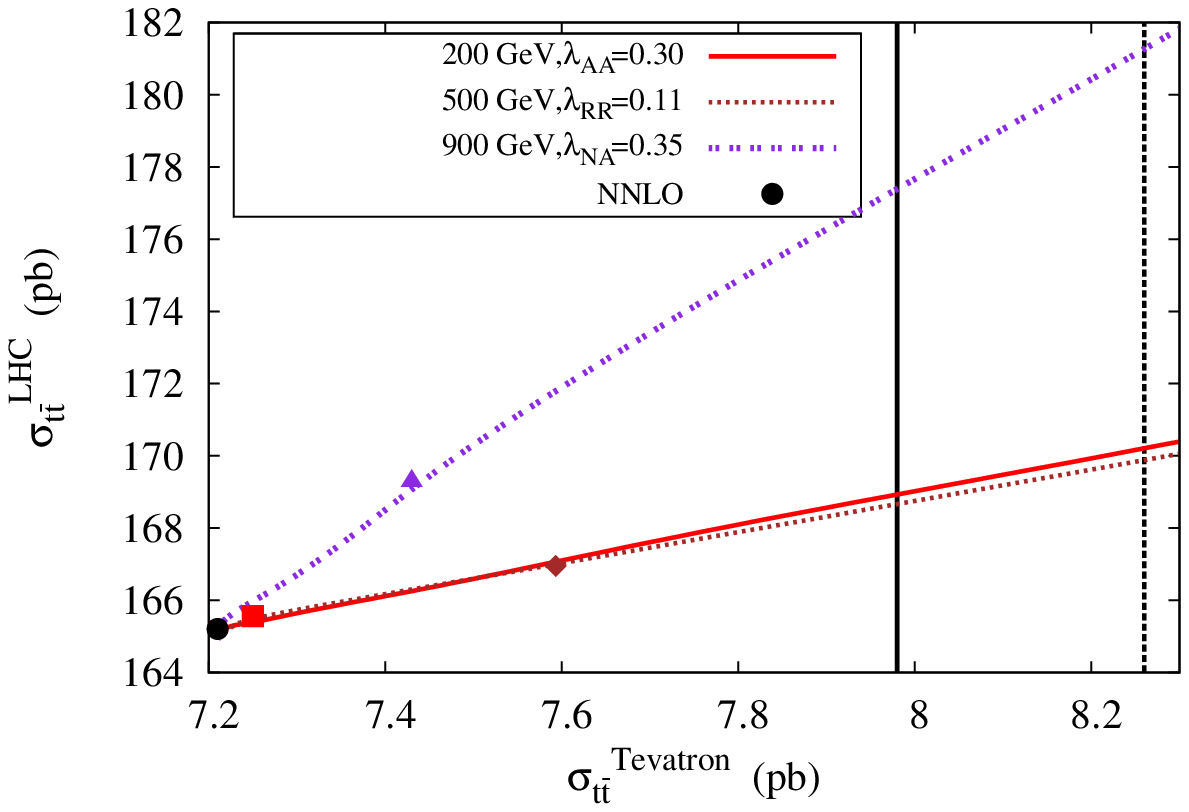}}
  \subfloat[FV]{\label{fig:xsecfv}\includegraphics[width=0.5\textwidth]{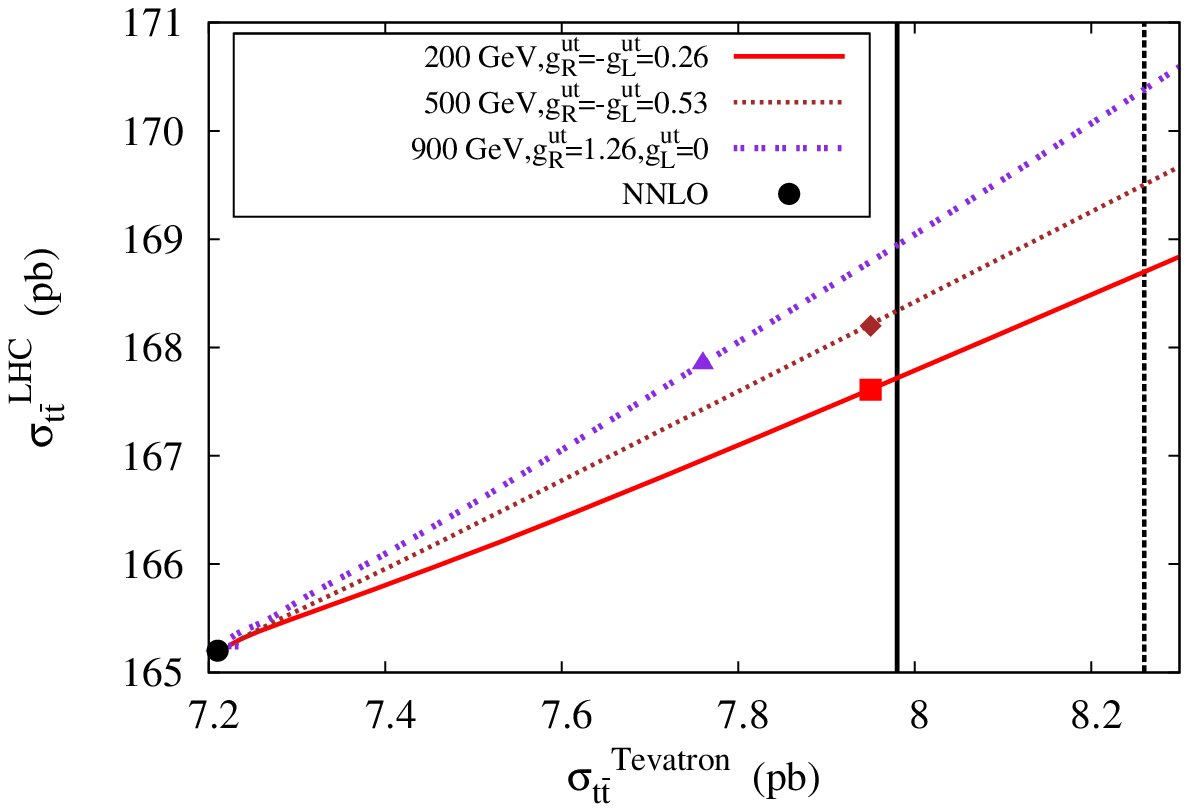}}
\caption{\small \em { The $x$ and $y$ coordinates on the curve depicts the production cross sections $\sigma_{t\bar t}$ at the Tevatron and the LHC, respectively corresponding to a fixed value of the coupling and the resonant mass for (a) flavor conserving and (b) flavor violating  cases. The highlighted colored data points corresponds to the focus points mentioned in Tables \ref{afbmttdata_FC1} and \ref{afbmttdata_FV}.  Vertical lines depicts the 1-$\sigma$ boundary for CDF \cite{{cdf:9913}} and D$\O$ \cite{Abazov:2011cq}, respectively as given in the text. Black point  corresponds to the SM NNLO approximate value of  at the LHC (165.2 pb) and  the Tevatron (7.2 pb) \cite{Kidonakis:2012db}.  }}
\label{xsec}
\end{figure*}

 We  explore the subspace of the parameters of the color-octets which can explain the observed forward-backward asymmetry at the Tevatron  as discussed  in section \ref{chi2section} and thus attempt to examine the admissibility of these data points with respect to the recently observed the LHC data. As a first step, we compare the  cross section of the Tevatron and the LHC for the $t\bar t$ production corresponding to the same values of the couplings with a given   resonant mass and the nature of the exchange current. This is shown in Fig. \ref{fig:xsecfc}
where the $t\bar t$ cross section induced by the flavor conserving couplings within the allowed experimental limits for Tevatron and the LHC are depicted on 
the $x$ and $y$ axes, respectively. To highlight the chosen data-points we mark the focus points in the figure along with prediction from SM.  The two vertical lines in the figure corresponds to the 1-$\sigma$ boundaries of the maximum allowed $\sigma_{t\bar t}$ at the Tevatron \cite{cdf:9913,Abazov:2011cq}. It is clear from the figure that the observed $\sigma_{t\bar t}$ at the Tevatron completely lies within the experimentally observed range of $\sigma_{t\bar t}$ at the LHC \cite{Comb:xsec}. In Figure \ref{fig:xsecfv} the same is plotted for the flavor violation case.
\par
\begin{figure*}[!ht]
  \centering
  \subfloat[]{\label{fig:pt}\includegraphics[width=0.5\textwidth]{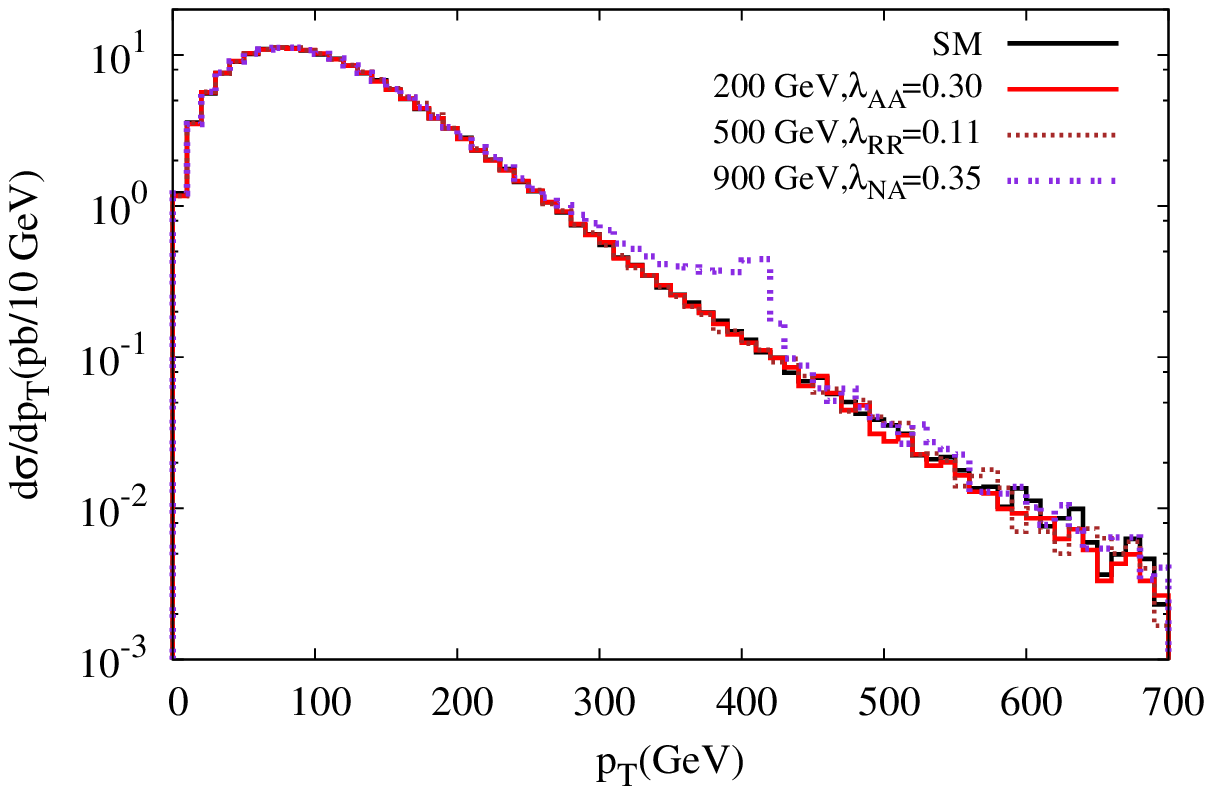}}
  \subfloat[]{\label{fig:mtt}\includegraphics[width=0.5\textwidth]{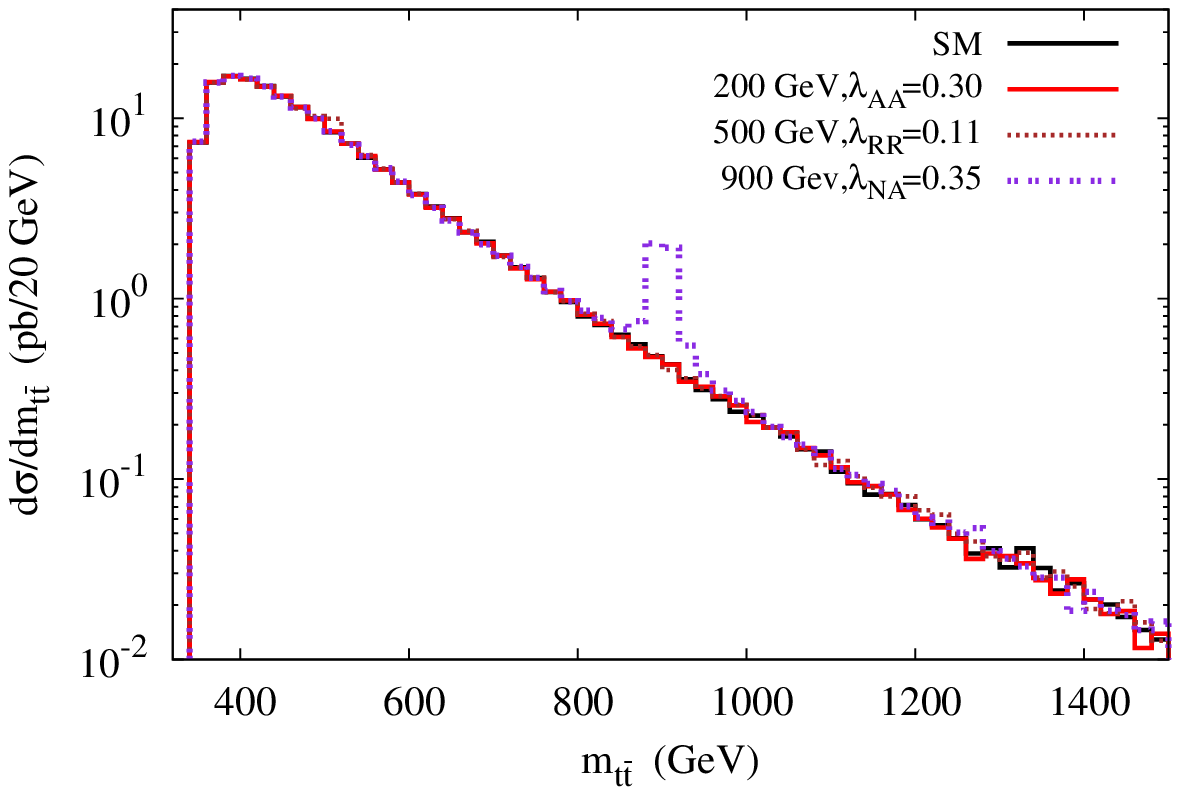}}
\caption{\small \em{ The differential distribution of (a) transverse momentum of the top $p_T$ with bin width 10 GeV and (b)  top-pair invariant mass with bin width 20 GeV at LHC with $\sqrt s = 7$ TeV corresponding to the SM and the flavor conserving focus points mentioned  in Table \ref{afbmttdata_FC1}.  }}
\label{mttfc}
\end{figure*}

 Unlike the Tevatron, the LHC possess a rich potential for the new physics resonant searches both for the threshold and the boosted production of the unlike sign top-pairs. The differential cross-sections for the $t\bar t$ production are studied in the Ref. \cite{:2012qk} along with $Z^\prime$ and other new physics resonant searches in Ref. \cite{:2012rq}. No significant deviations from the SM are observed. We investigated  the one dimensional distribution of the  transverse momentum of the top and the invariant mass of the top-pairs. Any large deviation that might occur due to the color-octet contribution in these distributions will exclude the corresponding resonant mass and the couplings. In Figure \ref{fig:pt} and \ref{fig:mtt}   we show $p_T$ and $m_{t\bar t}$ distribution  at the LHC for the preferred values of parameters required to obtain the experimentally observed $\afbt$. The $p_T$ distribution as well as the $t\bar t$ invariant mass   distribution for 7 TeV the LHC data show a clear narrow resonance for $M_{V_8}$ = 900 GeV on top of the SM background. Since ATLAS and CMS  \cite{:2012qk,:2012rq} have not yet observed any kind of such resonance effect for the  $p_T$ and $\mttb $ distribution, the  octet vector boson model with $M_{V_8}$=900 GeV can be excluded with the coupling constant of 0.35. 
\par In the flavor violation case $t\bar t$ production proceeds through $V_8^0$ exchange in the $t$-channel and therefore no resonance effect is expected.
\subsection{Charge Asymmetry and spin-correlation}
We found that the $t\bar t$ production data at the Tevatron shows a relatively large $\afbt$, the LHC data on the other hand exhibits a small 'charge asymmetry' $A_C$ given by
\begin{eqnarray}
A_C = \frac{N(\Delta \left\vert y
\right\vert > 0) - N(\Delta \left\vert y
\right\vert < 0)}{N(\Delta \left\vert y
\right\vert > 0) + N(\Delta \left\vert y
\right\vert < 0)},
\end{eqnarray}
where $\Delta \left\vert y \right\vert = \left\vert y_t\right\vert-\left\vert y_{\bar t}\right\vert $ is the difference between absolute rapidities of the top and antitop quarks. In the SM both the asymmetries $\afbt$ and $A_C$ are generated at the NLO of QCD. The most recent results from the CMS \cite{Chatrchyan:2011hk} and the ATLAS \cite{ATLAS:2012an} collaborations at the LHC give $A_C^{ATLAS} = -1.9 \pm 2.8 (stat.) \pm 2.4 (syst.) \%$ and $A_C^{CMS} = -1.3 \pm 2.8 (stat.)^{+2.9}_{-3.1} (syst.) \%$. These values are consistent with the SM prediction, $A_C = 1.15 \pm 0.06 \%$ within the experimental uncertainty \cite{Kuhn:2011ri}.
\par
\begin{figure*}[!ht]
\centering
 \subfloat[FC]{\label{fig:afbacfc}\includegraphics[width=0.5\textwidth]{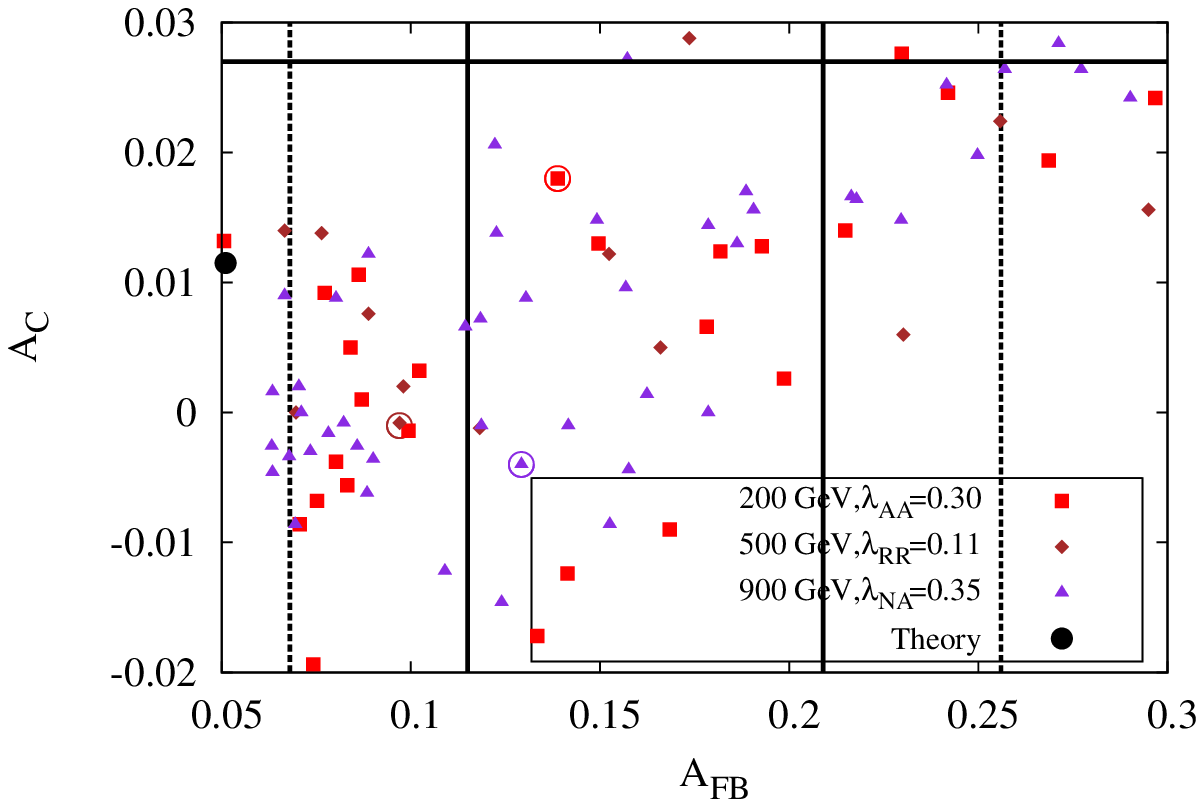}}
 \subfloat[FV]{\label{fig:afbacfv}\includegraphics[width=0.5\textwidth]{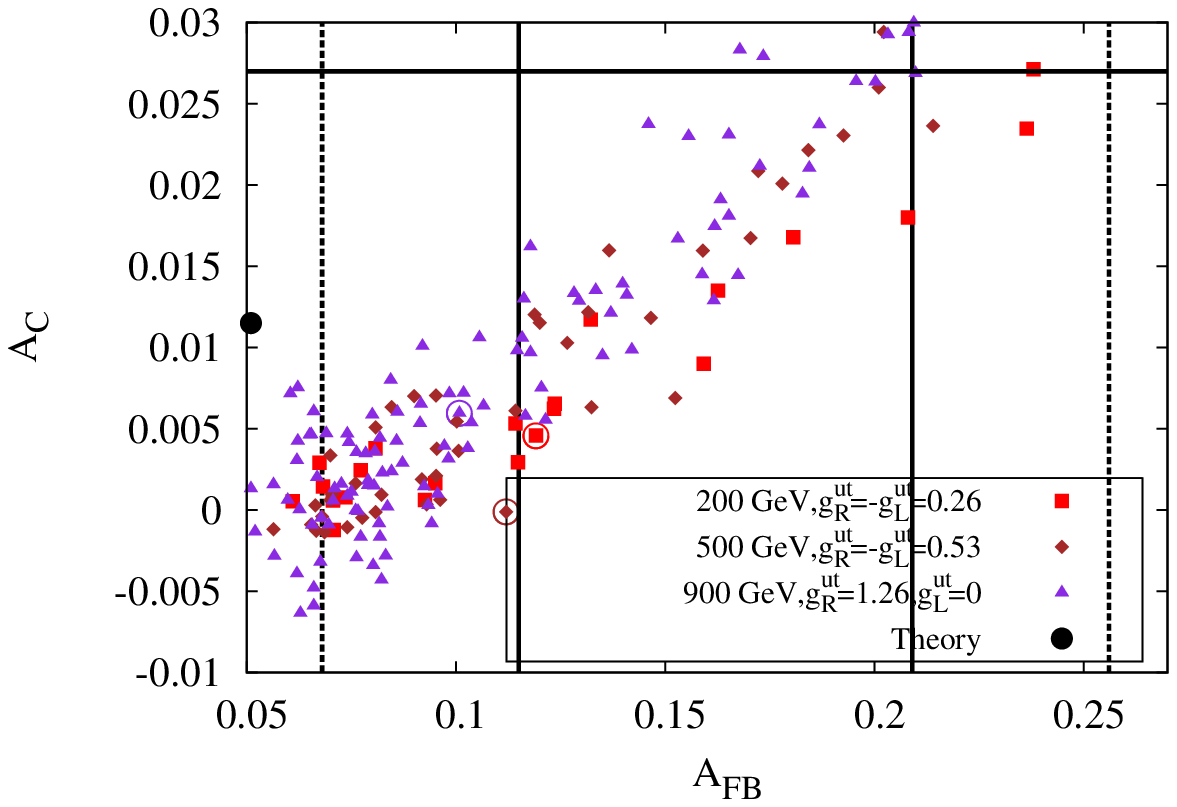}}
\caption{\small \em {The $x$ and $y$ coordinates of the points in the scatter plot depict the \afbt at the Tevatron and $A_C$ at the LHC, respectively corresponding to a fixed value of the coupling and the resonant mass for (a) flavor conserving and (b) flavor violating  cases. The encircled colored data points corresponds to the focus points mentioned in Tables \ref{afbmttdata_FC1} and \ref{afbmttdata_FV}.  The inner and outer pair of vertical lines are 1- and 2-$\sigma$ boundaries of the experimental forward-backward asymmetry at the Tevatron \cite{CDF:10807} while horizontal line shows the 1-$\sigma$ boundary of the experimental charge asymmetry at the LHC \cite{Chatrchyan:2011hk}.}}
\label{afbac}
\end{figure*}

 Both the asymmetries depend on the coupling of color vector bosons with light and top quarks i.e. on the $q\bar q \to t \bar t$ process. We provide a scatter plot in Figs. \ref{fig:afbacfc} and \ref{fig:afbacfv} in order to study the correlation between \afbt at the Tevatron and $A_C$ at the LHC for three different vector boson masses corresponding to FC and FV interactions. In these figures $x$  and $y$ coordinates depict the \afbt in the Tevatron and  $A_C$ at the LHC, respectively for a fixed value of the resonant mass and the coupling. The range of the couplings on the $x$ axis are chosen such that it generates the appropriate \afbt observed in the $t\bar t$ production at the Tevatron. The favorable points mentioned in Tables \ref{afbmttdata_FC1} and \ref{afbmttdata_FV} are encircled in the figure. 
The inner and outer pair of vertical lines corresponds to the  1- and 2-$\sigma$ boundaries of the experimental forward-backward asymmetry at the Tevatron \cite{CDF:10807} while horizontal line shows the 1-$\sigma$ boundary of the experimental charge asymmetry at the LHC \cite{Chatrchyan:2011hk}.
\par
\begin{figure*}[!ht]
  \centering
  \subfloat[FC]{\label{fig:spcrfc}\includegraphics[width=0.5\textwidth]{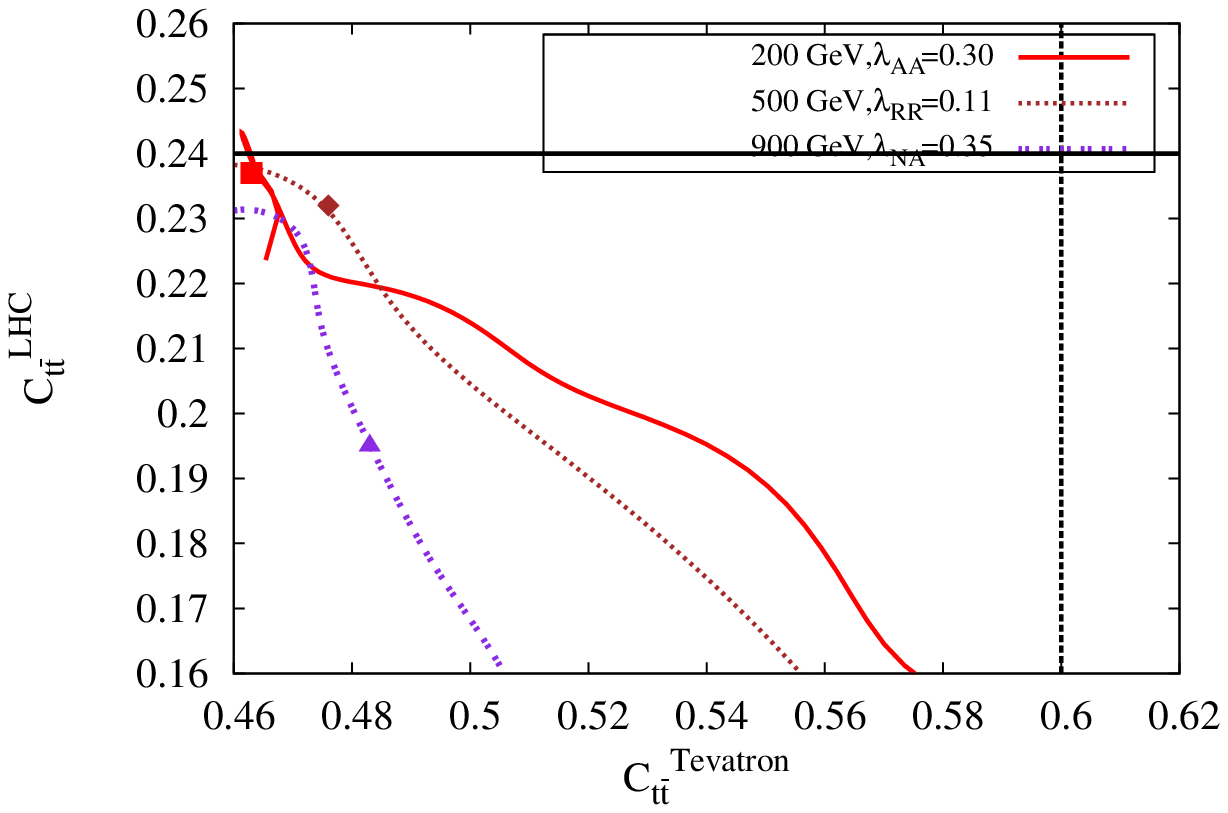}}
  \subfloat[FV]{\label{fig:spcrfv}\includegraphics[width=0.5\textwidth]{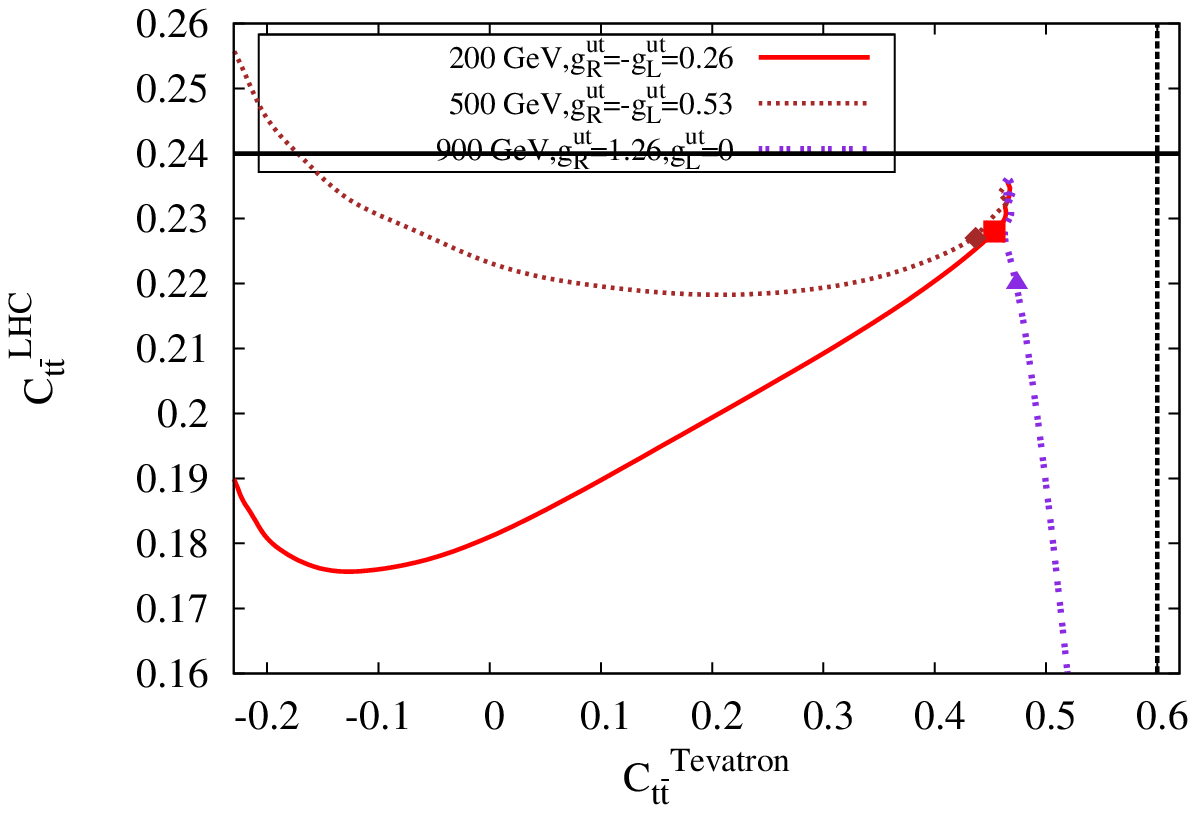}}
\caption{\small \em { The $x$ and $y$ coordinates on the curve depicts the spin-correlation coefficients \spincorr at the Tevatron and the LHC, respectively corresponding to a fixed value of the coupling and the resonant mass for (a) flavor conserving and (b) flavor violating  cases. The highlighted colored data points corresponds to the focus points mentioned in Tables \ref{afbmttdata_FC1} and \ref{afbmttdata_FV}.  
The horizontal and vertical lines corresponds to the central values  of \spincorr in the helicity basis at  the LHC \cite{CMS:12004} and the Tevatron \cite{Aaltonen:2010nz}, respectively.}}
\label{spcr}
\end{figure*}

 The LHC provides a unique platform to study the spin and polarization distribution of top and antitop for both the threshold and boosted events. We study and compare the contribution of the color-octets to the spin-correlation coefficient \spincorr. To constrain the parameter space we plot the curves in the Figs. \ref{fig:spcrfc} and \ref{fig:spcrfv} for the flavor conserving and violating cases, respectively.
Each point $(x,y)$ on the curve estimates the contribution to \spincorr at the Tevatron and the LHC, respectively corresponding to a fixed value of the resonant mass and the coupling. The vertical and the horizontal line depicts the experimental central values of \spincorr at the Tevatron and the LHC, respectively.
We observe that our focus points which are highlighted in the figure are in good agreement with the experimental values within $1\sigma$ error estimations.  
\subsection{Dijet resonance searches}
\label{dijetlhcsection}
Recent searches for dijet resonances in 7 TeV $pp$ collisions at ATLAS and CMS \cite{Chatrchyan:2011ns} provides exclusion limits for axigluon/coloron masses. CMS data exclude axigluons and colorons with mass less than 2.47 TeV at 95 $\%$ confidence level while ATLAS exclusion limits is between 0.60 and 2.10 TeV for the same resonances.\\   
The color-octet vector bosons produced from the $q\bar q$ initial state will give rise to dijet events by decaying into the $q\bar q$ states. The dijet cross section thus depends on the same parameters namely $M_{V_8}, \Gamma_{V_8}, g_L^q $ and $g_R^q$ as the other observables considered in the study. 
Whereas the $t\bar t$ cross section depends on the product of the couplings of the color-octet vector bosons with light and top quarks $g^q_i$ $g^t_j (i,j=L,R)$, the dijet cross section depends only on $(g^q_{L/R})^2$ and therefore can provide a more stringent bounds on these couplings from the direct resonant searches. As discussed in the Introduction, CMS and ATLAS collaborations have performed a search of narrow dijet resonances.
 The dijet resonance searches are based on the narrow width approximation and therefore they do not constraint vector bosons with large width. 
\par We study the $p_T$ distribution of the jet with highest $p_T$ and invariant mass distribution of  the two highest $p_T$ jets in SM and then compare the distribution with the specific choices of resonant mass along with their couplings as mentioned in Table \ref{afbmttdata_FC1} for flavor conserving case. It is to be noted that there is no contribution from flavor violating couplings involving the first and third generation quarks.
\par
\begin{figure*}[!ht]
  \centering
  \subfloat[]{\label{fig:ptj}\includegraphics[width=0.5\textwidth]{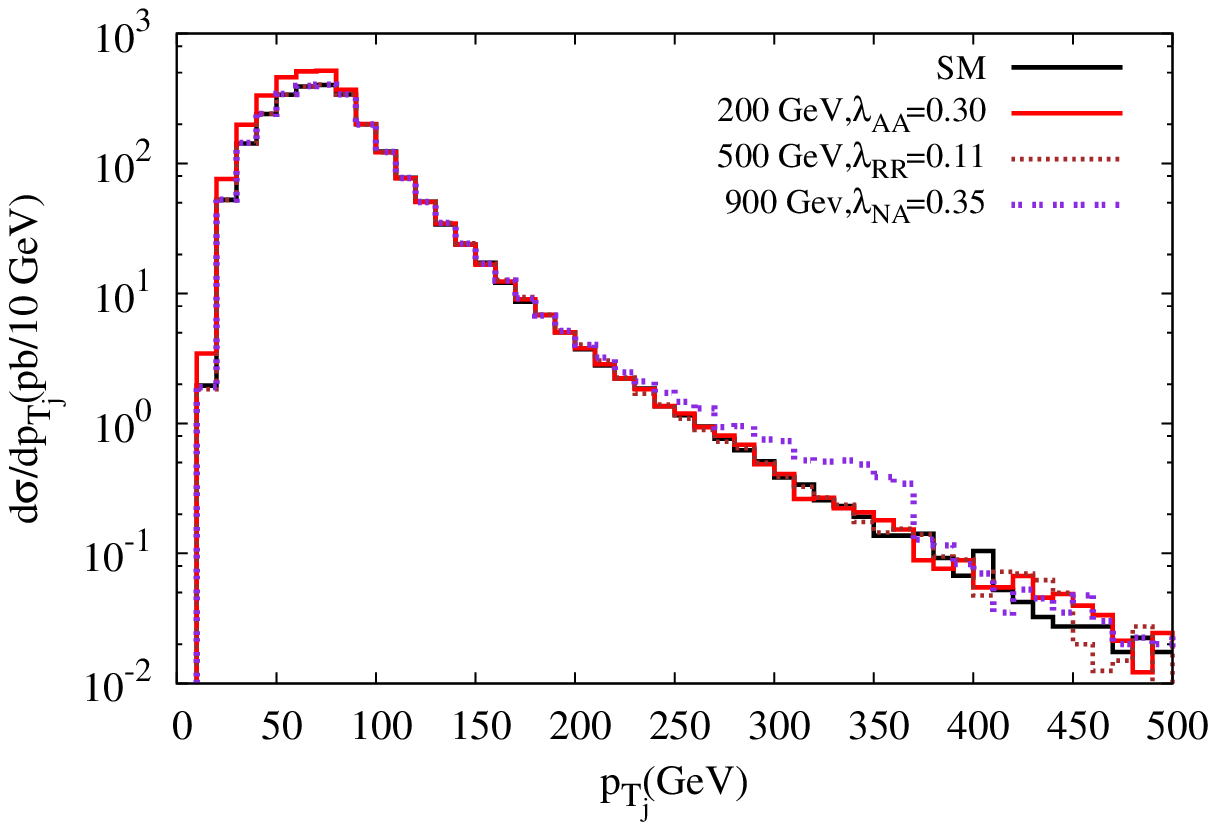}}
  \subfloat[]{\label{fig:mjj}\includegraphics[width=0.5\textwidth]{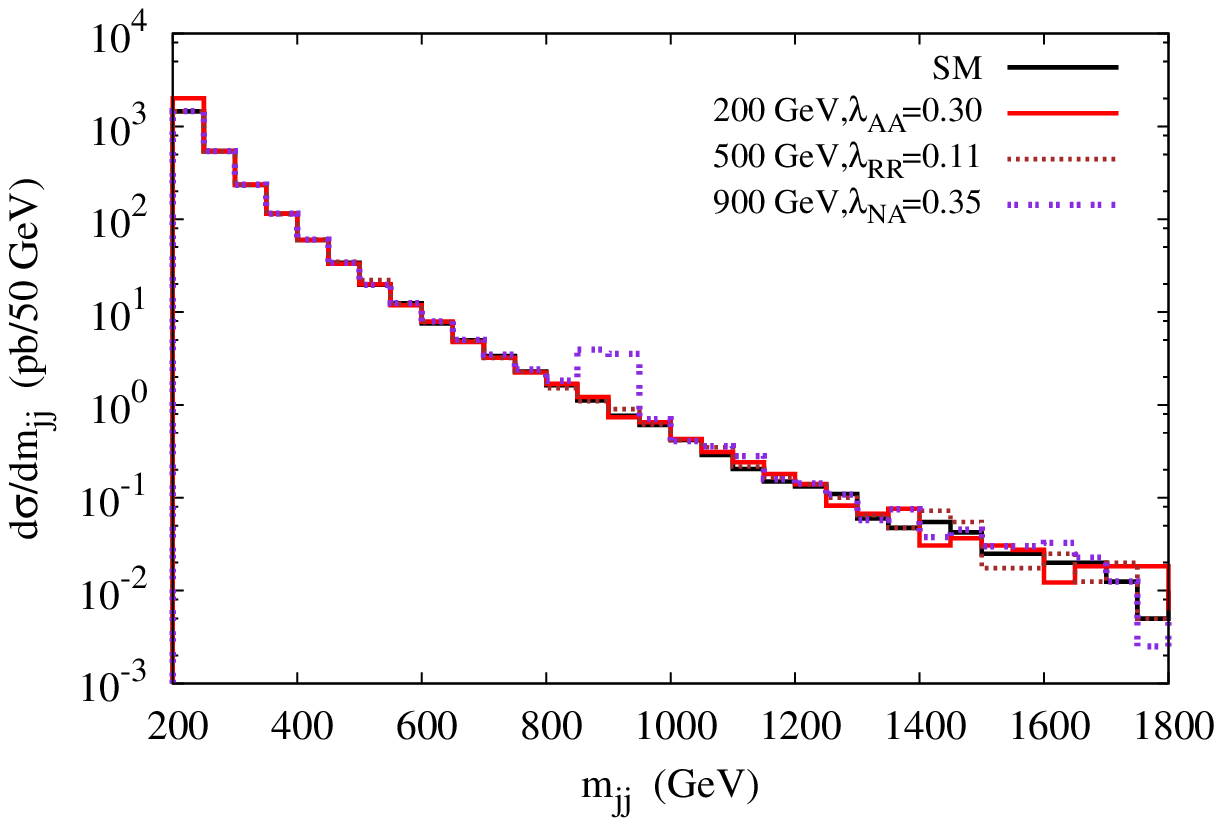}}
\caption{\small \em{ The differential distribution of (a) transverse momentum of the highest  $p_T$ jet with bin width 10 GeV and (b)  dijet invariant mass $m_{jj}$ of the two highest $p_T$ jets with bin width 50 GeV at the LHC with $\sqrt s = 7$ TeV corresponding to the SM and the focus points mentioned  in Table \ref{afbmttdata_FC1} for flavor conserving case.}}
\label{dijet}
\end{figure*}

  We have imposed the standard acceptance cuts for these distributions. The minimum dijet invariant mass $m_{jj}$  is taken to be 200 GeV along with the required pseudorapidity separation $\left\vert \Delta \eta \right \vert \le 1.3$ and both jets satisfying  $\left\vert \eta\right\vert \le 2.5$. We show the $p_T$ distribution  in  Fig.  \ref{fig:ptj} with the bin width 10 GeV and the invariant mass distribution of the the two highest $p_T$ jets in Fig. \ref{fig:mjj} with bin width 50 GeV.
\par The $d\sigma/dm_{jj}$ distribution  are all in good agreement with the SM QCD background within the experimental error bars except for the case of  $M_{V_8}$ = 900 GeV. Therefore $M_{V_8}$ = 900 GeV resonance can be excluded based on the resonant searches not only from the $t\bar t$ production  but also from the dijet searches as well. 
\section{Summary and Conclusion}
\label{sec:summary}
We have revisited and extended the analysis of the color-octet vector boson model in  the top sector at the Tevatron and the LHC for FC and FV couplings. We have considered the effect of decay width of the color-octet vector bosons  throughout our analysis and  configured our study to constrain the parameter space of the  model from the observed differential distribution of the forward-backward asymmetry with 8.7 fb$^{-1}$ full data set of CDF collaboration at teh Tevatron  \cite{CDF:10807}   and charge asymmetry data set from the LHC at 7 TeV \cite{Chatrchyan:2011hk}, which was missing in the literature \cite{Grinstein:2011dz,axig,Wang:2011hc}. We have made an attempt to find a parameter space which is also consistent with the spin-correlation observation at the Tevatron \cite{Aaltonen:2010nz,CDFNote10719}, single top production at the Tevatron \cite{CDF:10793}, same-sign top production at the Tevatron \cite{CDF:10466} and the LHC \cite{Chatrchyan:2011dk,Aad:2012bb} and dijet invariant mass distribution at the LHC \cite{Chatrchyan:2011ns}.
The observed features of our analysis  for top quark physics at the Tevatron and the LHC are enumerated as follows:
 \begin{enumerate}
\item  We notice  appreciable contribution to \afbt and spin-correlation coefficient from the axial current and the right-handed chiral current without  transgressing the production cross-sections within experimentally allowed one sigma region. 
\item We scanned the parameter space of the model to explain the anomaly observed in one dimensional \mttb and \dyy distributions of \afbt. We predict few focus points based on the $\chi^2$ minimization at $\chi^2_{\rm min.}$ which are likely to satisfy these constraints. This is summarized in Tables \ref{afbmttdata_FC1} \& \ref{afbdydata_FC} and in Tables \ref{afbmttdata_FV} \& \ref{afbdydata_FV} for the flavor conserving and flavor changing neutral currents, respectively. \mttb distribution of \afbt corresponding to  focus points  are also depicted in  Figs. \ref{fig:TP_FC_mtt_A},\ref{fig:TP_FC_mtt_NA}, \ref{fig:TP_FC_mtt_R} and \ref{fig:TP_FV_mtt_A1}, \ref{fig:TP_FV_mtt_A2}, \ref{fig:TP_FV_mtt_R}  for FC and FV couplings, respectively. Similarly the agreement with respect to \dyy distribution is shown in Figs.  \ref{fig:TP_FC_dy_A}, \ref{fig:TP_FC_dy_NA}, \ref{fig:TP_FC_dy_R}  and  \ref{fig:TP_FV_dy_A1}, \ref{fig:TP_FV_dy_A2}, \ref{fig:TP_FV_dy_R} for flavor conserving and violating cases, respectively. 
\item  We  verified that the top quark couplings corresponding to these focus points evade the lower bounds on the chiral couplings required  to form  top quark condensates \cite{Hill:1991at}.
\item Single top quark production through massive color charged vector boson is studied for the $s$ and $t$ channel separately with distinguishable final states as in SM. We observe that a large  parameter region is allowed by the one and two sigma bands corresponding to $s$ and $t$ channels, respectively from CDF \cite{CDF:10793}. Since we have performed our analysis with $\left\vert V_{tb}\right\vert ^2=1$, we need to be careful about the interplay of new physics parameters and allowed deviation for $\left\vert V_{tb}\right\vert$ from unity.
\item The introduction of flavor changing neutral current for the $t\bar t$ production also induces the single top production in $s$ and $t$ channels with the same final states. So, we compared our results with the observed  combined cross-sections from $s$ and $t$ channels at the Tevatron.  We find that the cross section of the single top quark production is comparatively more sensitive to the new physics couplings in comparison to the $t\bar t$ model. We are able to constrain the product of FC and FV couplings of the neutral current from this process.
\par  As discussed in Sec. \ref{sec:model}, the inclusion of nonuniversal couplings in the up quark sector help the product of  flavor violating couplings   to evade the low energy stringent  bounds from $B$ and $D$ physics. The benchmark points obtained by us  are consistent with all the observables discussed above and are in broad agreement with those obtained in Ref. \cite{Grinstein:2011dz}.
\item Consistency of the color-octet  vector boson model with respect to focus points are examined in the light of recent the LHC data in Sec. \ref{lhc}. We probed the admissibility of the constrained parameter space at the LHC which explained the required \afbt at the Tevatron as mentioned in Sec. \ref{chi2section}. We observe  that the focus points do not transgress the cross section of the top-pair production \cite{Comb:xsec} as well as the measured charge asymmetry \cite{Chatrchyan:2011hk} and spin-correlation \cite{CMS:12004} at the LHC.
\par To  have more insight on the implication of the new physics parameter space we study the $p_T$ and $ m_{t\bar t}$ distributions of the top-pair. We find that all our FC focus points as mentioned in Table \ref{afbmttdata_FC1} except for the higher resonance mass of 900 GeV are consistent with the observations at the LHC.
\par The correlation study of the \afbt at the Tevatron and $A_C$ at the LHC in Figs. \ref{fig:afbacfc} and \ref{fig:afbacfv} shows that the large \afbt at the Tevatron be accommodated by the recent observations at the LHC within 2 $\sigma$ limit. The spin-correlation coefficient predicted with constrained parameter space of the color excited states at the LHC are also ‭found to lie within the one $\sigma $ limit of the experimental values \cite{Aaltonen:2010nz,CMS:12004}.
\item The ballpark estimate of the production cross-sections of the color exotics involving the light quark color-octet vector interactions are given in Table \ref{productioncrosssection}. Since the strongest bound for the light quark couplings to the color exotics comes from the dijet searches, we studied the transverse momentum and  invariant dijet mass distribution at the LHC corresponding to the parameter space which generated a large \afbt at the Tevatron. We observe  an appreciable deviation for the color-octet at 900 GeV similar to that observed for $m_{t\bar t}$ distribution.

\item We also studied the production cross section of same-sign top and anti top-pairs {\it via} FV couplings at the Tevatron. We imposed the constraints of non-observability of large same-sign dilepton events at the Tevatron and provided the 95 \% exclusion contours in the Fig.~\ref{tevcont} on the plane of chiral couplings. Exclusion contours at 95 \% are also computed from recent results at CMS and ATLAS for the same-sign top production only which are depicted in Figs.~\ref{cmscont} and ~\ref{atlascont}, respectively. The constraints from the LHC restrict the allowed parameter space of FV to a narrow allowed region. We observe that all focus points except one ( 900 GeV with $g_{R}^{ut}=1.26$) corresponding to FV couplings as shown in Fig. \ref{atlascont} lies within this narrow allowed region.
\end{enumerate}
Our analysis for the top quark physics in  vector color-octet model  based on the recent observations at the Tevatron and the LHC   has shrunk the allowed parameter space to a great extent. We  propose four focus data points  (two each from flavor conserving and violating couplings) which can explain the \afbt anomaly at the Tevatron and are also consistent with  the $t\bar t$,  same-sign top and dijet production cross-sections and associated observables at the LHC. 
\section*{Acknowledgements}
The authors would like to thank Amitabha Mukherjee, Debajyoti Choudhury, Mamta Dahiya and Rashidul Islam for fruitful discussions. SD and MK likes to thank Fabio Maltoni and Rikkert Frederix for illuminating discussions on MadGraph and single top analysis. We acknowledge the partial support from DST, India under grant SR/S2/HEP-12/2006. SD and AG would like to acknowledge the UGC research award and CSIR(ES) award, respectively for the partial financial support . We also thank RECAPP, HRI for local hospitality where the part of the work was done.
\appendix
\def\theequation{\thesection.\arabic{equation}}
\setcounter{equation}{0}
\section{Computation of Helicity Amplitudes}
\subsection{\bf  Helicity amplitudes for  $q \bar q \to t \bar t$ via flavor conserving  vector octets}
All couplings are in units of $g_s$. $g_L^q=g_L^t=g_R^q= g_R^t=1$ for SM.
\begin{eqnarray}
{\cal  M}_{+-\pm\pm}^{V_8^0} &&= {\cal F}_s g_R^q (g_L^t+ g_R^t)\frac{\hat s}{2}\sqrt{1-\beta^2} \, \sin\theta \cr &&=  {\cal  M}_{-+\mp\mp}^{V_8}(L\leftrightarrow R,\,\, R\leftrightarrow L ) \label{A1} \\ 
{\cal  M}_{+-\pm\mp}^{V_8^0} &&= {\cal F}_s g_R^q [(g_L^t + g_R^t) \mp \beta (g_L^t - g_R^t)]\frac{\hat s}{2} (1 \pm \cos\theta) \cr &&={\cal  M}_{-+\mp\pm}^{V_8}(L\leftrightarrow R,\,\, R\leftrightarrow L ) \label{A2} \\
\text{where,} \, {\cal F}_s &&= \frac{g_s^2 T^aT^a}{(\hat s - M_{V_8^0}^2) + i M_{V_8^0} \Gamma_{V_8^0}} \,  \nonumber
\end{eqnarray}
and $T^a$ is the SU(3) matrices.
\subsection {\bf  Helicity amplitudes for  $q \bar q \to t \bar t$ via flavor violating vector octets}
\begin{eqnarray}
{\cal  M}_{++\pm\pm}^{V_8} =&& {\cal F}_t g_R^{ut} g_L^{ut} {\hat s} (1\pm\beta) \cr =&& {\cal  M}_{--\mp\mp}^{V_8}(L\leftrightarrow R,\,\, R\leftrightarrow L ) \label{A3} \\ 
{\cal  M}_{+-\pm\pm}^{V_8} =&& {\cal F}_t {g_R^{ut}}^2 \frac{\hat s}{2} \sqrt{1-\beta^2}\, \sin\theta \cr =&& {\cal  M}_{-+\pm\pm}^{V_8} (L\leftrightarrow R,\,\, R\leftrightarrow L ) \label{A4} \\ 
{\cal  M}_{+-\pm\mp}^{V_8} =&& {\cal F}_t {g_R^{ut}}^2 \frac{\hat s}{2} (1\pm\beta) (1\pm\cos\theta)  \cr =&& {\cal  M}_{-+\mp\pm}^{V_8}(L\leftrightarrow R,\,\, R\leftrightarrow L ) \label{A5} \\
\text{where,} \, {\cal F}_t &&= \frac{g_s^2 T^aT^a}{(\hat t - M_{V_8^0}^2) + i M_{V_8^0} \Gamma_{V_8^0}} \,  \nonumber
\end{eqnarray}

%%%

\end{document}